\documentclass[twoside,twocolumn,9pt]{article}
\usepackage{extsizes}
\usepackage[super,sort&compress,comma]{natbib} 
\usepackage[version=3]{mhchem}
\usepackage[left=1.5cm, right=1.5cm, top=1.785cm, bottom=2.0cm]{geometry}
\usepackage{balance}
\usepackage{mathptmx}
\usepackage{sectsty}
\usepackage{graphicx} 
\usepackage{lastpage}
\usepackage[format=plain,justification=justified,singlelinecheck=false,font={stretch=1.125,small,sf},labelfont=bf,labelsep=space]{caption}
\usepackage{float}
\usepackage{fancyhdr}
\usepackage{fnpos}
\usepackage[english]{babel}
\addto{\captionsenglish}{%
  
}
\usepackage{array}
\usepackage{droidsans}
\usepackage{charter}
\usepackage[T1]{fontenc}
\usepackage[usenames,dvipsnames]{xcolor}
\usepackage{setspace}
\usepackage[compact]{titlesec}
\usepackage{hyperref}
\usepackage{mathrsfs}
\usepackage{bm}
\usepackage{color}
\usepackage{braket}
\usepackage[normalem]{ulem}

\usepackage{epstopdf}

\definecolor{cream}{RGB}{222,217,201}

\begin{document}

\pagestyle{fancy}
\thispagestyle{plain}
\fancypagestyle{plain}{
\renewcommand{\headrulewidth}{0pt}
}
\newcommand{\ha}{\hat{a}}
\newcommand{\hb}{\hat{b}}
\newcommand{\hc}{\hat{c}}
\newcommand{\hd}{\hat{d}}

\newcommand{\hA}{\hat{P}}
\newcommand{\hE}{\hat{E}}
\newcommand{\hH}{\hat{H}}
\newcommand{\hR}{\hat{R}}
\newcommand{\hr}{\hat{r}}
\newcommand{\hT}{\hat{T}}

\newcommand{\htau}{\hat{\tau}}
\newcommand{\hP}{\hat{P}}
\newcommand{\hp}{\hat{p}}
\newcommand{\hq}{\hat{q}}
\newcommand{\hV}{\hat{V}}
\newcommand{\hU}{\hat{U}}

\newcommand{\bbf}{\mathbf{f}}

\newcommand{\bA}{\mathbf{A}}
\newcommand{\bB}{\mathbf{B}}
\newcommand{\bb}{\mathbf{b}}
\newcommand{\bD}{\mathbf{D}}
\newcommand{\bE}{\mathbf{E}}
\newcommand{\bd}{\mathbf{d}}
\newcommand{\bJ}{\mathbf{J}}
\newcommand{\bk}{\mathbf{k}}
\newcommand{\bK}{\mathbf{K}}
\newcommand{\bP}{\mathbf{P}}
\newcommand{\br}{\mathbf{r}}
\newcommand{\bR}{\mathbf{R}}
\newcommand{\bu}{\mathbf{u}}
\newcommand{\bX}{\mathbf{X}}
\newcommand{\bx}{\mathbf{x}}
\newcommand{\blambda}{\boldsymbol{\lambda}}
\newcommand{\bg}{\mathbf{g}}

\newcommand{\MF}{\mathcal{F}}
\newcommand{\ML}{\mathcal{L}}
\newcommand{\MM}{\mathcal{M}}
\newcommand{\MP}{\mathcal{P}}
\newcommand{\mq}{\mathcal{Q}}

\newcommand{\YZ}[1]{{\color{blue}YZ:#1}}

\makeFNbottom
\makeatletter
\renewcommand\LARGE{\@setfontsize\LARGE{15pt}{17}}
\renewcommand\Large{\@setfontsize\Large{12pt}{14}}
\renewcommand\large{\@setfontsize\large{10pt}{12}}
\renewcommand\footnotesize{\@setfontsize\footnotesize{7pt}{10}}
\makeatother

\renewcommand{\thefootnote}{\fnsymbol{footnote}}
\renewcommand\footnoterule{\vspace*{1pt}%
\color{cream}\hrule width 3.5in height 0.4pt \color{black}\vspace*{5pt}} 
\setcounter{secnumdepth}{5}

\makeatletter 
\renewcommand\@biblabel[1]{#1}            
\renewcommand\@makefntext[1]%
{\noindent\makebox[0pt][r]{\@thefnmark\,}#1}
\makeatother 
\renewcommand{\figurename}{\small{Fig.}~}
\sectionfont{\sffamily\Large}
\subsectionfont{\normalsize}
\subsubsectionfont{\bf}
\setstretch{1.125} 
\setlength{\skip\footins}{0.8cm}
\setlength{\footnotesep}{0.25cm}
\setlength{\jot}{10pt}
\titlespacing*{\section}{0pt}{4pt}{4pt}
\titlespacing*{\subsection}{0pt}{15pt}{1pt}

\fancyfoot{}
\fancyfoot[RO]{\footnotesize{\sffamily{1--\pageref{LastPage} ~\textbar  \hspace{2pt}\thepage}}}
\fancyfoot[LE]{\footnotesize{\sffamily{\thepage~\textbar\hspace{4.65cm} 1--\pageref{LastPage}}}}
\fancyhead{}
\renewcommand{\headrulewidth}{0pt} 
\renewcommand{\footrulewidth}{0pt}
\setlength{\arrayrulewidth}{1pt}
\setlength{\columnsep}{6.5mm}
\setlength\bibsep{1pt}

\makeatletter 
\newlength{\figrulesep} 
\setlength{\figrulesep}{0.5\textfloatsep} 

\newcommand{\topfigrule}{\vspace*{-1pt}%
\noindent{\color{cream}\rule[-\figrulesep]{\columnwidth}{1.5pt}} }

\newcommand{\botfigrule}{\vspace*{-2pt}%
\noindent{\color{cream}\rule[\figrulesep]{\columnwidth}{1.5pt}} }

\newcommand{\dblfigrule}{\vspace*{-1pt}%
\noindent{\color{cream}\rule[-\figrulesep]{\textwidth}{1.5pt}} }

\makeatother

\twocolumn[
  \begin{@twocolumnfalse}
\vspace{1em}
\sffamily
\begin{tabular}{m{0.5cm} p{17.5cm} }

& \noindent\LARGE{\textbf{Theory and modeling of light-matter interactions in chemistry: current and future$^\dag$}} \\
\vspace{0.3cm} & \vspace{0.3cm} \\

 & \noindent\large{Braden M. Weight{$^{a,b}$}, Xinyang Li\textit{$^{a}$}, and Yu Zhang$^{\ast\ddag}$\textit{$^{a}$}} \\

& \noindent\normalsize{
Light-matter interaction not only plays an instrumental role in characterizing materials' properties via various spectroscopic techniques but also provides a general strategy to manipulate material properties via the design of novel nanostructures. This perspective summarizes recent theoretical advances in modeling light-matter interactions in chemistry, mainly focusing on plasmon and polariton chemistry. The former utilizes the highly localized photon, plasmonic hot electrons, and local heat to drive chemical reactions. In contrast, polariton chemistry modifies the potential energy curvatures of bare electronic systems, and hence their chemistry, via forming light-matter hybrid states, so-called polaritons. The perspective starts with the basic background of light-matter interactions, molecular quantum electrodynamics theory, and the challenges of modeling light-matter interactions in chemistry. Then, the recent advances in modeling plasmon and polariton chemistry are described, and future directions toward multiscale simulations of light-matter interaction-mediated chemistry are discussed.
} \\
\end{tabular}

 \end{@twocolumnfalse} \vspace{0.6cm}
  ]

\renewcommand*\rmdefault{bch}\normalfont\upshape
\rmfamily
\section*{}
\vspace{-1cm}


\footnotetext{\textit{$^{a}$~Theoretical Division, Los Alamos National Laboratory, Los Alamos, NM, 87545, USA. Tel: +1 505 606 2149; E-mail: zhy@lanl.gov}}

\footnotetext{\textit{$^{b}$~Department of Physics and Astronomy, University of Rochester, Rochester. NY, 14627, U.S.A. }}

\section{Introduction}
In 2007, a report compiled for the Office of Science of the Department of Energy (DOE) identified five grand challenges in basic energy science~\cite{ratner2011physicstoday}, including 1) controlling material processes at the level of electrons, 2) designing and perfecting atom- and energy-efficient syntheses of new forms of matter with tailored properties, 3) understanding and controlling the remarkable properties of matter that emerge from complex correlations of atomic and electronic constituents, 4) mastering energy and information on the nanoscale to create new technologies with capabilities rivaling those of living things, and 5) characterizing and controlling matter away—especially far away—from equilibrium. One of the emerging techniques to address (some of) these challenges is light-matter interactions, which can be used to monitor, manipulate, and design materials' properties through the complex interplay between electrons, photons, and phonons.\cite{Anton2019NatRevPhys,Revera2020NatRevPhys}

In conventional chemistry, chemists modify the functionalities of molecules based on different functional groups. For centuries, such a vision has been widely used in chemical synthesis to modify molecules or molecular materials' chemical and physical properties. However, in the current era of technology,\cite{anderson1972more}, previously ``standard'' chemistry, now in a complex electromagnetic (EM) environment, has emergent properties due to multiple new types of couplings between the light and matter degrees of freedom that can be used to control chemistry. Light-matter interactions have been instrumental in many branches of physics, chemistry, materials, and energy science~\cite{novotny2006principles,Revera2020NatRevPhys,weiner2012light}. In most previous applications, the magnitude and scale of the light-matter interaction fall within the so-called weak coupling regime and can usually be treated at the lowest order in quantum electrodynamics via many-body perturbation theory~\cite{mahan2000many}. Such treatments are widely used in different types of spectroscopy techniques~\cite{Stiles2008annurev,Mukamel2000annurev}, optoelectronics~\cite{zhang2014quantum,yam2015multiscale,meng2015multiscale,meng2017multiscale, wang2015quantum}, quantum sensing~\cite{Degen2017rmp}, quantum information~\cite{Flamini_2018}, light harvesting~\cite{zhou2018interlayer,Mirkovic2018cr}, and beyond.

Alternatively, light-matter interactions open multiple new avenues for manipulating matter through novel emerging elementary excitations~\cite{Forndaz:2019rmp}, including plasmons and polaritons. Plasmons (either surface plasmons or localized surface plasmons) are the collective oscillations of conduction electrons in nanostructures, which can be excited when the frequency of external light matches the plasmon resonant energies. Plasmon excitation results in significantly amplified absorption and scattering cross-sections. Moreover, plasmon excitations can overcome the diffraction limit and concentrate the incident light into a highly localized volume, enhancing the electromagnetic (EM) field (or photons) by several orders of magnitude in the near field. The subsequent plasmon decay process generates hot electrons or heat through electron-electron and electron-phonon scatterings at different length and time scales~\cite{Brongersma:2015vd}. Nevertheless, the resulting locally enhanced EM field, hot electrons, and heat can stimulate chemical reactions through various mechanisms~\cite{Zhan:2023tc, Kazuma2019angew}. In fact, plasmon-mediated chemical reactions (namely plasmon chemistry) have become a promising strategy to drive chemical processes over the past decade~\cite{Zhan:2023tc}.

On the other hand, when molecules are collectively and resonantly coupled with plasmon excitations, a new quasiparticle (namely polaritons) can be formed in the strong coupling regime. The term "strong" coupling is relative, meaning that the light-matter coupling is large enough to compete with or overcome dissipation or dephasing (\textit{i.e.}, when the coherent energy exchange between a confined light mode and the quantum matter is faster than the decay and decoherence time scales of each part). Such strong coupling can be achieved when either a plasmonic mode is coupled with a few molecules~\cite{Jeremy2022nl} or many molecules are collectively coupled to a single cavity mode~\cite{Ribeiro2018CS}. In the strong coupling regime, photons and electronic/excitonic excitations in matter become equally important and are strongly coupled on an equally quantized footing. As a consequence, individual "free" particles no longer exist. Instead, the fundamental excitations of the light-matter interacting system are polaritons, which are hybrid light-matter excitations (superpositions of quantized light and matter) (Fig 1a)\cite{Revera2020NatRevPhys} and possess both light and matter characteristics/topologies. Experiments have shown that matter properties can be modified with the formed polaritons, resulting in different photophysics and photochemistry\cite{Ebbesen2016ACR}. Since photon energies and light-matter coupling strengths are relatively tunable through cavity control, light-matter interaction in the strong coupling regime provides a fundamentally new way to manipulate matter properties for various desired applications, including lasing~\cite{Kena-Cohen:2010ue,Kang:ws}, electronics~\cite{Orgiu:2015us}, long-range energy transfer~\cite{Zhong:2017vq, Georgiou2021angwew, Want2021natcomm, Coles:2014wm}, Bose-Einstein condensates~\cite{Dusel:2020wt, Zasedatelev:2019um, Kavokin:2022tn}, chemical reactions~\cite{pavosevic_ClickChem_Arxiv2022, pavosevic_PTQED_JACS2022, pavosevic_PTQED_JACS2022, schafer_shining_NatComm2022, schafer_EmbeddingRadReaction_JPCL2022, Cave1997JCP, MartinezMartinez2017AP, Yang2021JPCL, Climent2020PCCP, Wang2022JPCL, Imperatore2021JCP, Galego2016NC, CamposGonzalezAngulo2019NC, Philbin2022JPCC, Efrima1974CPL, Phuc2020SP, Davidsson2020JCP, Galego2017PRL, Mauro2021PRB, Vurgaftman2020JPCL, Hiura2021C, Hiura2019-rate, weight_abQED_JPCL2023},
internal conversion~\cite{Curchod2016JCP, Wu2022JCP, Avramenko:2020uz}, singlet fission~\cite{Climent2022CRPS, Gu2021JPCL, MartinezMartinez2018JPCL}, manipulation of conical intersection~\cite{Cho2022JACS, Csehi2022NJP, Gu2020JPCL, Gu2020CS, Szidarovszky2018JPCL, Fabri2022JPCL, Natan2016PRL, Hofmann2001CPL, Farag2021PCCP, Arnold:2018vz,Ulusoy:2019vy}, etc.
Nevertheless, in polariton chemistry, light and matter cannot be treated as separate entities, as the strong coupling between them dresses each of them. Consequently, previously developed quantum chemistry methods for treating the electronic structure of matter alone become invalid. Brand new theoretical and modeling capabilities for describing polariton chemistry are required.

\begin{figure}[!htb]
    \centering
    \includegraphics[width=0.4\textwidth]{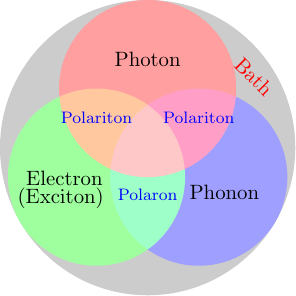}
    \caption{\label{fig:lm1}
    Light-matter interactions are a fundamentally multiscale/multiphysics problem. It involves multiple interactions among electrons, photons, and phonons at different time and length scales. As a result, multiple quasiparticles may be excited due to the light-matter interactions, including exciton (due to electron-hole interaction), (excitonic/vibrational) polariton (due to coupling between photon and exciton/phonon), polaron (due to electron-phonon coupling), and polaron-polaritons. In addition, all quantum systems are fundamentally open systems, leading to dissipation/dephasing due to the bath degrees of freedom (DOFs).
    }
\end{figure}

Despite the attractive applications of strong light-matter interaction, there are many open questions and fundamental problems about the mechanisms of physics and chemistry mediated by the strong light-matter interaction. The experimental progress should be complemented by theoretical advances that can offer atomistic insights into the problems that are not accessible by experiments alone. Progress in understanding the coupling between photons and elementary quasiparticles (plasmons, phonons, and excitons) in materials requires a generalized treatment of photons as one of the core DOFs in light-matter interactions. Unfortunately, the strong light-matter interaction is fundamentally a multiscale and multiphysics problem that involves multiple interactions between many DOFs and their complex interplay with environments across different spatial and time scales. The traditional perturbation methods in the weak coupling regime are not applicable to the strong coupling regime, making it urgent to develop new theoretical and modeling techniques, especially multiscale methods, to understand and ultimately predict strong light-matter interaction-mediated physics and chemistry.

This perspective reviews recent theoretical efforts toward understanding the underlying mechanisms of plasmon and polariton chemistry due to the complex light-matter interactions and discusses our thoughts on future work. The article is structured as follows: Section~\ref{sec:basics} introduces the framework and basic mathematical structure of molecular quantum electrodynamics theory (QED), Section~\ref{SEC:PLASMON} explores the theoretical development in simulating plasmonic cavities and their strong interactions with molecules, Section~\ref{SEC:POLARITON} surveys recent progress in polariton chemistry in Fabry-Pérot-like cavities, and Section~\ref{SEC:CONCLUSION} concludes the discussion and provides a final perspective on future work in all realms of strong light-matter interaction.

\section{Brief introduction to light-matter interactions and molecular QED theory}\label{sec:basics}

This section briefly introduces the quantum theories of light-matter interactions across different coupling regimes in the nonrelativistic limit~\cite{craig1998molecular}. In principle, all light-matter interactions arise from the interplay of matter DOFs (degrees of freedom) (electrons or spins, nuclei) and an EM environment. Hence, a full ab initio theory for light-matter interactions fundamentally requires the electronic structure theory of matter and principles of electrodynamics. Quantum electrodynamics (QED) is the indispensable and most precise theory for describing the interactions of charged particles and the dynamics of the EM field in mutual interaction. Hence, we start with a brief introduction to the universal molecular QED theory, followed by its derivation of ab initio methods for plasmon and polariton chemistry (in different coupling regimes), respectively. 

\subsection{Molecular quantum electrodynamics theory}
The minimally coupled Coulomb Hamiltonian governs the nonrelativistic dynamics of matter in an EM environment, 
\begin{equation}\label{EQ:QED_HAM_GENERAL}
    \hH=\sum^{N_e+N_n}_i \frac{1}{2m_i}\left[\hP_i - z_i \bA(\br_it)/c\right]^2 + \hat{V} + \hH_{EM},
\end{equation}
with the Coulomb gauge $\nabla\cdot\bA=0$. $\hP_i= -i\hbar\nabla_i$ and $z_i$ are the momentum operator and charge of particle $i$, respectively. $N_{e}$ and $N_n$ are the number of electrons and nuclei in the systems.
$\hat{V}=\hat{V}_{ee}+ \hat{V}_{nn} + \hat{V}_{en}+V_{ext}$ contains all of the Coulomb interactions ($\frac{1}{8\pi\epsilon_0}\frac{q_i q_j}{|\br_i-\br_j|}$) between the electronic and nuclear DOFs and another external potential. The Hamiltonian that describes the EM fields is
\begin{equation}\label{EQ:H_PH}
    \hH_{EM} = \frac{\epsilon_0}{2}\int d\br\left[
    \hat{\bE}^2(\br) + c^2\hat{\bB}^2(\br)\right].
\end{equation}
It obeys Maxwell's equations of motion and couples with the Schr\"odinger equation self-consistent through the vector potential $\bA(\br_i)$.

The general quantization of electromagnetic field involves three steps: 1) classical mode description of the electromagnetic field, 2) discretization of classical modes via standard mode decomposition techniques, and 3) quantization via the correspondence principle~\cite{Archambault2010prb}. 
After the quantization of the electromagnetic field, each field (including the vector potential $\hat{\bA}$) can be expressed as a sum over all possible modes~\cite{Archambault2010prb,andrew1989irpc}
\begin{align}
    \hat{\bA}(\br,t)=&\sum_\alpha \bu_\alpha \Big(\hat{A}_\alpha e^{i(\bK\cdot\br-\omega_\alpha t)} + \hat{A}^\dag_\alpha e^{i(\bK\cdot\br+\omega_\alpha t)}\Big)
    \\
    \hat{\bE}(\br,t)=&i\sum_\alpha \omega_\alpha \bu_\alpha \Big(\hat{A}_\alpha e^{i(\bK\cdot\br-\omega_\alpha t)} + \hat{A}^\dag_\alpha e^{i(\bK\cdot\br+\omega_\alpha t)}\Big)
    \\
    \hat{\bB}(\br,t)=&i\sum_\alpha (\nabla\times \bu_\alpha) \Big(\hat{A}_\alpha e^{i(\bK\cdot\br-\omega_\alpha t)} + \hat{A}^\dag_\alpha e^{i(\bK\cdot\br+\omega_\alpha t)}\Big)
\end{align}
where $\nabla\times \bu=\bK\times\bu$, $\alpha \equiv \lambda\bK$ denotes the photon mode with momentum $\bK$ and polarization $\lambda \in \{-1,1\}$, $\bu$ is the unit vector denoting the direction of the vector potential, and $\hat{A}_\alpha$ ($\hat{A}_\alpha^\dag$) is the mode decomposition coefficients that create (annihilate) the $\alpha^\mathrm{th}$ radiation mode. Note that $\hat{\bE}=-\frac{1}{c}\partial_t \hat{\bA}(\br)$ and $\hat{\bB}=\frac{1}{c}\nabla\times \hat{\bA}(\br)$ in the Coulomb gauge (\textit{i.e.}, $\nabla\cdot\hat{\bA}=0$). The photon modes in a given nanophotonic/nanoplasmonic structure can be readily calculated via standard mode decomposition techniques~\cite{SnyderLove:1983}.

In the second quantization, the photonic Hamiltonian can then be written as,
\[
\hH_p=\sum_\alpha \frac{\hbar\omega}{2}\left[\ha_\alpha\ha^\dag_\alpha+h.c.\right]=\sum_\alpha\hbar\omega(\ha^\dag_\alpha\ha_\alpha+1/2).
\]
with the equivalence,
\begin{align}
    A_\alpha \equiv \sqrt{\frac{\hbar}{2\epsilon_0\omega V}} \ha_\alpha,
    \ \ \ \ \ \
    A^*_\alpha \equiv \sqrt{\frac{\hbar}{2\epsilon_0\omega V}} \ha^\dag_\alpha,
\end{align}
Thus, the fields are quantized by the isomorphic association of a quantum mechanical harmonic oscillator to each radiation mode $\alpha$. Introducing the canonical position and momentum operators as, $\hq_\alpha=\sqrt{\frac{\hbar}{2\omega_\alpha}}(\ha^\dag_\alpha+\ha_\alpha)$, $\hp_\alpha=i\sqrt{\frac{\hbar\omega_\alpha}{2}}(\ha^\dag_\alpha-\ha_\alpha)$. The photonic Hamiltonian can now be rewritten as,
\begin{equation}
    \hH_p = \frac{1}{2}\sum_\alpha (\hp^2_\alpha +\omega^2\hq^2_\alpha). 
\end{equation}
Then the full light-matter Hamiltonian in Coulomb gauge can now be fully constructed and written as,
\begin{equation}\label{EQ:H_QED_MinCoup}
    \hH_c = \sum^{N_e+N_n}_i \frac{1}{2m_i}\left(\hP_i-z_i\hat{\bA}(\br_i)/c\right)^2
    +\hat{V} + \hH_p,
\end{equation}
which is often referred to as the minimal coupling Hamiltonian.

We now aim to apply a unitary transformation on Eq.~\ref{EQ:H_QED_MinCoup} to achieve an expression where the momenta of the molecular DOFs ($\hp_i$) are decoupled from the vector potential ($\hat{\bA}(\br_i)$). In other words, we will shift the light-matter coupling from momentum fluctuations into displacement fluctuations. This transformation is written as $\hat{U}(\bD) = exp[ -\frac{i}{\hbar}\hat{\bD}\cdot\hat{\bA} ]$ where $\hat{\bD}=\sum_i^{N_n} z_i\hat{\bR}_i -\sum_i^{N_e} e\hat{\br}_i$ is the molecular dipole moment~\cite{Woolley:2020prr}. This transformation $\hU(\bD)\hH_c\hU^\dag(\bD)$ is nothing but a reduction in matter momentum such that $\hp_i-z_i\bA\rightarrow \hp_i$ and a boost in photonic momentum by $\hp_\alpha \rightarrow\hp_\alpha+\sqrt{2\omega/\hbar}\bD\cdot\bA_0$. Applying the transformation results in the QED Hamiltonian, referred to as the dipole gauge Hamiltonian,
\begin{equation}
    \hH=\hT + \hV + \hH_{ep}
\end{equation}
where 
\begin{align}\label{EQ:H_PH_SHIFTED}
\hH_{ep}=\hat{U}^\dag\hH_p\hat{U}&=\frac{1}{2}\sum_\alpha \left[(\hp_\alpha+\blambda_\alpha\cdot\bD)^2 + \omega^2_\alpha(\hq_\alpha)^2\right] \\
    &=\frac{1}{2}\sum_\alpha \left[\hp^2_\alpha + \omega^2_\alpha(\hq_\alpha-\frac{\blambda_\alpha}{\omega_\alpha}\cdot\bD)^2\right].\nonumber
\end{align}
The second line is reached via canonical transformation between coordinate and momentum operators, $\hp \rightarrow -\omega \hq,   \hq\rightarrow 1/\omega \hp$.
Here, $\blambda_\alpha=\sqrt{\frac{1}{\epsilon V}}\bu_\alpha$ and $g_\alpha\equiv\blambda_\alpha\cdot\bD$ defines the light-matter coupling strength, which depends on the volume of quantized photon $V$ in radiation mode $\alpha$. The total light-matter Hamiltonian can now be \textit{re-partitioned} after the unitary transformation as,
\begin{align}\label{EQ:H_PF}
    \hH_\mathrm{PF} &= \hH_\mathrm{M} + \hH_\mathrm{p} + \hH_\mathrm{ep} + \hH_\mathrm{DSE} \\ &= \hH_\mathrm{M} + \sum_{\alpha} \Big[ \omega_\alpha (\ha^\dag_\alpha\ha_\alpha+\frac{1}{2}) + \sqrt{\frac{\omega_\alpha}{2}} \blambda_\alpha \cdot \hat{\bD} (\ha^\dag_\alpha + \ha_\alpha) + \frac{1}{2} (\blambda_\alpha \cdot \hat{\bD})^2 \Big].
\end{align} 
Here, $\hat{H}_\mathrm{M} = \hat{T}_\mathrm{n} + \hat{T}_\mathrm{e} + \hat{V}$ is the bare molecular Hamiltonian which includes all Coulomb interactions $\hat{V}$ between electrons and nuclei as well as the kinetic energy operators of both, $\hat{T}_\mathrm{e}$ and $\hat{T}_\mathrm{n}$, respectively. This Hamiltonian is often referred to as the Pauli-Fierz (PF) Hamiltonian.

Now it is clear that the light-matter coupling strength is determined by the two quantities: the molecular dipole strength and the quantized cavity volume. Hence, there are two general strategies to enter the strong coupling regime: (I) reduce the cavity volume and (II) increase the number of molecules (increasing the total dipole moment). Therefore, there are two major experimental nanocavity designs for strong light-matter coupling. The first is the nanophotonic cavity that leverages a large number of molecules to achieve strong coupling. The other is the nanoplasmonic cavity that leverages a locally enhanced electric field confined in a small volume to enhance the coupling. This local field is effectively confined to the nearby surrounding of a spherical nanoparticle residing on a lattice of nanoparticles on the scale of ~nm$^3$ or even ~\AA$^3$ (namely picocavities~\cite{Chikkaraddy2016nature,Jeremy2022nl,Benz2016science}).

The derivation of Eq.~\ref{EQ:H_PH_SHIFTED} assumes dipole approximation. However, the dipole approximation may fail in the ultra-confined nanoplasmonic cavities where the EM fields are confined within nanometric volumes~\cite{Neuman2018nl,Wang:2021jpcc}. Consequently, the size of molecules becomes comparable to the cavity volumes, and the widely used dipole approximation breaks down. Position-dependent coupling strength that requires the spatial distribution of excitonic and photonic quantum states is found to be a key aspect in determining the dynamics in ultrasmall cavities both in the weak and strong coupling regimes~\cite{Neuman2018nl}. 

\subsection{Weak and strong coupling}
The light-matter interactions in nanoplasmonic environments can be split into weak and strong coupling regimes. The weak coupling regime is associated with the Purcell enhancement of spontaneous emission. This effect is particularly strong when the molecule is placed next to a metallic surface or nanostructure~\cite{Kelly:2002jpcb}. The plasmon can tune photophysics and photochemistry (i.e., plasmon chemistry) via various possible pathways in such a regime, leading to a promising approach for accelerating and manipulating chemical reactions, which will be the topics of Section~\ref{SEC:PLASMON}. In this regime, the light-matter interaction can be treated semiclassically via the classical treatment of the electromagnetic field,
\begin{align}
  H_{EM} = &\frac{\epsilon_0}{2}\int d\br\left[
  \bE^2(\br) + c^2\bB^2(\br)\right].
  \\
  \hH_{ep} = & \int d\br E(\br)\cdot (e\hat{\br}).
\end{align}
Where $\bE(\br)$ and $\bB(\br)$ are classical electromagnetic fields.
Afterward, the conventional quantum chemistry method can be utilized and extended to describe the plasmon-induced chemical processes. But it should be noted that, even in the weak coupling regime, there is growing interest in exploring the quantum properties of plasmon for applications~\cite{Varas:2016tf, Tame2013np, Marinica2012nl, Archambault2010prb}. More details will be discussed in Section~\ref{SEC:PLASMON}.

The other regime is the strong coupling regime, where the light-matter interaction cannot be treated perturbatively. The strong coupling is characterized by a reversible (coherent) exchange of energy (known as Rabi oscillation) between the matter and the cavity photon because the coupling is strong enough to compete with the dissipation. In this regime, polariton formation requires theoretical methods that treat the matter and photonic DOFs on equal footing and describe the multiple interactions between photons, electrons, and nuclei on different length and time scales (polariton chemistry). The theoretical advances towards understanding polariton chemistry and our perspective on future multiscale modeling of polariton chemistry are discussed in Section~\ref{SEC:POLARITON}.

\section{Plasmon chemistry} \label{SEC:PLASMON}

Although confined plasmonic modes result in localized and enhanced fields, plasmons usually suffer from strong dissipation, making it hard to enter the strong coupling regime in plasmonic systems. Nevertheless, even without strong coupling, plasmonics provides a unique setting for manipulating light via the confinement of EM (below the diffraction limit). Such extreme concentration of EM field~\cite{Schuller2010NatMat} has led to a wide range of applications, such as plasmon-enhanced molecular spectroscopy~\cite{Stiles2008annurev, Jiang2003jpcb, Brus2008acr, Nie1997sci, Zhan:2018tv}, photovoltaics~\cite{NatMater2010,zhang2016fundamental}, nanophotonic lasers and amplifiers~\cite{Ma2013laser, Berini2012natphotonics, Guan2021am, Guan2022cr}, quantum information~\cite{Flamini_2018}, and many others~\cite{Cushing016jpcl, Mirkovic2018cr, Li2015natphotonics}. 

Following the ultrafast plasmon excitation, nonradiative plasmon decay leads to the formation of energetic electron–hole pairs (namely hot-carriers)~\cite{Zhang2021JPCA, Clavero2014,bernardi2015,acsnano5b06199,nn502445f}, which are highly nonthermal and can have considerably higher energies than those rising from thermal equilibrium. The hot electrons (HEs) (and their corresponding holes) redistribute their energies quickly as a result of electron–electron scattering~\cite{Zhang2021JPCA, Clavero2014}, reaching a quasi-thermal equilibrium but with an effective high temperature. Further cooling of the hot electrons takes place via energy dissipation into the phonon modes of the nanoparticle, and the energy is ultimately dissipated to the surroundings via thermal conduction. Nevertheless, investigations in the last two decades have found that chemical reactions can be stimulated by localized EM fields (or photons), and electronic and/or thermal energies (that result from plasmon decay) via various pathways (more details in Sec.~\ref{sec:plasmonchemdetails}). This leads to an emerging field of plasmon chemistry that designs nanostructure-based surface plasmons as mediators to redistribute and convert photon energy in various time, space, and energy scales to drive chemical reactions~\cite{Zhan:2018tv, Zhan:2023tc, Yuan:2022tb, Brongersma:2015vd, Zhou:2018tu, wu2023chemsci, zhang2018plasmonic, wu2020mechanistic, Kazuma2019angew, Kazuma2018science, ZhangYuchao2018chemrev, Tesema2019jpcc, Adleman2009nl, Boergter2016natcomm, zhang2019atomistic}.
One extraordinary feature of Plasmon chemistry in the realm of chemical reaction mechanisms is its blend of facets from thermochemistry, photochemistry, and photocatalysis~\cite{Zhan:2018tv}. Unlike traditional reactions that typically focus on singular mechanisms, plasmon chemistry showcases the interaction between different mechanisms in complex electronic and optical environments. At the same time, the mixture of different mechanisms makes it very challenging to comprehend the underlying principles of plasmon chemistry, making it considerably more complicated than transition chemistry. This unique interplay often results in a diverse distribution of reactive zones on substrates~\cite{Zhan:2018tv}. To fully understand plasmon chemistry's nuances, an interdisciplinary approach is essential, which should consider multi-scale processes, the current state of the field, and the need for advanced experimental methods.

\subsection{Experimental Endeavours in Plasmon-Mediated Chemical Reactions: The Need for Theoretical Insights}

Plasmon chemistry stands at the intersection of innovation and discovery for boosting chemical transformation. However, despite the spreading interest, inherent complexities make it still very challenging for efficiency optimization. By leveraging knowledge from established domains such as plasmon-enhanced spectroscopy, many experimental strategies to bolster plasmon chemistry efficiency can be extrapolated. In this section, we discuss several typical experimental investigations into plasmon chemistry across diverse applications and illustrate the essential role of theoretical/numerical modeling in understanding the underlying mechanisms for further improvement. For a more expansive disquisition on the plasmon-mediated chemical reactions for various applications, enthusiasts are redirected to several recent reviews~\citenum{Gelle:2020uh,Zhan:2019wp, ZhangYuchao2018chemrev, Li:2023ux}.

\begin{figure*}[!htb]
    \centering
    \includegraphics[width=0.95\textwidth]{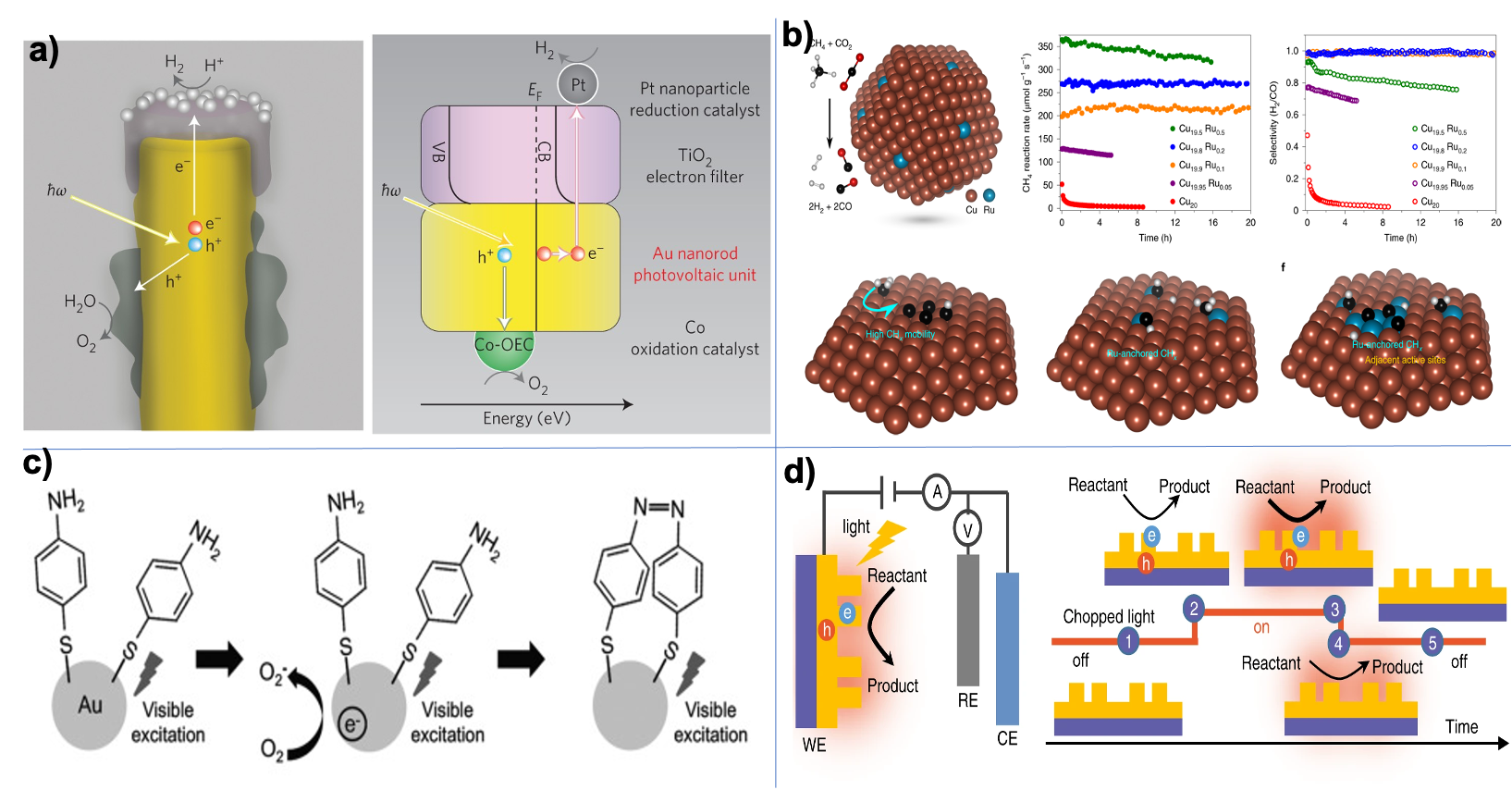}
    \caption{
    Experimental examples: 
    a) Structure and mechanism of operation of the autonomous plasmonic solar water splitter. Adapted with permission from Ref.~\citenum{Mubeen:2013tk}. Copyright 2013 Springer Nature Limited.
    b) Light-driven CO$_2$ reduction with single atomic site antenna-reactor plasmonic photocatalysts. Adapted with permission from Ref.~\citenum{Zhou:2020tn}. Copyright 2020 the author(s), under exclusive license to Springer Nature Limited.
    c) Schematic diagram of the proposed mechanism for the oxidation of PTAP to DMAB. Hot electrons generated from the plasmon transfer to the adsorbed O$_2$ molecules, generating $^3$O$_2$ that participated in the PATP oxidation to DMAB. Adapted with permission from Ref.~\citenum{Wang:2015uk}. Copyright 2015 John Wiley and Sons.
    d) Schematic of the three-electrode electrochemical system for distinguishing thermal from hot electron effects by utilizing photoelectrochemical characterization under the chopped light. Adapted from Ref.~\citenum{Zhan:2019wc} under Creative Commons CC BY.
    }
    \label{fig:exp}
\end{figure*}

\textbf{Artificial photosynthesis}
One of the most enticing applications of plasmon chemistry echoes the wonders of photosynthesis~\cite{Mirkovic2018cr, Mubeen:2013tk, Xiao:2017uw, Yu:2019ut}. The goal is to convert solar energy into useful chemicals, a task intricate due to its kinetic and thermodynamic demands. However, in advancements in the realm of solar conversion, a groundbreaking study has showcased the potential 
 of autonomous plasmonic catalysts capable of water splitting under visible light~\cite{Mubeen:2013tk}, as shown in Figure~\ref{fig:exp}a, present promising trajectories for renewable energy solutions.
This paradigm shift focuses on deriving charge carriers from these plasmonic structures.
The research highlights the introduction of a solar water-splitting device engineered from a gold nanorod array. Unlike conventional methods, this device leverages the power of ``hot electrons" initiated by stimulating surface plasmons within the gold nanostructure. A standout feature is each nanorod's autonomous operation, eliminating the need for external wiring and its impressive capability to produce H2 molecules under sunlight. The device's resilience and long-term stability further cement its potential.
While the initial solar-to-hydrogen efficiency aligns with that of early semiconductor-based water splitters, it remains on the lower side for real-world applications. Nevertheless, the research suggests prospective structural enhancements, emphasizing the modification of the nanorod configuration, that could potentially escalate its efficiency. An important note for consideration, despite the current efficiency metrics, is the device's remarkable longevity, which overtakes some of the top-performing semiconductor-based alternatives. This study marks a pivotal step towards the innovative utilization of plasmonic devices in solar conversion.

\textbf{CO$_2$ Reduction.} 
Given the increasing global CO$_2$ levels, plasmon chemistry offers a promising solution. Plasmonic structures have shown potential in boosting the photo-reduction of CO$_2$~\cite{Singh:2023wt, Zhou:2020tn, Li:2023ux, Dhiman2023chemsci}, which is crucial for environmental cleanup and advancing green chemistry. In a new development, researchers have presented a plasmonic photocatalyst that blends a copper (Cu) nanoparticle ``antenna" with individual atomic ruthenium (Ru) sites (Figure~\ref{fig:exp}b). This design allows methane dry reforming to happen at room temperature, using the energy from light. Unlike traditional thermocatalytic methods, this photocatalyst works effectively with light at normal conditions and has notable stability and selectivity in its actions. This difference from standard thermally-driven reactions is due to the creation of hot carriers, which improve the rate of carbon-hydrogen activation on Ru sites and speed up hydrogen release. It should also be noted that, different from conventional plasmonic structure, the design of antenna–reactor systems utilize plasmonic metal (the antenna) to collect and concentrate visible light energy and transfer that energy to a catalytic metal (that is, the reactor) to drive a chemical reaction~\cite{Aslam:2017tw, Zhang:2016vx, Swearer:2016pnas}. Because the reactors were not in direct contact with the plasmonic nanoparticles, it was argued that the energy transfer resulting in increased reaction rates could only take place via a field effect where the field from the plasmonic metal was felt by the catalytic reactors resulting in the excitation of charge carriers in reactors.

\textbf{Organic transformation.}  Beyond the catalysis of inorganic reactions, as previously addressed, plasmonic NPs have showcased commendable performance in catalyzing organic transformations. These hold potential applications, spanning from the synthesis of basic commodity molecules to intricate pharmaceutical compounds. This approach boasts benefits such as superior selectivity, accelerated reaction rates, and comparatively lenient reaction conditions. An illustrative instance is the oxidation of p-aminothiophenol (PATP) or p-nitrothiophenol (PNTP) to dimercaptoazobenzene (DMAB). This reaction has been the focal point of myriad studies in recent years~\cite{Wang:2015uk, Silva:2016tp, Huang:2010uf, Zhan:2019wp, Huang:2014ux}. The hypothesized microscopic mechanism posits that plasmon-excited electrons transfer from plasmonic nanostructures to O$_2$ molecules, resulting in the emergence of activated oxygen species, which, in turn, catalyze the formation of DMAB, as shown in Figure~\ref{fig:exp}(c). 
Such a mechanism found experimental validation through experiments toggling specific conditions. The obvious evidence of PATP's transformation into DMAB, when air permeates the PATP and plasmonic nanostructured system, is corroborated by the ascendant Raman modes of DMAB over time. In contrast, in the absence of O$_2$ or electron transfer (attained by coating the plasmonic NPs with an insulator like SiO$_2$), merely the Raman signals of PATP are~\cite{Huang:2014ux}. In addition, it was found that the plasmon-induced hot holes can trigger the oxidation of PATP, even in the absence of oxygen~\cite{Zhao:2014vj}. Notably, despite the absence of a tangible PATP transformation on gold nanoparticles due to inefficient charge transfer from the plasmonic nanostructure directly to the adsorbates, PATP oxidation has been witnessed either in the presence or absence of O$_2$, specifically with Au nanostructures fortified with a TiO$_2$ shell. These reactions outperform their counterparts in sheer gold nanoparticle systems in terms of efficiency~\cite{Zhan:2019wp}. Fascinatingly, it was unveiled that achieving the selective oxidation of PATP to either DMAB or PNTP is possible, depending on whether the plasmonic NPs are reinforced by TiO$_2$ and the presence of UV-illumination~\cite{Wang:2015uk}. Under UV exposure on the bare plasmonic NPs, PATP's oxidation to DMAB is observed. In contrast, there is no catalytic effect with the TiO$_2$ support. However, after applying UV light exposure, a transition to PNTP is detected with the TiO$_2$ support. For an expansive disquisition on the plasmon-mediated organic transformation, enthusiasts are redirected to a contemporary review in Ref.~\citenum{Gelle:2020uh}.

\textbf{Experimental Limitations.} Though there are many demonstrated applications of plasmon chemistry as a promising approach to accelerate or manipulate chemical processes, the underlying mechanisms are still unclear. 
The advances in experimental techniques have played a pivotal role in elucidating plasmon chemistry mechanisms. Techniques, such as transient absorption spectroscopy, offer insights into the domain of plasmonic hot-carriers~\cite{Reddy:2020sci}. Nonetheless, the vast expanse of these reactions remains to be traversed for a holistic comprehension. 
Traversing the landscape of plasmon chemistry requires a comprehensive understanding of multiple coherent and dissipative processes, including the enhancement of electromagnetic near-fields, local heating effects, and charge-carrier excitation/transfer. These processes occur across varied scales, from femtoseconds to nanoseconds. Since all these effects can drive chemical processes simultaneously, the individual contributions of these elements to overarching reactions are challenging to discern. On the other hand, decoupling individual contributions is vital for designing better plasmon catalysts that synchronize the orchestration of these effects to improve efficiency. In particular, distinguishing the thermal effect from the hot electron effect has become an important topic and is the subject of many debates~\cite{}. Despite the complexity of distinguishing these effects, there are growing efforts to solve these key questions via various designs~\cite{Zhan:2019wc, Yu:2018vh, Zhou:2018aa, Zhang:2018ug, Ou:2020vk, Baffou:2020ts}. One of these efforts utilized the unique photoelectrochemical behavior of a plasmonic Au nanoelectrode array as shown in Figrue~\ref{fig:exp}d~\cite{Zhan:2019wc}. The plasmonic photocurrent can be intensified at negative and positive potentials, with its direction determined by the selected potential. This allows the electrode to favor either reduction or oxidation reactions. The photocurrent is composed of a fast photoelectronic component and a slower photothermal component. The photoelectronic current aligns with the plasmon absorption spectrum and intensifies with light intensity, whereas the photothermal current displays a linear relationship only within specific light intensities. Consequently, this proposed design can differentiate between the photoelectronic and photothermal effects. However, the photoelectronic response measured in Ref~\ref{Zhan:2019wc} happens at a time scale several orders of magnitude larger than the hot-electron lifetime and thermal conduction time scales. More advanced time-resolved techniques are required to provide more comprehensive evidence.

Overall, the distinctiveness of plasmon chemistry is rooted in the interplay between molecules, incident photons, and plasmonic nanostructures. Though many recent breakthroughs have been made to demonstrate plasmon chemistry's application in various areas, further improvements require a deep microscopic understanding of the synergy of multiple effects at the electron level. Such understanding will play an indispensable role in the strategic design and fabrication of plasmonic nanostructures, optimization of surface or interface intermediaries, and their harmonized integration. To elucidate the intricate mechanism that governs plasmon chemistry, comprehensive investigations encompassing various time, spatial, and energy scales are essential, which cannot be achieved by experiments alone. It's crucial that advanced experimental methods, boasting high space-time resolutions, are synergized with microscopic theoretical modeling. 
Specifically, theoretical models capable of addressing all potential mechanisms on equal footing provide invaluable perspectives into quandaries that remain elusive to experimental endeavors alone. Intuitively, the convenience of modeling allows for the toggling of individual effects to scrutinize their ramifications on holistic reactions. The real theoretical challenge, however, lies in considering these multifaceted effects equitably, spanning varied time and length scales. In the following sections, we will review our theoretical efforts toward simulating all conceivable mechanisms on an equal footing, with an aspiration to understand plasmon chemistry at the electron level.

\subsection{Ab initio method for quantum plasmonics}
In most scenarios, the nanoplasmonic properties are obtained by solving Maxwell's equations with proper dielectric function and boundary conditions. However, Maxwell's equations fail to describe the quantum effects (such as the tunneling in charge transfer mode and quantum confinement~\cite{Marinica2012nl}), and quantum theory for plasmon (quantum plasmonics) is required in nanoscale. 

From a quantum mechanical point of view, a plasmon is nothing but a specific collective excitation. Hence, the quantum chemistry method for excited states can be used to predict the optical properties of plasmonic nanostructures~\cite{Varas:2016tf}. The many-body perturbation theory, time-dependent density functional theory (TDDFT), or time-dependent density functional tight-binding (TDDFTB) are methods of choice to compute the plasmon excitations~\cite{Perera:2020jpcc, Rossi:2017jctc, Asadi:2020jpcc, Pandeya:2021jpca, Alkan:2018ug, Alkan:2021vq, DellaSala:2022vo, DAgostino:2018tg, You:2019vn}. Without losing generality, here we use TDDFT theory as an example to demonstrate the computation of plasmon excitations. 

TDDFT is the formal extension of the Hohenberg-Kohn-Sham density functional theory (DFT). Within the DFT, The Hohenberg-Kohn (HK) theorem~\cite{HK1964} states that the ground-state electron density unambiguously defines the many-electron ground state for an $N-$electron system under the influence of an external potential. And there exists an energy functional that guarantees the variational principle can reach the ground state energy, though the exact functional is an unsolved problem,
\begin{equation}
    E[\rho]=T+V_H(\br)+E_{xc}[\rho]+\int V(\br)\rho(\br)d\br.
\end{equation}
Where $V_H[\rho]=\int \frac{\rho(\br')}{|\br-\br' |}d\br'$ is the Hartree potential, $ E_{xc}$ is the exchange-correlation functional, which contributes $<10\%$ to the total energy, but describes the most critical correlation effects~\cite{Parr1995annuphys}. But DFT is a ground-state theory. Light-matter interactions usually result in many elementary excitations within the matter. Among various many-body methods, TDDFT (the formal extension of DFT theory) is the preferred tool to evaluate the excited states and optical properties in extended molecular or condensed matter systems. 

In the linear response regime, the induced electron density due to the light-matter interaction $\delta V(\br)$ is given by 
\begin{equation}\label{eq:rho1}
    \delta\rho(\br) = \int \chi(\br,\br')\delta V_{ext}(\br')d\br'.
\end{equation}
Where $\chi(\br,\br')$ is the polarizability describing the density response of the many-body ground state with respect to an external perturbation. In addition, according to Runge-Gross theorem~\cite{RG1984}, $\delta\rho(\br)$ is also the induced density of the KS system, but due to a perturbation
\begin{equation}\label{eq:rho2}
    \delta\rho(\br)=\int\chi_{KS}(\br,\br')\delta V_{KS}(\br')d\br'.
\end{equation}
$\chi_{KS}$ here is the linear response of the KS electrons, which can be trivially evaluated from the KS orbitals and energies.   

\begin{equation}\label{eq:vks}
    \delta V_{KS}(\br) = \delta V_{ext}(\br) + V_H[\delta\rho(\br)] + \delta V_{xc}(\br).
\end{equation}
The induced XC potential is given by $\delta V_{xc}(\br)=\int f_{xc}(\br,\br')\delta\rho(\br')d\br'$, where $f_{xc}$ is the dynamical XC kernel $\frac{\delta V_{xc}(\br)}{\delta n(\br')}$~\cite{Casida:2012ti},
\begin{equation}\label{eq:tddftkernel}
f_{xc}(\br,\br')=\frac{\delta V_{xc}(\br)}{\delta\rho(\br')},
\end{equation}
Hence, a linear equation for the induced density can be derived from Equations~\ref{eq:rho1}-\ref{eq:tddftkernel},
\begin{equation}\label{eq:LRTDDFT}
    [1-\chi_{KS}(\omega) f_{hxc}(\omega)]\chi(\omega)\delta V_{ext}(\omega)=\chi_{KS}(\omega)\delta V_{ext}(\omega).
\end{equation}
where $f_{hxc}(\br,\br')=\frac{1}{|\br-\br'|}+f_{XC}(\br,\br')$. Casting Equation~\ref{eq:LRTDDFT} into the matrix of coupled KS single excitations, it is possible to calculate the excitation energies $\omega_s$ of the system (the poles of the response function) and transition densities (and oscillator strengths). Consequently, the response function $\chi(\omega)$ can be rewritten as
\begin{equation}\label{eq:chi}
\chi(\br,\br',\omega)=2\sum_s\rho_s(\br)\rho_s(\br')\zeta_s(\omega),
\end{equation}
where $\zeta_s(\omega)=\frac{1}{\omega-\omega_s+i\eta}-\frac{1}{\omega+\omega_s-i\eta}$ and $\eta$ represents a positive inifinitesimal. The factor of 2 comes from summation over spin indices. The excitation energies in the system are denoted as $\omega_s$, which are calculated from the Casida method~\cite{Casida:2012ti}. 
$\rho_s(\br)$ is the transition density, which can be expanded in terms of electronic transitions between occupied state $i$ to unoccupied states $a$~\cite{PRB73205334}, 
\begin{equation}
\rho_s(\br)=\sum_{ia}X^s_{ia}\psi_i(\br)\psi_a(\br)\left(\frac{\epsilon_a-\epsilon_i}{\omega_s}\right)^{1/2}, 
\end{equation}
where $\epsilon_i$ are the KS eigenvalues, and the corresponding molecular orbitals are $\psi_i$. And $X^s_{ia}$ are the Casida transition coefficient from the occupied $i$ state to the unoccupied $a$ state of the $s^{th}$ excitation. By examining the nature of $X^s_{ia}$, it is possible to distinguish between normal excitation and plasmonic excitation. Plasmonic excitations generally are characterized by collective transitions from occupied to virtual orbitals~\cite{Castellanos, wu2022jcp_relaxation}.

\subsection{Hot electron generation and relaxation}
Once the plasmon excitation is obtained, the electron-plasmon interaction can be described by the Hamiltonian 
\begin{equation}
\hH_{int}=\frac{e}{2m_e}\int d\br \hat{\Psi}^\dag V_{\textit{eff}}(\br) \hat{\Psi},    
\end{equation}
where $V_{\textit{eff}}(\br)$ is the effective potential induced by the excitation, which can be calculated from the transition density $\rho_s(\br)$ by following Lundquvist's approach~\cite{Castellanos, PRB73205334}. The polarizability of the NP upon external excitation is given by Equation~\ref{eq:chi}. And the induced potential is given by~\cite{PRB73205334}, 
\begin{equation}
    V_{\textit{eff}}(\br)=\epsilon^{-1}(\br,\br')\delta V_{ext}(\br'),
\end{equation}
where the dielectric function $\epsilon^{-1}(\br,\br')$ is given by
\begin{equation}
\epsilon^{-1}(\br,\br')=\delta(\br,\br')+\int d\br_1 f_{hxc}(\br,\br_1)\chi(\br_1,\br',\omega).
\end{equation}
Thus, within the second quantization, electron-excitation Hamiltonian $\hH_{eps}$ reads
\begin{equation}\label{eq:coupling}
\hH_{eps}=\sum_{ij}\left[\MM^s_{ij} \hc^\dag_{i}\hc_{j} \hb_s+\text{h.c.}\right],
\end{equation}
where $\MM^s_{ij}$ is the electron-excitation coupling strength, describing the scattering of quasiparticles from state $i$ into state $j$ via the emission or absorption of excitation in the state $s$. These elements are given by (see SI for details)
\begin{equation}\label{eq:eplasmoncoup}
\MM^s_{ij}=  \zeta_s(\omega)\bE\cdot \bbf^s\langle\psi_i(\br_1)|V^s_H(\br_1)|\psi_j(\br_1)\rangle
\end{equation}
where $\bbf^s=\sqrt{\frac{2m_e\omega_s}{2\hbar^2}}\sum_{ia}X^s_{ia}\left(\frac{\epsilon_a-\epsilon_i}{\omega_s}\right)^{1/2}\bd_{ia}$ is the oscillator vector and $|\bbf^s|^2=\frac{2m_e}{3\hbar^2}\omega_s |\langle\Psi_0|\br|\Psi_s\rangle|^2$ is the oscillator strength.
$V^s_H(\br)=\int d\br'\frac{\rho_s(\br')}{|\br-\br'|}$ is the Hartree potential induced by the transition density.
Detailed derivation can be found in the SI. Above equation actually includes both plasmon excitation ($\zeta_s(\omega)\bE\cdot \bbf^s$), screening effect and hot electron-hole pairs generation ($\langle\psi_i(\br_1)|V^s_H(\br_1)|\psi_j(\br_1)\rangle $). Neglecting the screening effect would result in a much larger electron-plasmon coupling matrix and shorter HC lifetime distribution~\cite{acsph7b00881}. In contrast, our model connects the widely used semiempirical model and the recently developed quantum model with the plasmon excitation and HC generation treated on equal footing. Equation~\ref{eq:coupling} is the general formalism that describes the coupling between electrons and photoexcitation. If the external field matches the plasmon energy, plasmon resonance will be excited, and Equation~\ref{eq:coupling} reduces to the electron-plasmon coupling. 

{\bf HC generation.}
The electron-plasmon coupling describes the HC generation following plasmon decay. After the electron-plasmon coupling matrix is obtained, the HC generation can be readily calculated from the Fermi golden rule~\cite{nn502445f,Zhang2021JPCA},
\begin{equation}\label{eq:hcgen}
\Gamma^{ex}_{i\rightarrow a}=\frac{4}{\hbar}\sum_s|\MM^s_{ia}|^2 \frac{\gamma_{ex}}{(\epsilon_i-\epsilon_a+\omega_s)^2+\gamma^2_{ex}}\delta(\omega-\omega_s).
\end{equation}
$\gamma_{ex}$ refers to the linewidth of the excitation. $\omega$ is the excitation energy. $\delta(\omega-\omega_s)$ describes the generalized photoexcitation. Hence, Equation~\ref{eq:hcgen} describes the photoexcitation and HC generation. When the energy of the external field $\omega$ matches the plasmon energy, Equation~\ref{eq:hcgen} describes the HC generation from plasmon decay. Otherwise, it describes the HC generation from regular excitation.

{\bf HC relaxation.}
After generation from plasmon decay, the HCs will undergo a relaxation process mainly due to the electron-electron and electron-phonon scatterings. The dynamics of the HCs are investigated by propagating the density matrix $\rho(t)$. The equation of motion (EOM) of $\rho$ is subject to the following quantum Liouville Von-Neumann equation~\cite{breuer2002theory},
\begin{equation}\label{mastereom}
i\partial_t\rho(t)=i\partial_t\rho(t)|_{\text{coh}}+\ML_{ee}[t]+\ML_{ph}[t]+\ML_{s}[t].
\end{equation}
The above equation's first part on the right-hand side (RHS) describes the coherent evolution of the density matrix between different states. The second and third parts on the RHS of Equation~\ref{mastereom} represent the dissipations induced by the electron-electron and electron-phonon scattering effects, respectively. Finally, the last term on the RHS of Equation~\ref{mastereom} describes the extraction of HCs. In general, in the presence of electron-electron and electron-phonon interaction, the dissipation can be described by a Liouville equation by employing the many-body perturbation theory (MBPT)~\cite{haug1998}, $\ML(t)=\mq(t)-\mq^\dag(t)$. Where the dissipation matrix $\mq_t$ can be written in terms of Green's functions $G$ and self-energies $\Sigma$~\cite{haug1998,Zhang2021JPCA}, i.e., 
\begin{equation}
\mq(t)=i\int^t dt'_{-\infty}\left[\Sigma^<(t,t')G^>(t',t) - \Sigma^>(t,t)G^<(t',t)\right].
\end{equation}
Hence, the key point is to develop approximations to the self-energies $\Sigma^{<,>}(t,t')$ and efficient numerical methods to compute the $\mq(t)$. The detailed formalisms of $\mq(t)$ for electron-electron and electron-phonon scatterings can be found in Ref.~\cite{Zhang2021JPCA}. But it should be noted that the approximation used for electron-electron scattering should conserve both particles and energy.

\subsection{Plasmon-mediated chemical reactivities and theoretical challenges}\label{sec:plasmonchemdetails}

\begin{figure}[!htb]
    \centering
    \includegraphics[width=0.5\textwidth]{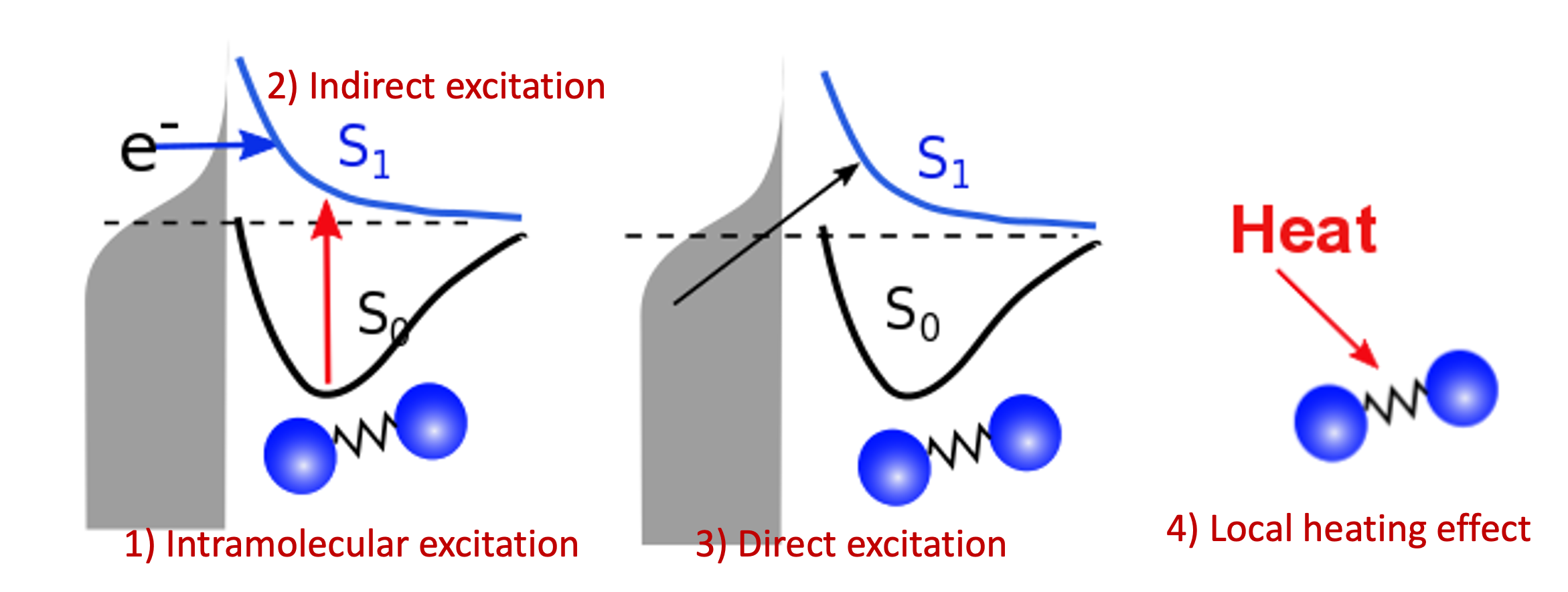}
    \caption{Proposed mechanisms of plasmon-mediated reactivities. 1) Enhanced intramolecular excitation, 2) direct charge transfer, 3) indirect charge transfer, and 4) local heating. 
    }
    \label{fig:plasmonchem}
\end{figure}

However, the computations of plasmon excitation and distribution of hot electrons (from plasmon decay) are insufficient to give insights into plasmon-mediated chemistry because plasmon-induced photocatalysis is a complex dynamical process that involves multiple interactions and mechanisms. To date, four major microscopic mechanisms have been proposed to explain how plasmonics facilitates various chemical reactions through the concentration of light and hot-carrier dynamics (Figure.~\ref{fig:plasmonchem}). The first mechanism involves the extreme concentration of light, which significantly promotes the excitation of electrons within the adsorbed molecule. Such an excitation takes place within the adsorbate only but is significantly enhanced by plasmonics, which is referred to as the enhanced intramolecular excitation mechanism~\cite{Tesema2019jpcc, Kazuma2018science, Tesema2017jpcc}. The second mechanism involves the transfer of plasmonic hot carriers. Due to the hybridization of the adsorbate–nanoparticle system, the generated hot electrons (or holes) can transfer to the absorbed molecules. Such transfers result in the placement of an adsorbate–nanoparticle system onto a manifold of excited potential energy surfaces (PESs), where the adsorbed molecule experiences strong forces that activate its internal vibrational excitations and, ultimately, chemical transformation. This mechanism is attributed to an indirect HE transfer from metal nanoparticles to the adsorbed molecule~\cite{ZhangYuchao2018chemrev, Kazuma2019angew}. Third, due to the hybridization, direct excitation of electrons from metal states near the Fermi level to the unoccupied molecular orbital (LUMO) of the adsorbed molecule can occur when the plasmon frequency is resonant with the excitation energy between the metallic occupied orbitals and hydrated metal-molecular orbitals. Such a reaction pathway is referred to as the direct charge transfer (CT) mechanism. Such excitation circumvents the thermalization of HEs~\cite{Boergter2016natcomm, Boerigter2016acsnano, Longrun2014jacs, Long2012jacs}, but requires matching between the plasmon energy and the energy gap between the metal states and the molecular LUMO. Finally, local heating resulting from hot-carrier relaxation can thermally activate a reaction~\cite{Bora2016scirep, Adleman2009nl, Sivan2019farady}. 

These mechanisms share some common features that render their clear distinction very challenging. Indeed, different mechanisms have been proposed even for the same chemical reaction, leaving a very confusing situation. For example, The chemical decomposition of methylene blue molecules on the surface of plasmonic NPs was reported by different research groups~\cite{Fusco:2022jmcc, Boergter2016natcomm, Boerigter2016acsnano, Chen:2012ur, Tesema2017jpcc}, direct hot electron transfer mechanism was proposed in Ref.~\citenum{Boergter2016natcomm, Boerigter2016acsnano} and indirect hot electron transfer is proposed in Ref.~\citenum{Chen:2012ur}. The same situation also applies to the plasmon-mediated reduction of PNTP, where both the nonthermal and local heating effects were proposed~\cite{Golubev:2018vr, Keller:2018tl}.
Hence, in order to obtain a comprehensive understanding of plasmon-mediated chemistry, it's essential for the theoretical methods to fulfill the following requirements:
\begin{enumerate}
    \item Efficient computation of a dense manifold of excited states because plasmonic excitations in metallic nanostructures are usually not low-lying states. Even for a small nanoparticle (2~nm diameter, for example), thousands of excited states are to be computed to reach the plasmon excitations~\cite{wu2023chemsci}.
    
    \item The ability to capture the nonadiabatic transition between excited states since the indirect transfer mechanism involves the transition between the hot electron state (excitations localized within the plasmonic system only) and the charge transfer state~\cite{wu2020mechanistic}.
    
    \item Inclusion of electron-vibrational couplings that lead to the hot electron relaxation and local heating.

    \item Inclusion of electron-electron scattering that leads to the redistribution of hot electrons~\cite{Brown2017prl, acsnano5b06199}.
\end{enumerate}
In particular, recent debates on the thermal impact underpin the necessity of atomistic and dynamical insights via nonadiabatic simulations of the HE generation, transfer, and relaxation processes on equal footing.  

\subsection{Nonadiabatic simulation of plasmon-mediated chemical reactivities}

{\bf Ehrenfest dynamics and Surface Hopping.}
With the Born-Oppenhermier approximation, the electronic wave function $\Theta(\mathbf{r},\mathbf{R})$ is expanded on the basis of adiabatic BO states, which depend on the electronic coordinates $\mathbf{r}$ and the nuclear coordinates $\mathbf{R}(t)$ according to
\begin{equation}
    \Theta(\textbf{r},\textbf{R}) = \sum_{n=1}^{N_{st}} c_n(t)\ket{\phi_{n}(\textbf{r},\textbf{R}(t))}. \label{eq:1}
\end{equation}
Here $N_{st}$ is the total number of adiabatic electronic states, $\ket{\phi_{n}(\textbf{r},\textbf{R}(t))}$ is the adiabatic electronic wavefunction of state n and $c_n(t)$ are the time-dependent complex expansion coefficients. $\textbf{R}(t) = \left\{\textbf{R}_{A}(t)\right\}_{A=1}^{A=N_{A}}$ ($N_A$ is the total number of atoms in the system) are the nuclear trajectories which are obtained by solving the classical Newton's equations of motion (EOMs) with the mixed quantum-classical framework~\cite{Nelson2020ChemRev},
\begin{equation}
    M_{A}\frac{d^2\textbf{R}_{A}}{dt^2} = -\nabla_{\textbf{R}_{A}}E(\textbf{R}), \label{eq:2}
\end{equation}
where $M_A$ is the mass of $A^{th}$ atom. $E(\textbf{R})$ is the potential energy surface (PES), which can be an averaged one $E(\bR)=\sum_n C_n(t) E_n(\bR)$ within the mean-field Ehrenfest dynamics or a single PES of a certain state $E_k(\bR)$ within the Surface hopping (SH) framework~\cite{tully1990molecular, Tully2012jcp, Nelson2020ChemRev}. 

By substituting Eq.~\ref{eq:1} into the time-dependent Schr\" {o}dinger equation and keeping only the first-order nonadiabatic coupling terms,
a set of EOMs for the coefficients $c_n(t)$ along a given classical trajectory can be obtained~\cite{Nelson2020ChemRev,curchod2013trajectory}
\begin{equation}
    i\hbar\frac{\partial c_{n}(t)}{\partial t} = c_{n}(t)E_{n}(\textbf{R}) - i\hbar\sum_{m}c_{m}(t)\dot{\textbf{R}}\cdot \textbf{d}_{nm}. \label{eq:3}
\end{equation}
Here the orthogonal condition of adiabatic states $\langle \Psi_n |\Psi_m \rangle = \delta_{nm}$ is used. $\mathbf{d}_{nm} = \langle \Psi_n|\nabla_{\textbf{R}}|\Psi_{m} \rangle$ is the nonadiabatic derivative coupling term (or nonadiabatic coupling vector, NACR).
A key variable in Eq.~\ref{eq:3} is the time-derivative nonadiabatic coupling scalar (NACT) between two adiabatic states
\begin{equation}\label{eq:4}
    \dot{\textbf{R}}\cdot \textbf{d}_{nm} = \langle \Psi_{n}|\frac{\partial}{\partial t}|\Psi_{m}\rangle, 
\end{equation}
which is responsible for the nonadiabatic transitions between different adiabatic states and can be easily calculated with many ab initio methods for excited states.

With the SH algorithm, the state that governs the dynamics of nuclei is determined by a stochastic process. The time-dependent coefficients $c_{n}(t)$ obtained in Eq.~\ref{eq:3} are used to calculate the hopping probabilities between different electronic excited states within the framework of the FSSH algorithm. The hopping probabilities between excited states $n$ and $m$ are given by\cite{malone2020nexmd}
\begin{equation}\label{eq:7}
    g_{n \rightarrow m}(\textbf{R},t) = \frac{\int_{t}^{t+N_{q}\delta t}dt\  b_{mn}(\textbf{R},t)}{a_{nn}(t)},
\end{equation}
where $N_q=\frac{\Delta t}{\delta t}$, with $\Delta t$ and $\delta t$ correspond to the time steps for evolving motions of nuclei and electrons in Eq.~\ref{eq:2} and Eq.~\ref{eq:3}, respectively. The chosen value of $\delta t$ must be small enough to resolve strongly localized peaks in NACT in order to avoid underestimation of transition probabilities. This is particularly important when crossings between adiabatic states are encountered in the trajectory, in which $\delta t$ will be further refined. $a_{nn}(t) = c_n(t)c_n^*(t)$ defines the time-dependent density matrix elements, and $b_{mn}(\textbf{R},t) = -2Re(a_{nm}^*\ \dot{\textbf{R}}\cdot \textbf{d}_{nm})$. Note that $g_{n\rightarrow m} = -g_{m \rightarrow n}$ and $g_{n\rightarrow n} = 0$ since $\textbf{d}_{nm}$ are antisymmetric. Hopping between adiabatic states is determined stochastically by comparing $g_{n\rightarrow m}$ to a random number $\xi (\xi \in (0,1))$. A hop from state $n$ to state $m$ is performed if
\begin{equation}
  \sum_{l=1}^{m-1} g_{n \rightarrow l} < \xi \leq \sum_{l=1}^{m} g_{n \rightarrow l},
    \label{eq:8}
\end{equation}
where states are assumed to be ordered with increasing transition energy. On the other hand, the system remains in state n when $\sum_{l=1}^{N_{st}} g_{n \rightarrow l} < \xi < 1$. If $g_{n\rightarrow m} < 0$, the hop is unphysical, and the probability is set to zero. Finally, if a hop to a higher energy state is predicted, there must be sufficient nuclear kinetic energy along the direction of NACR. Otherwise, the hop is rejected. After a successful hop, the total electron-nuclear energy is conserved by rescaling the nuclear velocity in the direction of the NACR according to the procedure described in reference~\cite{Tully2012jcp, Fabiano2008chemphys}. In addition, during the dynamics, we monitor the relative phase of the ground to excited state transitions and maintain the same phase (sign) to avoid a sudden sign change in the NACT. This is done by enforcing the sign of the largest component of the Casida eigenvectors to the same along the trajectory.

Despite broad popularity in the community, either Ehrenfest dynamics or the surface hopping (SH) approach have well-known limitations such as generating artificial electronic coherence or giving incorrect long-time population~\cite{Tully2012jcp}. To address these challenges and provide accurate dynamics, a multiconfigurational Ehrenfest (MCE) dynamics approach~\cite{makhov_MCEh_ChemPhys2017} and ab initio multiple cloning (AIMC)~\cite{makhov_AIMS_JCP2014,makhov_AIMS_PCCP2015,makhov_AIMS_FaraDisc2016,song2021ab} are developed accordingly. 

{\bf Multiconfigurational Ehrenfest (MCE).} MCE generalizes EHR formalism by representing the wave function as a linear combination of Ehrenfest configurations. Each MCE configuration moves along its own Ehrenfest (mean-field) trajectory. Within the MCE formalism, the molecular wavefunction $\ket{\Psi}$ is expressed in the trajectory-guided Gaussian basis functions (TBF) representation ($\ket{\psi_n}$),
\begin{equation}\label{eq_mcewf}
    \ket{\Psi(t)}=\sum c_n \ket{\psi_n(t)}.
\end{equation}
And each configuration (or TBF) is described by the product of nuclear and electronic parts,
\begin{equation}
    \ket{\psi_n(t)} = \ket{\chi_n(\bR,t)} \sum_I a^{n}_I\ket{\phi^n_I(\br,\bR(t))}.
\end{equation}
Where $\ket{\phi^n_I(\br,\bR(t))}$ is the adiabatic state of configuration $n$. $\ket{\chi_n}$ are Gaussian nuclear basis functions
\begin{equation}
    \ket{\chi_n}=\left(\frac{2\alpha}{\pi}\right)^{N_d/4}e^{\left\{-\alpha(R-\bar{R})+\frac{i}{\hbar}P(R-\bar{R})+\frac{i}{\hbar}\gamma_n(t)\right\}}.
\end{equation}

The couplings between TBFs in the MCE approach are described by the EOM of $c_n(t)$, which can be readily obtained by substituting Eq.~\ref{eq_mcewf} into the Schr\ "odinger equation:
\begin{equation}
    i\hbar\sum_n S_{mn} \dot{c}_n = 
    \sum_n \left[H_{mn} - i\hbar\bra{\psi_m}\frac{d\psi_m}{dt}\rangle \right]c_n.
\end{equation}
where 
\begin{equation}\label{eq_hmn}
    H_{mn}=\sum_{I,J}(a^m_I)^*a^n_J
\bra{\chi_m\phi^m_I}T+V\ket{\chi_n\phi^n_J},
\end{equation}
and the overlap $S_{mn}$ is
\begin{equation}
    S_{mn}=\bra{\psi_m}\psi_n\rangle
    =\langle{\chi_m}\ket{\chi_n}\sum_{I,J}(a^m_I)^*a^n_J
    \langle{\phi^m_I}\ket{\phi^n_J}.
\end{equation}
The nuclear part of Eq.~\ref{eq_hmn} can be obtained analytically,
\begin{equation}
     \bra{\chi_m\phi^m_I}V\ket{\chi_n\phi^n_J}=
     \bra{\chi_m}-\frac{\hbar^2}{2}\nabla_{\bR}M^{-1}\nabla_{\bR}
     \ket{\chi_n}\langle\phi^m_I\ket{\phi^n_J}.
\end{equation}
While the electronic part (or the potential energy matrix elements) are approximated by~\cite{song2021ab}
\begin{align}
& \bra{\chi_m\phi^m_I}V\ket{\chi_n\phi^n_J}=\frac{1}{2}
 \bra{\phi^m_I}\phi^n_J\rangle\bra{\chi_m}\chi_n\rangle\times \nonumber\\
 &\Big\{ (V^m_I+V^n_J)+\frac{i}{4\alpha\hbar}(\bP_n-\hP_m)\cdot(\nabla_{\bR}V^m_I+\nabla_{\bR}V^n_J)\nonumber\\
  &-\frac{1}{2}(\bR_m-\bR_n)\cdot(\nabla_{\bR}V^m_I-\nabla_{\bR}V^n_J)\Big\}.
\end{align}
As shown above, within the MCE schemes, the electronic states are now different for different configurations. The overlaps between TBFs have to be calculated and taken into account.

{\bf Ab initio multiple cloning (AIMC).} The AIMC method combines the best features of ab initio Multiple Spawning (AIMS)~\cite{bennun_AIMS_JPCA2000} and Multiconfigurational Ehrenfest (MCE) methods. Similar to the MCE method, the individual trajectory basis functions (TBFs) of AIMC follow Ehrenfest equations of motion. However, the basis set is expanded in a similar manner to AIMS when these TBFs become sufficiently mixed. Consequently, AIMC avoids prolonged evolution on the mean-field potential energy surface (PES).

Within the MCE formalism, the Ehrenfest basis set is guided by an average potential, which is accurate for dynamical processes where the coupling between states persists in time between nearly parallel PES. But the mean-field treatment can be unphysical when the PES of two or more populated electronic states become different in shape, which leads to wave packet branching after leaving the nonadiabatic coupling region. To deal with these cases, the AIMC algorithm is applied to expand the original basis set of TBFs by "cloning" one TBF into two copies in a way that does not alter the original wave function~\cite{makhov_AIMS_FaraDisc2016,makhov_AIMS_JCP2014,makhov_AIMS_PCCP2015} This is done by creating one of the clones $\ket{\psi_{n_1}}$ in a pure state and the other clone $\ket{\psi_{n_2}}$, which includes contributions from all other electronic states. The corresponding MCE amplitudes $\{c_{n_1}, c_{n_2}\}$ are adjusted to conserve the original wavefnction~\cite{song2021ab, makhov_AIMS_FaraDisc2016}. 

It should be noted the MQC method can be complemented with any electronic structure solvers as long as the gradients and NACs are available. Practical applications need to balance accuracy and numerical efficiency. 

\begin{figure}[!htb]
    \centering
    \includegraphics[width=0.5\textwidth]{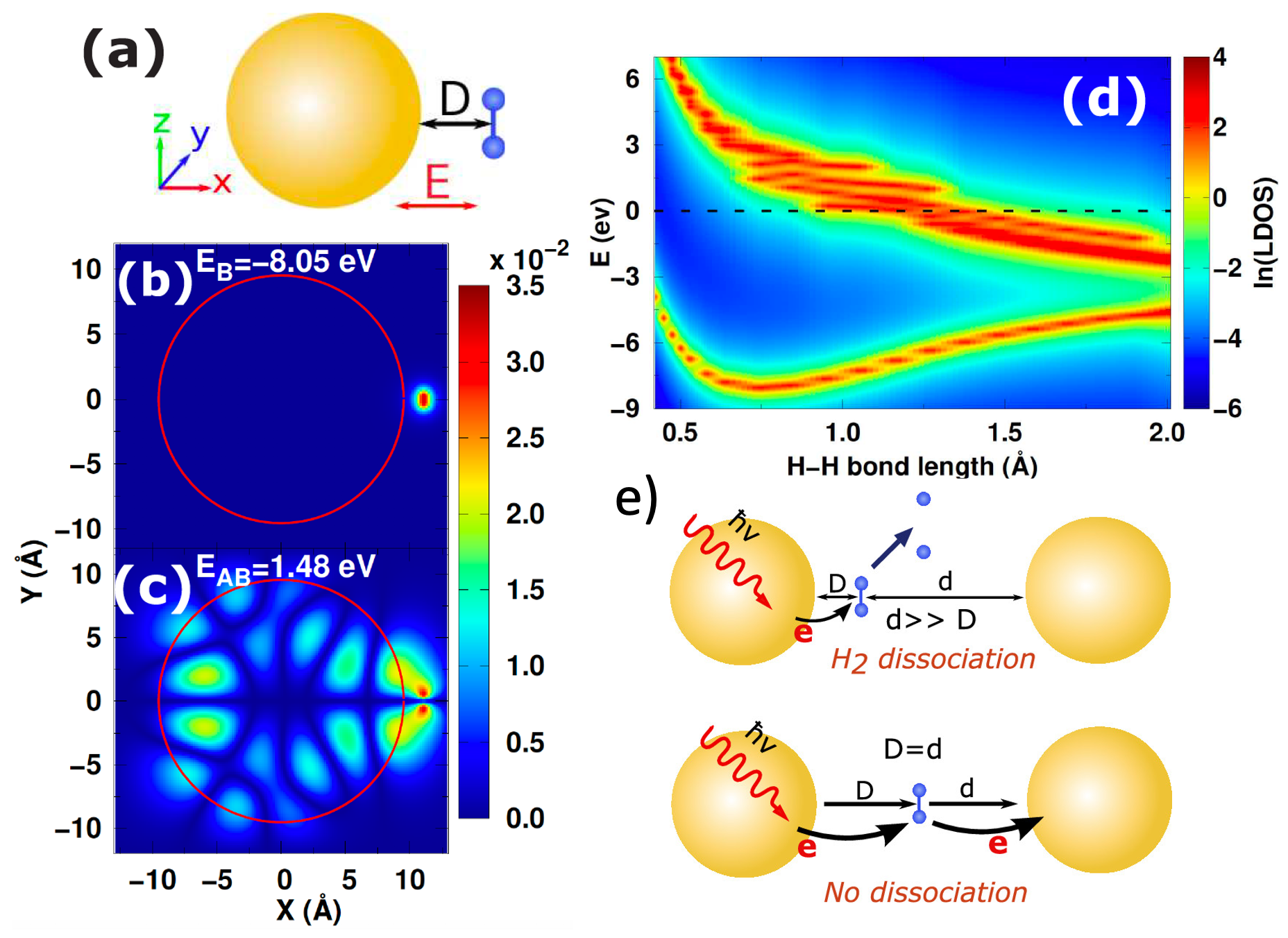}
    \caption{(a) Schematic diagram of H$_2$ molecule adsorbed on the NP surface. b)-c) Spatial distribution of the bonding and antibonding orbitals of H$_2$ on Jellium NP. d) Local DOS of bonding and antibonding orbitals of H$_2$ modules as a function of H-H bond length. e) Depending on the symmetry, the H$_2$ dissociation can be suppressed or restored in the plasmonic dimer. 
    Figures are adapted with permission from Ref.~\citenum{zhang2018plasmonic}. Copyright 2018 American Chemical Society. 
    }
    \label{fig:jellium}
\end{figure}

\subsection{Simulation of plasmon-mediated phenomena}
\subsubsection{Jellium model-based NAMD simulation} 
Compared to the costly atomistic ab initio calculations for plasmonic nanoparticles, the optical absorption of simple sp-metal nanoparticles described by the Jellium model can be easily calculated and analyzed~\cite{Besterio2017acsphotonics, Chang2019acs, Varas:2016tf, Yan:2016uz, nn502445f, zhang2018plasmonic}. In general, good agreement with experimental optical properties can be achieved by tuning the Jellium radius. Using the Jellium model, we investigated the generation and relaxation of plasmonic hot carriers~\cite{Zhang2021JPCA}, as well as the atomic-scale mechanism of plasmonic hot-carrier-mediated chemical processes such as H$_2$ dissociation~\cite{zhang2018plasmonic}. 
Our numerical simulations showed that after photoexcitation, hot carriers transfer to the antibonding state of the H$_2$ molecule from the nanoparticle, leading to a repulsive-potential-energy surface and H$_2$ dissociation (Figure~\ref{fig: Jellium}(b-d)). This process occurs when the molecule is close to a single nanoparticle. However, in a plasmonic dimer, dissociation can be inhibited due to sequential charge transfer that effectively reduces the occupation of the antibonding state, as shown in Figure~\ref{fig:jellium}(e). When the molecule is asymmetrically positioned in the gap, the symmetry is broken, and dissociation is restored by significantly suppressing additional charge transfer. Thus, these models illustrate the potential for structurally adjustable photochemistry through plasmonic hot carriers.

\subsubsection{TDDFT calculations.} 
However, the Jellium model oversimplifies the electronic structures of the plasmonic nanostructures, and it lacks atomistic details. Insights from PESs built on the ab initio atomistic model are essential for a more in-depth understanding of plasmon chemistry. Thus, we employed the linear response TDDFT (LR-TDDFT) calculations within the Casida formalism to compute the adiabatic PESs of the H$_2$ molecule adsorbed on an Au6 cluster (H2@Au6)~\cite{wu2020mechanistic}, in order to explore key pathways in LSPR-promoted chemical reactions. Despite the model system being too small to support plasmonic mode and thus cannot describe the dephasing of plasmons that produce hot electrons, the key point of using the model system is to capture key aspects of the later stages of plasmon-facilitated photocatalysis, thus providing mechanistic insights.

Our findings based on DFT calculations indicate that in the ground state, the Au6 cluster supports the adsorption of the H$_2$ molecule at the tip site. We further use LR-TDDFT calculations within the Casida formalism to determine the adiabatic excited states and corresponding oscillator strengths. We calculate two-dimensional PESs in the desorption and dissociation reaction coordinates and observe that the adiabatic excited PESs bear similarity to the ground state PES in the Franck-Condon region. Therefore, the H$_2$ adsorbate is comparatively stable, as corroborated by the relatively low dissociation and desorption probabilities from our quantum dynamics simulations. By developing the orbital wave function overlapping (OWO) diabaization scheme, we are able to divide the dense manifold of the excited state into two groups: 1) one group is dominated by electronic excitations confined to the Au$_6$ cluster, which can be likened to HE states in metal nanoclusters; 2) the other group of excited states has the antibonding $\sigma^*$ characteristics of the H2 adsorbate due to the hybridization between H$_2$ and Au$_6$ cluster, which are denoted as CT states. The crossings among the HE and CT states provide pathways leading from the excited HE states to CT states via nonadiabatic transitions. Quantum dynamics simulations on the diabatic PESs demonstrate that the CT diabatic states are able to drive H$_2$ dissociation efficiently and thus are responsible for the experimentally observed HD formation on Au nanoparticles. Our results nevertheless give a clear physical picture of photoinduced H$_2$ dissociation on Au clusters. 

\begin{figure}[!htb]
    \centering
    \vspace{-10pt}
    \includegraphics[width=0.5\textwidth]{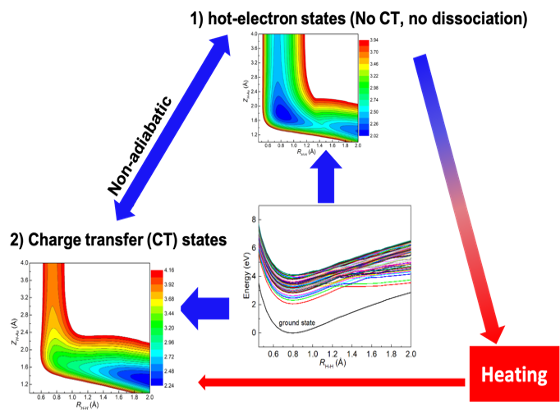}
    \caption{
    Schematic diagram of competition among different pathways in plasmon-mediated chemical reactions on a dense manifold of excited states. The excited states can be divided into HE and CT states. The CT states are responsible for the chemical reaction, which can be triggered by the nonadiabatic transition between HE and CT states. Such nonadiabatic transitions have to compete with hot carrier relaxations that lead to local heating.
    }
    \label{fig:pathway}
\end{figure}

The presence of HE-CT crossings is not unique to plasmonic catalysis. Such features have also been found in other nonplasmonic catalysis, such as those on the surface of semiconductors~\cite{akimov_theoretical_ChemRev2013}. However, plasmonic materials are unique in generating a high concentration of HE due to their larger absorption cross sections. The combination of a high concentration of HE and HE-CT crossings is thus the distinctive feature of plasmonic catalysis.

\subsubsection{TDDFTB-based NAMD simulations}
As shown in Figure~\ref{fig:pathway}, plasmon-mediated chemistry involves a dense manifold of excited states, nonadiabatic transitions between HE and CT states, and their competition with various dissipation channels. Such complicated processes require nonadiabatic simulations that treat the HE generation, relaxation, and HE-CT transitions treated on equal footing. To this end, larger clusters should be used in modeling plasmonic catalysis, and more efficient semiempirical methods, such as those based on time-dependent density functional tight binding (TDDFTB)~\cite{trani_TDDFTB_JCTC2011}, might be needed for the efficient electronic structure calculations. To this aim, We recently developed an efficient NAMD method by combining the TSH algorithm and LR-TDDFTB method.~\cite{niehaus2001tight, malone2020nexmd, song2020first, wu2022nonadiabatic} LR-TDDFTB is the tight-binding extension of TDDFT,~\cite{hourahine2020dftb+, niehaus2001tight} which has been successfully applied to investigate the optical properties of plasmonic NPs.~\cite{douglas2019plasmon, berdakin2019interplay} It enables calculations of plasmonic excitation along with several hundred excitation states per time step. The relaxation processes of plasmon induced by electron-phonon interactions are treated by the TSH algorithm. With the NAMD-DFTB method, we demonstrate the plasmon relaxation of an Au$_{20}$ cluster. Our simulations show that Au$_{20}$ can support plasmon-like excitation, which includes the superposition of multiple single-particle excitation components. 

The numerically efficient LR-TDDFTB method allows us to address a dense manifold of excited states to ensure the inclusion of plasmon excitation. Starting from the photoexcited plasmon states in Au$_{20}$ cluster, we find that the time constant for relaxation from plasmon excited states to the lowest excited states is about 2.7~ps, mainly resulting from a step-wise decay process caused by low-frequency phonons of the Au$_{20}$ cluster. Furthermore, our simulations show that the lifetime of the phonon-induced plasmon dephasing is $\sim$10.4~fs, and such a swift process can be attributed to the strong nonadiabatic effect in small clusters. Our simulations demonstrate a detailed description of the dynamic processes in nanoclusters, including plasmon excitation, hot carrier generation from the plasmon excitation dephasing, and the subsequent phonon-induced relaxation process.

The NEXMD-DFTB method was employed to investigate the plasmon-induced bond activation of CO adsorbed on Au${20}$ cluster~\cite{wu2023chemsci}. The simulations provide a comprehensive and accurate depiction of the multiple dynamic processes involved in plasmon-mediated chemistry, including plasmon excitation, HEs relaxation, direct/indirect HE transfer, and the activation of CO vibration mode induced by HE transfers. These simulations reveal the critical role of charge transfer (CT) states in plasmon-induced CO activation. The CT states excite both direct and indirect HE transfer, which leads to the activation of CO stretching mode, an essential component of plasmon energy relaxation. Our simulations demonstrate the high efficiency of CO vibrational mode activation, achieving a success rate of approximately 40\%. Notably, the HE transfer occurs at a faster rate than the conventional scattering process of the Au${20}$, completing within approximately 100 fs, while the energy relaxation takes place over a timescale of approximately 1 ps. Furthermore, our direct atomistic simulations provide detailed insights into the potential energy evolution during the plasmon-mediated chemical transformation, thereby enabling a comprehensive understanding of the energy relaxation and HE transfers during the reaction, as well as elucidating the reaction pathways.

\section{Polariton chemistry}\label{SEC:POLARITON}

Since there are several recent experimental and theoretical reviews on polariton chemistry~\cite{mandal_ChemRev2022, foley:2023arxiv, Herrera:2020jcp, Simpkins:2023tm, Dunkelberger:2022uo, Ebbesen2016ACR},  only a brief introduction to the recent theoretical development of polariton chemistry will be reviewed in this perspective. Instead, we will focus on our perspective on the theoretical challenges and ongoing and future development of quantum many-body and multiscale methods toward solving the mysteries of polariton chemistry and beyond.

\subsection{Introduction and background}

As introduced in Sec.~\ref{sec:basics}, in the strong light-matter coupling limit (strong enough to compete with individual DOF's dissipation rate), the formed quasiparticles (polaritons) can alter the potential energy landscape and consequently manipulate the chemical processes. 
Generally, the strong coupling regime is characterized by the coherent energy exchange between the photon field and the electronic emitter (\textit{i.e.} Rabi oscillations). In this regime, the electronic and photon subsystems can no longer be treated separately, making the accurate simulation of polariton dynamics very challenging. In this section, we explore various theoretical approaches in modeling the light-matter interactions present in optical cavities, such as Fabry–P\'{e}rot, where the coherent exchange in energy between all DOFs directly affects the resulting chemistry. Here, the electronic and photonic DOFs must be treated quantum mechanically on equal footing.

Recent experiments have shown a propensity to change chemistry via the coupling of quantized radiation and various molecular DOFs, namely electronic and vibrational strong coupling. The Fabry–P\'{e}rot-like cavities offer an extremely tunable cavity frequency (via the effective length between the cavity mirrors) while exhibiting widely varying coupling strengths highly dependent on the experimental setup~\cite{Hutchison2012ACIE, Hutchison2012ACIE, Mauro2021PRB, George2016PRL, Vergauwe2019ACIE, Lather2019ACIE, Sau2021ACIE, Hirai2021CS, Takele2021JPCC, Thomas2019S, Lather2022CS}. There are many open questions regarding these experiments, such as the collective effects (\textit{i.e.}, many-molecule or many-mode effects), which are present in many of the recent works due to the complexity of performing single-molecule experiments.

\subsection{Theoretical and Computational Challenges} 

There are conceptual and technical challenges in performing simulations of polaritonic systems. Like the situation in plasmon chemistry, the challenges in understanding polariton chemistry still lie in the multiple coherent and dissipative processes across various time/length scales.
\begin{enumerate}
 \item The first theoretical hurdle is that of the Hamiltonian itself. For the Pauli-Fierz Hamiltonian (Eq.~\ref{EQ:H_PF}), the light-matter interaction requires explicit knowledge of the molecular dipole operator $\hat{D}$ in the working basis. It turns out that a significant simplification can be made if one neglects all the contributions from the dipole matrix except for the ground-to-excited transition dipole matrix elements. There are two primary reasons for making this approximation: (I) The entirety of this quantity is not usually printed by default when executing standard electronic structure software for electronically excited states; however, the usual information given from these calculations is the ground-to-excited transition dipole moment, $D_{0J}$, which, for example in linear response calculations (\textit{e.g.}, TD-HF or TD-DFT), is a trivial quantity to achieve since the result of such schemes is the ground-to-excited transition density.\cite{tretiak2002CR} (II) Historically, the field of quantum optics was narrowly focused on the light-matter interaction between single atoms and the quantized cavity field. In such cases, the diagonal elements of the dipole operator are zero by construction, and the electronic energy differences between excited states were large, leading to neglecting the excited-to-excited dipole coupling. With these approximations, one would find that a reduced Hamiltonian can be achieved, commonly referred to as the Jaynes-Cumming (JC) Hamiltonian.
 \item Generally, the strong coupling relies on a larger number of molecules collectively coupled to a cavity, introducing significant challenges to computing the hybridized polaritonic states of many ($10^6$ or even larger) molecules in the cavity.
 \item While plasmonic cavities are able to achieve strong coupling in a few (or even single) molecule limits, these cavities are highly heterogeneous and dispersive. Consequently, multiple models, dispersion of the cavities, coupling beyond the dipole approximation, and the inherent strong dissipation should be taken into account, introducing different computational complexity. Besides, the molecules inside the nanoplasmonic cavities can also interact with the ones outside the cavity. For example, a recent 2DES investigation of J-aggregates in the plasmonic cavity reveals rich photophysical dynamics between the polaritonic states inside the cavity and uncoupled exciton states outside the cavity~\cite{timmer_plasmon_2023}, which indicates that the molecules outside the cavity cannot be trivially ignored.
 \item The damping of the polariton state is usually ignored or introduced by an empirical lifetime of cavity photon (cavity leakage). In principle, cavity leakage depends on the photonic density of states, which can be computed from the first principles. Besides, the light-matter coupling could affect the photonic DOF and, in return, the photon lifetime. Hence, one open question is to what extent the strong coupling changes the photon lifetime, which requires a multiscale method to address the feedback of molecular systems on the cavities and first-principles calculations of cavity leakage.
 \item Polariton-mediated phenomena are fundamentally kinetic procedures involving the dynamic interplay between electronic, nuclear, and photonic DOFs at various time and length scales. Indeed, the accurate simulation of the quantum dynamics of a molecule itself is already a challenging task~\cite{Curchod2018CR, Crespo-Otero:2018wa, Nelson2020ChemRev}. The inclusion of photonic DOFs in polariton dynamics adds further complexity. To accurately simulate polariton dynamics, efficient and precise calculations of the gradients and derivative coupling of polariton states are essential. Despite the recent advancements in accurately computing gradients of polariton states~\cite{mandal_ChemRev2022}, more efforts are needed for large-scale ab initio simulations of polariton dynamics of many molecules.
 \item Polariton-mediated phenomena are fundamentally kinetic procedures involving the dynamic interplay between electronic, nuclear, and photonic DOFs at various time and length scales. Indeed, the accurate simulation of the quantum dynamics of a molecule itself is already a challenging task. The inclusion of photonic DOFs in polariton dynamics adds further complexity. To accurately simulate polariton dynamics, efficient and precise calculations of polariton states' gradients and derivative coupling are essential. Despite the recent advancements in accurately computing gradients of polariton states~\cite{mandal_ChemRev2022}, more efforts are needed for large-scale ab initio simulations of polariton dynamics in multiple molecules.
\end{enumerate}
Hence, simulating polariton chemistry is non-trivial, and state-of-the-art theoretical models and efficient numerical methods are crucial for understanding the polariton dynamics of many molecules.

\subsection{Common Approximations Toward Historic Quantum Optics Models}

The PF Hamiltonian (as written in Eq.~\ref{EQ:H_PF}) is derived by applying the {\bf dipole approximation}, which assumes the wavelength of the cavity fields is substantially larger than the matter system so that the spatial dependence of the transverse fields is neglected. In addition, many of the recent works on the simulation of \textit{ab initio} polaritons have relied on approximate versions of the PF Hamiltonian that stem from historical applications in the quantum optics community.\cite{yang2021JCP,zhang2019jcp,Tichauer2021JCP,Groenhof2018JPCL,Luk2017JCTC,Groenhof2019JPCL,Tichauer2022JPCL} In these heavily approximated Hamiltonians, usually a large truncation of the electronic and photonic subspaces is also performed such that only the ground and a single electronic excited state, $|g\rangle$ and $|e\rangle$ are included while only including the vacuum and singly excited Fock states, $|0\rangle$ and $|1\rangle$ in a single-mode cavity $\alpha = 0$. The total basis for this simple model is then confined to $\{|g,0\rangle,|g,1\rangle,|e,0\rangle,|e,1\rangle\}$. Note here that another approximation is that there is no permanent dipole in the ground $|g\rangle$ or excited $|e\rangle$ electronic states, \textit{i.e.}, $\bD_{gg} = \bD_{ee} = 0$. The most commonly used Hamiltonian for modeling \textit{ab initio} polaritons is the {\bf Jaynes-Cummings (JC)} Hamiltonian $\hH_\mathrm{JC}$, which can be written as,
\begin{equation}\label{EQ:H_JC}
    \hH_\mathrm{JC} = \hH_m + \hH_p + \sqrt{\frac{\omega_\mathrm{c}}{2}} \blambda \cdot \bD_{ge} (\hat{\sigma}\ha^\dag + \hat{\sigma}^\dag\ha),
\end{equation}
where $\hat{\sigma}^\dag$ ($\hat{\sigma}$) is the creation (annihilation) operator for the molecule excitation between the ground $g$ and excited $e$ states. And the {\bf Tavis-Cummings (TC)}~\cite{Tavis1968PR} model is the many molecules generalization of the JC model. 4) Besides, the feedback of molecular systems on the cavity modes (via the current density) is usually ignored. Here, two approximations have been made to derive JC/TC models: (I) the {\bf rotating wave approximation (RWA)} -- which is to say, neglecting the highly oscillatory $\hat{\sigma}^\dag\ha^\dag_\alpha$ and $\hat{\sigma}\ha_\alpha$ terms in the light-matter interaction -- and (II) neglecting the DSE $\hH_\mathrm{DSE}$. This Hamiltonian is valid at ultra-low coupling strengths where the splitting between the one-photon-dressed ground state $|g,1\rangle$ and the excited state with zero photons $|e,0\rangle$ exhibit linear splitting (\textit{e.g.} Rabi splitting) with an increase in the coupling strength $\blambda$. The other two basis states $|g,0\rangle$ and $|e,1\rangle$ are completely decoupled from the interaction.

Two other common approximations to the PF Hamiltonian are the {\bf Rabi model} $\hH_\mathrm{Rabi}$ after explicitly dropping the dipole self-energy (DSE) term and the RWA $\hH_\mathrm{RWA}$. These two approximate Hamiltonians can be written as,
\begin{align}\label{EQ:H_RABI__H_RWA}
    \hH_\mathrm{Rabi} &= \hH_\mathrm{M} + \hH_\mathrm{p} + \sqrt{\frac{\omega}{2}} \blambda \cdot \hat{\bD}(\hat{\sigma} + \hat{\sigma}^\dag) (\ha^\dag + \ha),\\
    \hH_\mathrm{RWA} &= \hH_\mathrm{M} + \hH_\mathrm{p} + \sqrt{\frac{\omega}{2}} \blambda \cdot \bD_{ge} (\hat{\sigma}_{ge}\ha^\dag + \hat{\sigma}_{ge}^\dag\ha) + \hH_\mathrm{DSE},
\end{align}
where the Rabi Hamiltonian $\hH_\mathrm{Rabi} = \hH_\mathrm{PF} - \hH_\mathrm{DSE}$ and the RWA Hamiltonian applies the RWA to the PF Hamiltonian. Here, $\hat{\sigma}_{ge} = |g\rangle\langle e|$ is the annihilation operator of the two-level electronic system. A more in-depth discussion on these Hamiltonians and the effects of the dipole self-energy term can be found in Refs.~\citenum{mandal_ChemRev2022},~\citenum{yang2021JCP},~\citenum{Rokaj2018JPB}, and~\citenum{Mandal_QED_eT_JPCB2020}.


Since the correct PF Hamiltonian in the dipole gauge has been known, the question remains of why the community returns to the approximated Hamiltonians for \textit{ab initio} as well as model calculations. There are many subtleties to using the full PF Hamiltonian. From the electronic structure perspective, the many-level dipole operator needs to be computed as well as its square. The DSE term provides a very complicated description of the system for large coupling strengths since the dipole matrix, $\hat{\bD}$, in realistic molecules is far from sparse with its square leading to further complications.\cite{mandal_ChemRev2022,weight_abQED_JPCL2023} This allows for strong coupling between arbitrary states that is not trivial to know \textit{a priori} based on chemical or physical intuition. Often, these Hamiltonians are parameterized based on cavity-free electronic structure calculations to obtain the energies and dipoles of the electronic adiabatic states (see more details in Sec.~\ref{SEC:Direct_DIAG}). In this case, the number of included electronic (as well as photonic) basis states should be treated as a convergence parameter. In this case, the DSE causes mixing between far-separated-in-energy electronic states, which leads to slow convergence in the basis set size for the matter DOFs,\cite{yang2021JCP} and in general poor results compared to benchmarks, especially at large light-matter coupling strengths when using a small electronic basis.\cite{Flick2020JCP,wang_dissipative_2021JCP, mctague_nhCIS_JCP2022} This is precisely why many authors are developing self-consistent formulations to construct the low-energy polaritonic states without the need to calculate all the high-energy electronic states (see Sec.~\ref{SEC:Direct_DIAG}).\cite{Flick2020JCP,Vu_enhanced_JPCA2022,mctague_nhCIS_JCP2022,haugland2020PRX,weight_abQED_JPCL2023,Haugland_intermolecular_JCP2021,deprince_IP_EA_2021JCP,liebenthal_EOMCC_2022JCP}. Further when considering the quantum dynamical propagation of polaritons in \textit{ab initio} systems, additional nuclear gradients are required compared to cavity-free simulations where only the gradients of the adiabatic state energies are required to propagation the quantum dynamics, namely the nuclear gradients on the adiabatic dipole matrix and its square (\textit{i.e.}, $\nabla_{R} \hat{\bD}$ and $\nabla_{R} \hat{\bD}^2$).\cite{zhang2019jcp}

\subsection{Theoretical modeling of polariton chemistry}

\subsubsection{Cavity Born-Oppenheimer approximation}
Within the Born-Oppenheimer (BO) approximation,\cite{Flick2017JCTC} the total electronic-photonic-nuclear can be factorized as,
\begin{equation}
    \Phi(\br,\bR,{\bf q}_\alpha)=\chi(\bR)\Psi(\br,{\bf q}_\alpha;\bR),
\end{equation}
where $\chi(\bR)$ and $\Psi(\br,{\bf q}_\alpha;\bR)$ are the nuclear and polaritonic (\textit{i.e.,} electronic and photonic) wavefunctions, respectively. Note here that the polaritonic wavefunction is parameterized by the nuclear positions, exactly like the case without photonic DOFs outside the cavity. Further, one can invoke the usual Born-Huang-like expansion over the Born-Oppenheimer factorization as,
\begin{equation}
    \Phi(\br,\bR,{\bf q}_\alpha)=\sum_{\mu}\chi_a(\bR)\psi_\mu(\br,{\bf q}_\alpha;\bR),
\end{equation}
where $\psi_\mu(\br,{\bf q}_\alpha;\bR)$ are the BO wavefunctions analogous to those as outputted in standard electronic structure packages for the ground and excited adiabatic states. In this basis, which we will call the adiabatic polaritonic basis to draw a direct connection to the bare electronic case, we will discuss various ways to calculate such polaritonic wavefunctions $\psi_\mu(\br,{\bf q}_\alpha;\bR)$ from \textit{ab initio} calculations in a variety of approaches and levels of approximation.

In the following three sections, we will explore ways to obtain rigorous nuclear-position-parameterized wavefunctions for the entangled adiabatic electron-photon states $\psi_\mu(\br,{\bf q}_\alpha;\bR)$. First, a brief description of a direct diagonalization approach (Sec.~\ref{SEC:Direct_DIAG}) with Hamiltonians parameterized with information from standard electronic structure, while the following sections (Secs.~\ref{SEC:scQED_SP} and~\ref{SEC:scQED_CCSD}) will focus on the self-consistent approach toward re-developing the standard many-body schemes in electronic structure theory for the QED Hamiltonian (\textit{e.g.}, QED-HF, QED-DFT, etc.).

\subsubsection{Direct Diagonalization}\label{SEC:Direct_DIAG}

The polaritonic state can be readily computed via exact diagonalization of polariton Hamiltonian (either JC, TC, Rabi, RWA, or rigorous PF) in a certain basis set. For instance, 
the matrix elements of the single-mode PF Hamiltonian in the widely used adiabatic-Fock basis 
can be written as,
\begin{align}
    (\hH_\mathrm{PF})_{IJ,nm} &= \bigg[ E_{I} + \omega_\alpha(n + \frac{1}{2})\bigg]\delta_{IJ}\delta_{nm} \\
    &+ \sqrt{\frac{\omega_\alpha}{2}} \blambda \cdot \bD_{IJ} (\sqrt{m+1}\delta_{n,m+1} + \sqrt{m}\delta_{n,m-1})\nonumber  \\
    &+ \frac{1}{2} \sum_{K}^{\mathcal{N}_\mathrm{el}} (\blambda \cdot \bD_{IK}) (\blambda \cdot \bD_{KJ}) \delta_{nm}\nonumber,
\end{align}
where $\mathcal{N}_\mathrm{el}$ is the number of adiabatic electronic states included in the basis. It is important to note that in this basis, the PF Hamiltonian is extremely sparse since the coupling elements only connect adjacent Fock states via the molecular dipole matrix since the matrix elements of the photonic coordinate $\hat{q}$ are that of the harmonic oscillator (\textit{i.e.}, only have nonzero super- and sub-diagonal elements). Further, the DSE contributions, in general, connect all of the electronic states of the system (with the same photon number) and is the most non-trivial aspect of this Hamiltonian and will vary strongly between molecular systems. 

Except for the adiabatic-Fock basis, other options for expanding the JC and PF Hamiltonians exist, such as the coherent states\cite{philbin2014AmerJourPhys,haugland2020PRX,riso_molecularQED_NatComm2022} or polarized Fock states.\cite{Mandal2019JPCL} In both of these bases, the photonic states are chosen such that the molecular dipole parameterizes the photonic state, thereby, in principle, reducing the convergence of the photonic basis. Further, this ``direct diagonalization'' approach to solving the PF Hamiltonian is dependent on this basis convergence, and the electronic basis is much more rigid since the adiabatic basis is ubiquitously used for its convenience. However, the convergence of this basis has yet to be thoroughly tested for a wide range of systems in solving the PF Hamiltonian, but it is expected to converge slowly due to the contributions from the DSE term.\cite{weight_abQED_JPCL2023,yang2021JCP} This evidences the need to move to a more rigorous self-consistent solution for the polaritonic adiabatic states as is done for the electronic adiabatic states themselves.

\subsubsection{Self-consistent Polaritonic Single-particle Approaches}\label{SEC:scQED_SP}

{\bf scQED Hartree-Fock and Density Functional Theories}\newline
The mean-field approach to the QED Hamiltonian can be first cast in an identical way as the standard Hartree-Fock procedure in a larger Hilbert space, including the photonic states~\cite{riso_molecularQED_NatComm2022,riso_QEDionization_JCP2022,deprince_IP_EA_2021JCP,liebenthal_EOMCC_2022JCP,mctague_nhCIS_JCP2022}. A useful basis, referred to as coherent states (CS), can be performed such that the light-matter interaction part of the PF Hamiltonian can be shifted away. 

For a given bare electronic ground state wavefunction ($|\mathrm{HF}\rangle$), the photonic Hamiltonian can be obtained by integrating out the electronic DOF,
\begin{align}\label{EQ:PF_HF_partialDIAG}
\hat{H}_p \equiv \langle \mathrm{HF}&| \hat{H}_\mathrm{PF} |\mathrm{HF}\rangle = E_\mathrm{HF} +  \omega_\alpha \big(\hat{a}^\dag_\alpha \hat{a}_\alpha+\frac{1}{2}\big)\\
&+\sqrt{\frac{\omega_\alpha}{2}} \langle \blambda_\alpha \cdot \hat{\bD} \rangle_\mathrm{HF} \cdot (\hat{a}^\dag_\alpha + \hat{a}_\alpha)+ \frac{1}{2} \langle (\blambda_\alpha \cdot \hat{\bD})^2\rangle_\mathrm{HF},\nonumber
\end{align}
where $\langle \cdots \rangle_\mathrm{HF} = \langle \mathrm{HF} | \cdots | \mathrm{HF} \rangle$ is the HF ground state expectation value of the electronic subsystem. 
The photonic Hamiltonian can be trivially diagonalized 
by introducing the CS transformation $\hat{U}({\bf z})\hat{H}_p\hat{U}^{\dagger}({\bf z})$,
\begin{equation}\label{EQ:CS_ROTATION}
    \hat{U}({z_\alpha}) = e^{z_\alpha\hat{a}_\alpha^\dag - z_\alpha^*\hat{a}_\alpha},
\end{equation}
where ${\bf z} = \{z_\alpha\}$ is a vector of complex numbers specific to each cavity mode,
\begin{equation}
    {\bf z} \rightarrow -\frac{\langle \blambda_\alpha \cdot \hat{\bD}\rangle_\mathrm{HF}}{\sqrt{2\omega_\alpha}}.
\end{equation}
Thus, considering a QED-HF ansatz as $\ket{HF}\otimes\hat{U}({\bf z})\ket{0}$, the QED-HF energy and corresponding Fock matrix can be derived. 
Compared to the HF theory for bare electrons, the QED-HF method introduces DSE-mediated one-body and two-body integrals~\cite{foley:2023arxiv}.

Similarly to the scQED-HF approach, the self-consistent QED density functional theory (scQED-DFT) is composed in a similar way, where the main elements of DFT remain, such as the exchange-correlation functional of the density. 
There are many ways to set up the scQED-DFT problem, such as employing a novel exchange-correlation functional to account for light-matter correlation effects\cite{pellegrini_OEP_PRL2015,ruggenthaler_QEDFT_2014,ruggenthaler_QEDSpectra_NatRevChem2018,flick_KSQED_PNAS2015} or working with electron-only exchange-correlation functionals using the coherent state basis for the photonic DOFs.\cite{Vu_enhanced_JPCA2022,yang2021JCP,haugland2020PRX} In any case, the resulting ground state is uncorrelated by the nature of the DFT formalism.

{\bf Excited States scQED-TD-(HF,DFT)}

The time-dependent analogues to the aforementioned single-particle approaches are powerful tools to probe non-equilibrium densities that give rise to electronic excited states. One of the most popular approaches is one of linear response (LR), resulting in LR-TD-HF and LR-TD-DFT for bare molecular systems in the random phase approximation (RPA). Although, it should be noted that the real-time propagation of the single-particle density matrix -- leading to the real-time TD-HF and real-time TD-DFT approaches -- is, in principle, a more robust approach but one that is usually more costly than that of linear response. Such schemes have already been developed for the simulation of molecular polaritons using classical photon DOFs.\cite{Li_QEDNEO_JCTC2022,li_QEDNEO_2023}  Here, we will focus our attention on the LR formalism, specifically using a Casida-like approach to writing the random phase approximation (RPA), originally formulated by Flick and co-workers\cite{Flick2020JCP} using the QEDFT (or scQED-DFT in the notation of this0 work) method in the language of Casida and further used by the groups of Shao\cite{yang2021JCP} and DePrince.\cite{Vu_enhanced_JPCA2022} It should be noted that other formulations of CIS-like excited states can be found in the community, such as the non-Hermitian CIS aimed at simulating cavity loss via a complex photon frequency\cite{mctague_nhCIS_JCP2022}.

As per usual, and following the notation of Ref.~\citenum{Vu_enhanced_JPCA2022}, the LR-TD-HF and LR-TD-HF eigenvalue equations using the Casida formalism can be written to include the QED components that satisfy the Pauli-Fierz Hamiltonian (Eq.~\ref{EQ:H_PF}).

An important distinction between various implementations of the QED-TD-DFT approaches in the community is whether the single-particle orbitals used in the formulation are ``relaxed'' in the presence of the cavity or are simply the bare electronic single-particle states. For example, in Ref.~\citenum{yang2021JCP}, the orbitals are not relaxed while in the Refs.~\citenum{Flick2020JCP} and~\citenum{Vu_enhanced_JPCA2022} the orbitals are relaxed. While it is clear that using a relaxed reference state for the basis of the RPA equations would provide a more rigorous result, it is not clear whether identical results can be obtained in the infinite basis limit of both approaches, \textit{i.e.}, including more single-particle states in the CIS-like expansion in excited Slater determinants. Since the RPA equations are iteratively solved, the expansion coefficients of the excited Slater determinants may result in the same excited state observables in the infinite basis limit, while for a finite basis, it may not. 

\subsubsection{Self-consistent Polaritonic Coupled Cluster Approaches (CC,EOM-CC)}\label{SEC:scQED_CCSD}

Despite computational efficiency, DFT or mean-field HF methods usually underestimate the correlations. In particular, the mean-field method cannot describe the electron-photon and photon-mediated electron-electron correlations, and the exchange-correlation function for electron-photon interaction is unknown. Consequently, the QED counterpart of coupled-cluster theory (QED-CC) is proposed.\cite{deprince_IP_EA_2021JCP,Haugland_intermolecular_JCP2021,liebenthal_EOMCC_2022JCP,mordovina_QEDCC_PRR2020,pavosevic_ClickChem_Arxiv2022,pavosevic_PTQED_JACS2022,fregoni_QEDCCPlasmon_NanoLett2021} Similar to the conventional CC theory, QED-CC employs an exponential wavefunction Ansatz to derive the ground state, 
\begin{equation}\label{eq_ansatz}
    \ket{\Psi_{CC}}=e^{\hT}\ket{\Phi_0},
\end{equation}
where $\ket{\Phi_0}$ is the reference wave function, which is usually chosen to be the tensor product of  Hartree-Fock (HF) determinant and photon vacuum state, i.e., $\ket{\Phi_0}=\ket{\varphi_0}\otimes\ket{0}$. $\hT$ is defined in excitation configurations. Within QED-CC theory, the generalized excitation operator includes three components, $\hT=\hT_e+\hT_p+\hT_{ep}$, including electronic ($\hT_e$), photonic ($\hT_p$), and coupled electronic-photonic ($\hT_{ep}$) excitations,
\begin{align}
T_e=&\sum_{ia} t^a_i \ha^\dag_a\ha_i + \sum_{ijab} t^{ab}_{ij} \ha^\dag_a\ha^\dag_b\ha_j\ha_i +\cdots\equiv \sum^{N_e}_{\mu} t_{\mu} \htau_\mu, 
\\ 
\quad \hT_{p}= & \sum_\alpha \gamma_\alpha \hb^\dag_\alpha +
\frac{1}{2}\sum_{\alpha\beta} \gamma_{\alpha\beta} \hb^\dag_\alpha\hb^\dag_\beta + \cdots \equiv \sum^{N_p}_n \gamma_n \hat{B}_n,
\\
\quad \hT_{ep}=&\sum_{ia,\alpha} t^a_i \ha^\dag_a\ha_i \hb^\dag_\alpha + \frac{1}{2}\sum_{ijab,\alpha\beta} t^{ab}_{ij} \ha^\dag_a\ha^\dag_b\ha_j\ha_i  \hb^\dag_\alpha\hb^\dag_\beta
+ \cdots \equiv 
\sum_{\mu,n} \chi_{\mu,n}\htau_\mu \hat{B}_n.
\end{align}
where $N_e$ and $N_f$ are the numbers of electrons and photon Fock states, respectively. $\htau_\mu=\prod^\mu_k \hE^{a_k}_{i_k}$ and $\hE^a_i=\ha^\dag_a \ha_i + h.c.$ are the $\mu$-body and single-body excitation operators, respectively. $\hat{B}_n=\prod^n_{\alpha} b^\dag_\alpha$ is the $n$-body photonic excitation. 
The parameters ($t_\mu, \gamma_n, \chi_{\mu,n}$) are the cluster amplitudes. By projecting CC wavefunction into a set of orthogonal excited configurations ($\{\ket{\mu}\}$),
\begin{equation}
    \ket{\mu}=\hat{\mu}\ket{\Phi_0},
\end{equation}
where $\mu\in\{\htau_\mu, \hat{B}_n, \htau_\mu \hat{B}_n\}$ is the cluster operator,
The cluster amplitudes can be determined from the projected equation by following the standard CC procedure,
\begin{equation}
    \Omega_\mu\equiv\bra{\mu}e^{-\hT}\hH e^{\hT}\ket{\Phi_0}=0
\end{equation}
Though the scQED-CC theory leads to the FCI solution if no truncation on the excitation operator $\hT$ is applied, the excitation operator is usually truncated at the doubles (CCSD) level in order to trade-off between accuracy and computational efficiency. Some tests have been performed on the level of truncation in the photonic excitations (up to 10) in the CC operator for model systems.\cite{mordovina_QEDCC_PRR2020} However, the scaling of scQED-CCSD is $\mathcal{O}(N_{el}^6 N^{M_p}_{p})$ in general, where $M_p$ is the number of photon modes. Such a scaling makes it a bit expensive to rigorously test in extended molecular systems that include many electrons and/or when many photon modes and Fock states are used in the calculations. To further reduce the computational cost, the coherence state~\cite{Philbin:2014jcp} may be used to reduce the photon basis in the scQED-CC calculations.

\begin{figure}[t!]
    \centering
    \includegraphics[width=0.4\textwidth]{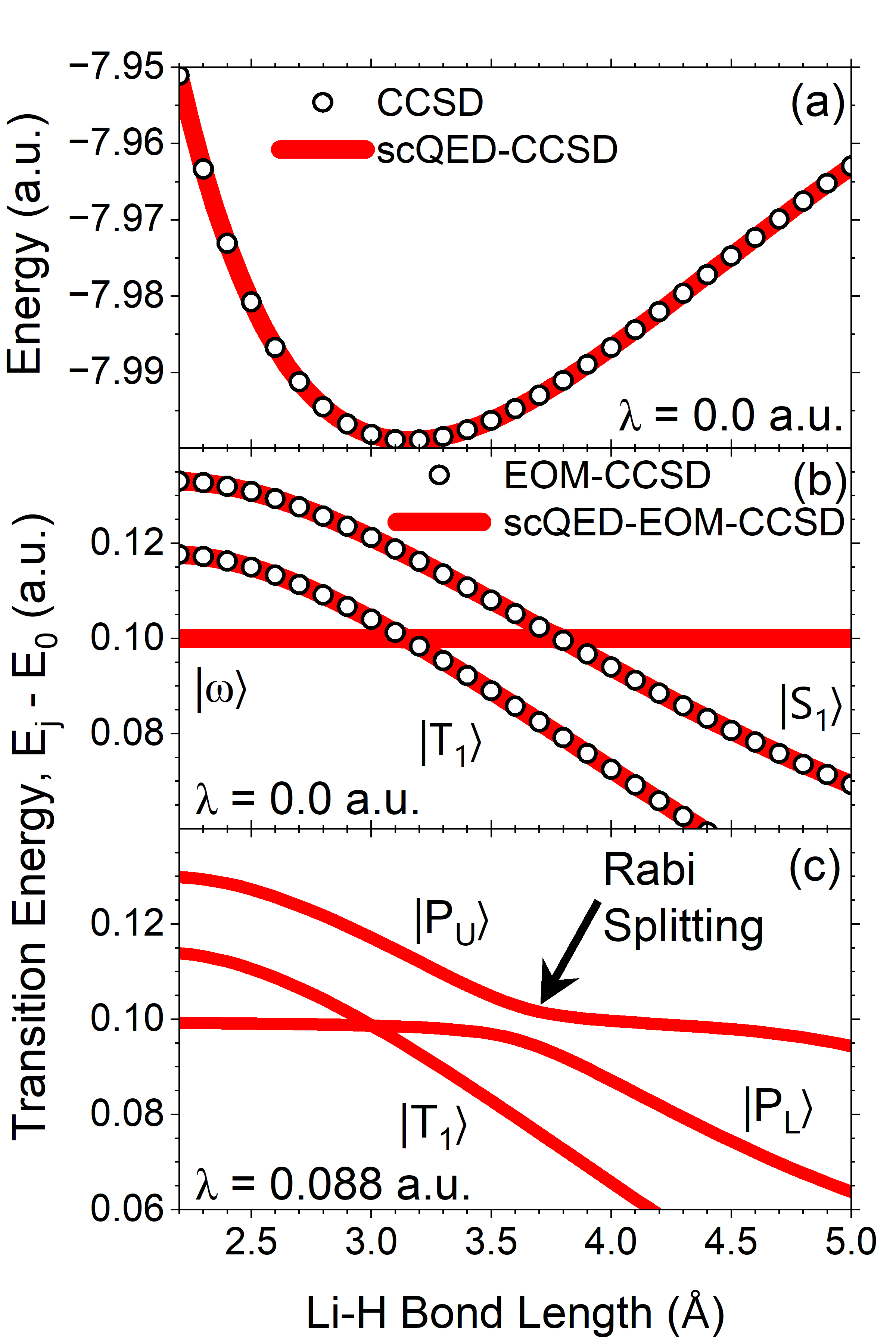}
    \caption{Polariton states of LiH as a function of Li-H bond length. (a) CCSD (black circles) and scQED-CCSD (solid red curve) ground states at zero light-matter coupling strength $\lambda = 0$ a.u. (b) EOM-CCSD (black circles) triplet $|T_1\rangle$ and singlet $|S_1\rangle$ states as well as the low-lying states ($|T_1\rangle$, $|\omega\rangle$, and $|S_1|rangle$) calculated at scQED-EOM-CCSD with light-matter coupling strength $\lambda = 0$ a.u. The plot clearly shows the photon line ($|\omega\rangle$) and perfect overlap of the $S_1$ states. (c) Upper ($|P_U\rangle$) and lower ($|P_L\rangle$) polariton states with $\lambda = 0.088$ a.u. with the triplet state $|T_1\rangle$ uncoupled.}
    \label{fig:lih_r}
\end{figure}

It should be noted that CC methods exhibit size extensivity even in their truncated forms, meaning that the sum of the energies of non-interacting subsystems is equal to the total energy. This is in contrast to CI approaches, where errors in extensivity can become increasingly large with an increasing number of subsystems. In addition, CC theory has the advantage of size extensivity in that the excitation energies of each subsystem do not vary with the size of the total system when the subsystems are non-interacting, making it an ideal candidate for computing the polaritonic eigenstates of many (non-interacting) molecules. 

In addition, the excited polariton states can be computed with the corresponding EOM formalism, which parameterizes a neutral or charged excitation by applying an excitation operator to the CC ground state~\cite{jcp0033132}
\begin{equation}
    \ket{\hR}=\hR\ket{\Phi_{cc}}=\hR e^{\hT}\ket{\Phi_0,0}.
\end{equation}
Where $\hR$ also include the electronic ($\hR_e$), photonic ($\hR_p$), and coupled electronic-photonic ($\hR_{ep}$) excitations.
Because the excitation operator, $\hR$, commutes with the excitation operators in $\hT$, solving this eigenvalue problem is equivalent to finding a right eigenvector of the similarity transformed Hamiltonian,
\begin{equation}
    \bra{\mu}\bar{H} \hR^n \ket{\Phi_0,0}= E_n \hR^n_\mu.
\end{equation}
Here, $E_n$ is the energy of the nth excited state, and $\mu$ indexes an element of the excitation operator $\hR$. The excitation operator, $\hR$, can be chosen to access charged or neutral excitations.

We recently implemented such QED-CCSD/EOM-CCSD method to compute the polariton states. With the diagrammatic technique, an auto code generator and optimizer are developed to generate the QED-CCSD/EOM-CCSD equations to arbitrary photon order. Fig.~\ref{fig:lih_r} is an example of the Li-H bond dissociation curves in a single-mode optical cavity. Fig.~\ref{fig:lih_r}a shows the ground state Born-Oppenheimer potential energy surface for the bare electronic system (black circles) and for the polaritonic system (red curve) at zero light-matter coupling strength ($\lambda = 0$). The excited states inside (red curves) and outside (black circles) of the cavity are shown at zero coupling strength ($\lambda = 0$), where two multiplicities (singlet $|S_1\rangle \equiv |S_1,0\rangle$ and triplet $|T_1\rangle \equiv |T_1,0\rangle$) of electronic states are shown as well as a single cavity state ($|\omega\rangle \equiv |S_0,1\rangle$). Here, we have used the notation $|S_J\rangle\otimes|n\rangle = |S_J,n\rangle$ where the left ket in the product signifies the electronic DOFs while the right signifies the photonic Fock state label (\textit{i.e.}, the number of photons in the photon-dressed electronic state). At finite light-matter coupling of $\lambda = 0.088$ a.u., the cavity state with one photon $|\omega\rangle$ couples strongly with the singlet excited electronic state with zero photons $|S_1\rangle$ while the excited electronic triplet state $|T_1\rangle$ is negligibly affected. The so-called Rabi splitting appears at the degenerate point between the singlet electronic state and the cavity state, $R_\mathrm{Li-H} \approx 3.75$, forming the upper ($|P_U\rangle$) and lower ($|P_L\rangle$) polaritonic states with mixed electronic and photonic character. 

\begin{figure}[t!]
  \vspace{-0.0 em}
    \centering
    \includegraphics[width=0.99\linewidth]{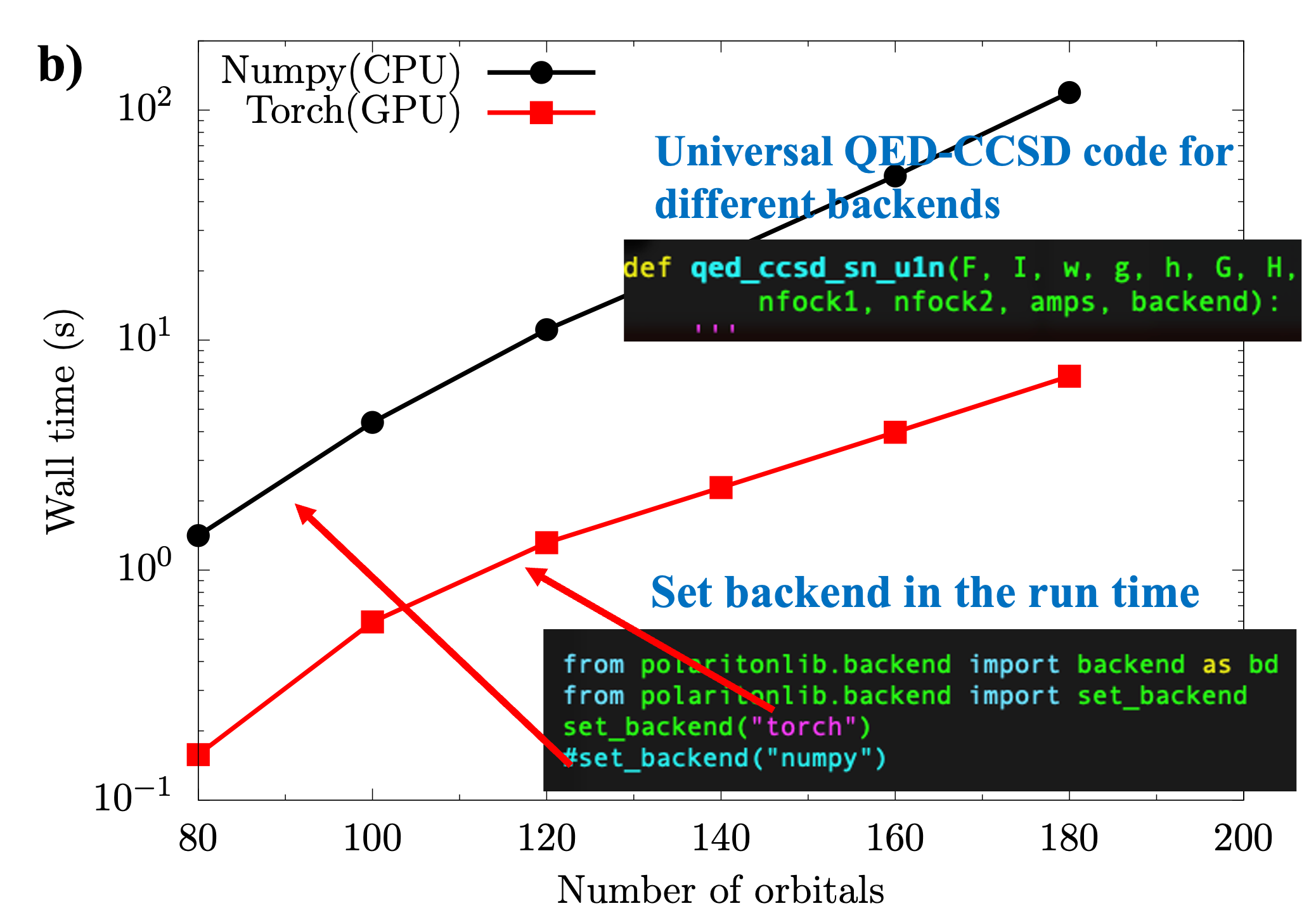}
    \caption{\small  Example of running calculations with our modular QED-CC code on different architectures (CPU vs. GPU on a desktop) by simply setting the backend (see the inset code). $10\times-20\times$ speed up is observed on Torch GPU.
    }
    \vspace{0 em}
    \label{fig:modularcode}
\end{figure}

The computational efficiency of many-body electronic (or polaritonic) structure codes is nearly as important as the method itself. We have implemented various backend options for our code (presented in Fig.~\ref{fig:lih_r}). The efficiency of Numpy \textit{einsum} (CPU-accessible) and Torch \textit{einsum} (GPU-accessible) functions are shown in Fig.~\ref{fig:modularcode} on a vertical log scale as a function of the number of orbitals (\textit{i.e.}, electrons) included in the calculation. The GPU hardware allows the QED-CC code to consistently operate with an order of magnitude less wall time than the CPU version. These results indicate that the conversion toward GPU hardware is required for exploring the frontier science of molecular polaritons.

\section{Perspective and summary}\label{SEC:CONCLUSION}

\subsection{Multiscale method for light-matter interaction}
Though there have been extensive developments in QED methods, most of the current implementations are based on the dipole approximation. However, the dipole approximation can fail in many cases. The nanoplasmonic cavities allow for light localization into deeply subwavelength dimensions, leading to effective mode volumes as small as a few nm$^3$ or even \AA$^3$ (picocavities)~\cite{Jeremy2022nl}. Consequently, the size of molecules becomes comparable to the cavity volumes, and the widely used dipole approximation breaks down. Moreover, the cavity confines the photon field in a certain direction when coupling molecules with many modes inside a cavity (either nanophotonic or nanoplasmonic ones). For a given probing angle $\theta$ (relative to the normal vector of the cavity), the photon energy has a certain dispersion function. Under this situation, the dipole approximation along the in-plane direction of the cavity no longer holds as well~\cite{Tichauer2021JCP, Li_MolPolRev_ARPC2021}.
Hence, a general method that can compute the multiple spatial-dependent cavity eigenmodes $\lambda_{n\bk}(\br)$ and light-matter couplings (from Maxwell's equations) is required. 

As argued above, the EM field within the cavity is, in principle, not homogeneous, and spatial dependence matters in light-matter interactions. The spatial distribution of the EM field can be solved via standard computational electromagnetic methods, such as Finite-difference time-domain (FDTD)~\cite{taflove2005computational}, for Maxwell's equations. Consequently, the light-matter coupling strength can stem from a delicate balance between the spatial dependence of the electronic wavefunctions and the photonic fields. Therefore, only a quantum model that fully incorporates the inhomogeneities of the exciton transition charge density can quantitatively describe this interplay. Using a fully first-principles methodology to describe the quantum chemistry of molecules placed inside the cavity, we can reveal the limitations of the point-dipole approach to address the exciton dynamics in weak and strong coupling regimes.
Besides, current methods usually don't consider the feedback of molecular systems on the cavities. In many situations, modeling of molecular systems only is insufficient as the molecular response can significantly affect the EM distributions, especially in nanoplasmonic cavities. Even though the \textit{ab initio} QED methods we previously developed take into account the interplay between the electronic and photonic DOFs, the molecular response on the EM environment is not considered. 

\begin{figure}[htb]
    \vspace{0.0 em}
    \centering
    \includegraphics[width=1.0\linewidth]{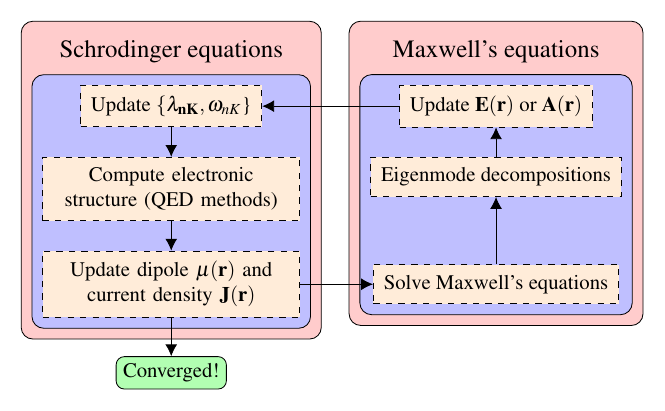}
    \vspace{-1.5em}
    \caption{Self-consistent loop between two sets of equations. The current density from the QED solvers is used to update EM fields.
    The self-consistency between Maxwell and Schr\"odinger equation can be bypassed (in the linear-response regime) by using the pre-calculated Dyadic Green's functions.
    }
    \vspace{0 em}
    \label{fig:flowchart}
\end{figure}

In principle, we could solve the Schr\"odinger and Maxwell equations simultaneously in order to obtain access to the radiated fields and, with it to the self-consistent evolution of light and matter, i.e., the cavity photon modes can be affected by the modular dipoles due to the Ampere's Law (Helmholtz's equation):
\begin{align}
    &\left[\nabla\times\frac{1}{\mu_r(\br\omega)}\nabla\times -\omega^2\mu_0\epsilon_0\epsilon(\br\omega)\right]
    \bE(\br)=-e\bJ(\br), 
    \text{ or }\\
    &  \left(\nabla^2-\frac{1}{c^2}\frac{\partial^2}{\partial t^2}\right)\bA=-\mu_0 \bJ.
\end{align}
Where $\bJ$ is the paramagnetic current density due to the molecular dipoles. According to Maxwell's equations, each current induces an electromagnetic field for which its precise spatial and polarization structure depends on the electromagnetic environment—oscillating charges emit light. 
Hence, a fully self-consistent QED method should consider the feedback of molecular dipoles on the cavity properties. The flowchart is demonstrated in Figure~\ref{fig:flowchart}.

Alternatively, the generated field can be expressed with the help of the dyadic Green's tensor,
\begin{equation}\label{eq:efildfromg}
    \bE(\br,\omega) = i\mu_0\omega\int d\br' G(\br,\br',\omega) [-e\bJ(\br',\omega)],
\end{equation}
where Green's function is the formal solution of Helmholtz's equation
\begin{equation}\label{eq:greenf}
    \left[\nabla\times\frac{1}{\mu_r(\br\omega)}\nabla\times -\omega^2\mu_0\epsilon_0\epsilon(\br\omega)\right]
    G(\br,\br',\omega)=\delta(\br,\br'),
\end{equation}
with linear media $\epsilon(\br)$. The Green's function in Eq.~\ref{eq:greenf} can be solved from the Frequency domain FDTD methods. 

Utilizing Dyadic Green's functions offers a powerful approach to simplify the description of certain components of a matter system while focusing on the computation of electronic structure only. Such a concept has been proposed in the past in a semiclassical treatment of light-matter interaction~\cite{Christian2022PRL,schafer_makingQEDFTFunc_PNAS2021}. Introducing Green's function is particularly advantageous in multicomponent systems that span different length scales, such as a microscopic molecule (which is described by the current density $\bJ$) and macroscopic solvent (represented by a parameterized dielectric function $\epsilon(\br)$). The EM environment embedded via Dyadic Green's function can be obtained numerically using many standard Maxwell solvers with certain boundary conditions, such as FDTD~\cite{taflove2005computational} and method of moments~\cite{peterson1998computational}, providing the photon mode structures that are coupled with QED electronic structure solvers. By computing $\mathbf{G}$ beforehand, the self-consistency is embedded solely via the current density, eliminating the need to treat QED electronic structures and Maxwell's equation simultaneously. The bypassing of the self-consistency between Maxwell-Schr\"odinger equations is a significant strength of Dyadic Green's function approach. This makes the approach more accessible and easier to implement in various electronic structure solvers. But, \textit{it should be noted that introducing linear media restricts the evaluation in the linear response regime.} 

Moreover, the multiscale methods described above can have a significant influence other than polariton chemistry. It could affect the study of light-matter interactions in chemistry, physics, materials science, and energy science at large, such as plasmon chemistry~\cite{Zhan:2023tc}, quantum plasmonics~\cite{Tame2013np}, quantum information transduction~\cite{Gaita-Arino:2019ti}, and photonic-microelectronics integration and polaritonic devices~\cite{Sanvitto:2016td}. The integration with open quantum system method, particularly time-dependent quantum transport theories~\cite{Zheng2007prb, zhang2013first,zhang2013dissipative}, will also open the door to simulating light-driven quantum transport phenomena~\cite{chen2018stark, Boolakee:2022tm} in either weak or strong coupling regimes.

\subsection{Summary}
Controlling chemistry or molecular properties has been a long-standing Holy Grail for many decades. Only recently, new possibilities have emerged in the context of light-matter interactions via either plasmon- or polariton-mediated chemistry, which holds the promise of providing fundamentally new strategies to control chemical reactions that are completely distinct from traditional ones (such as electrochemistry, thermochemistry, and photochemistry). Consequently, the application of light-matter interactions in manipulating chemistry has attracted increasing experimental and theoretical attention. However, the inherent multiscale nature of light-matter interaction problems poses significant challenges for both experimental and theoretical investigations of the underlying processes. Besides, multiscale processes at different length and time scales make it difficult to optimize the performance of light-matter interaction-mediated chemistry, and precise modeling of these processes is essential for obtaining a comprehensive understanding of the fundamental mechanisms or current experiments. Hence, the development of multiscale theoretical and computational models that can precisely describe the dynamical processes in light-matter interaction-mediated chemistry across different time and length scales, in conjunction with massively parallel algorithms for large-scale simulations, is essential for obtaining comprehensive insights into the fundamental mechanisms of current experiments. These simulations allow for the modeling of complex, multiscale processes that would be otherwise difficult or impossible to observe experimentally alone. Such development will ultimately facilitate the discovery of new reaction pathways enhanced by light-matter interactions.

\section*{Conflicts of interest}
There are no conflicts to declare.

\section*{Acknowledgements}
The authors acknowledge support from the Laboratory Directed Research and Development (LDRD) program of Los Alamos National Laboratory.%
(Grant No. 20220527ECR) and the support from the US DOE, Office of Science, Basic Energy Sciences, Chemical Sciences, Geosciences, and Biosciences Division under Triad National Security, LLC (“Triad”) contract Grant 89233218CNA000001 (FWP: LANLECF7). 
LANL is operated by Triad National Security, LLC, for the National Nuclear Security Administration of the U.S. Department of Energy (contract no. 89233218CNA000001).


\balance

\bibliography{rsc} 

\providecommand*{\mcitethebibliography}{\thebibliography}
\csname @ifundefined\endcsname{endmcitethebibliography}
{\let\endmcitethebibliography\endthebibliography}{}
\begin{mcitethebibliography}{255}
\providecommand*{\natexlab}[1]{#1}
\providecommand*{\mciteSetBstSublistMode}[1]{}
\providecommand*{\mciteSetBstMaxWidthForm}[2]{}
\providecommand*{\mciteBstWouldAddEndPuncttrue}
  {\def\EndOfBibitem{\unskip.}}
\providecommand*{\mciteBstWouldAddEndPunctfalse}
  {\let\EndOfBibitem\relax}
\providecommand*{\mciteSetBstMidEndSepPunct}[3]{}
\providecommand*{\mciteSetBstSublistLabelBeginEnd}[3]{}
\providecommand*{\EndOfBibitem}{}
\mciteSetBstSublistMode{f}
\mciteSetBstMaxWidthForm{subitem}
{(\emph{\alph{mcitesubitemcount}})}
\mciteSetBstSublistLabelBeginEnd{\mcitemaxwidthsubitemform\space}
{\relax}{\relax}

\bibitem[Fleming and Ratner(2008)]{ratner2011physicstoday}
G.~R. Fleming and M.~A. Ratner, \emph{Phys. Today}, 2008, \textbf{61},
  28--33\relax
\mciteBstWouldAddEndPuncttrue
\mciteSetBstMidEndSepPunct{\mcitedefaultmidpunct}
{\mcitedefaultendpunct}{\mcitedefaultseppunct}\relax
\EndOfBibitem
\bibitem[Frisk~Kockum \emph{et~al.}(2019)Frisk~Kockum, Miranowicz, De~Liberato,
  Savasta, and Nori]{Anton2019NatRevPhys}
A.~Frisk~Kockum, A.~Miranowicz, S.~De~Liberato, S.~Savasta and F.~Nori,
  \emph{Nat. Rev. Phys.}, 2019, \textbf{1}, 19--40\relax
\mciteBstWouldAddEndPuncttrue
\mciteSetBstMidEndSepPunct{\mcitedefaultmidpunct}
{\mcitedefaultendpunct}{\mcitedefaultseppunct}\relax
\EndOfBibitem
\bibitem[Rivera and Kaminer(2020)]{Revera2020NatRevPhys}
N.~Rivera and I.~Kaminer, \emph{Nat. Rev. Phys.}, 2020, \textbf{2},
  538--561\relax
\mciteBstWouldAddEndPuncttrue
\mciteSetBstMidEndSepPunct{\mcitedefaultmidpunct}
{\mcitedefaultendpunct}{\mcitedefaultseppunct}\relax
\EndOfBibitem
\bibitem[Anderson(1972)]{anderson1972more}
P.~W. Anderson, \emph{Science}, 1972, \textbf{177}, 393--396\relax
\mciteBstWouldAddEndPuncttrue
\mciteSetBstMidEndSepPunct{\mcitedefaultmidpunct}
{\mcitedefaultendpunct}{\mcitedefaultseppunct}\relax
\EndOfBibitem
\bibitem[Novotny and Hecht(2006)]{novotny2006principles}
L.~Novotny and B.~Hecht, \emph{Annu. Rev. Phys. Chem.}, 2006, \textbf{57},
  303--331\relax
\mciteBstWouldAddEndPuncttrue
\mciteSetBstMidEndSepPunct{\mcitedefaultmidpunct}
{\mcitedefaultendpunct}{\mcitedefaultseppunct}\relax
\EndOfBibitem
\bibitem[Weiner and Nunes(2012)]{weiner2012light}
J.~Weiner and F.~Nunes, \emph{Light-Matter Interaction: Physics and Engineering
  at the Nanoscale}, Oxford University Press, 2012\relax
\mciteBstWouldAddEndPuncttrue
\mciteSetBstMidEndSepPunct{\mcitedefaultmidpunct}
{\mcitedefaultendpunct}{\mcitedefaultseppunct}\relax
\EndOfBibitem
\bibitem[Mahan(2000)]{mahan2000many}
G.~Mahan, \emph{Many-Particle Physics}, Springer, 2000\relax
\mciteBstWouldAddEndPuncttrue
\mciteSetBstMidEndSepPunct{\mcitedefaultmidpunct}
{\mcitedefaultendpunct}{\mcitedefaultseppunct}\relax
\EndOfBibitem
\bibitem[Stiles \emph{et~al.}(2008)Stiles, Dieringer, Shah, and
  Van~Duyne]{Stiles2008annurev}
P.~L. Stiles, J.~A. Dieringer, N.~C. Shah and R.~P. Van~Duyne, \emph{Annu. Rev.
  Anal. Chem.}, 2008, \textbf{1}, 601--626\relax
\mciteBstWouldAddEndPuncttrue
\mciteSetBstMidEndSepPunct{\mcitedefaultmidpunct}
{\mcitedefaultendpunct}{\mcitedefaultseppunct}\relax
\EndOfBibitem
\bibitem[Mukamel(2000)]{Mukamel2000annurev}
S.~Mukamel, \emph{Annu. Rev. Phys. Chem.}, 2000, \textbf{51}, 691--729\relax
\mciteBstWouldAddEndPuncttrue
\mciteSetBstMidEndSepPunct{\mcitedefaultmidpunct}
{\mcitedefaultendpunct}{\mcitedefaultseppunct}\relax
\EndOfBibitem
\bibitem[Zhang \emph{et~al.}(2014)Zhang, Meng, Yam, and Chen]{zhang2014quantum}
Y.~Zhang, L.~Meng, C.~Yam and G.~Chen, \emph{J. Phys. Chem. Lett.}, 2014,
  \textbf{5}, 1272--1277\relax
\mciteBstWouldAddEndPuncttrue
\mciteSetBstMidEndSepPunct{\mcitedefaultmidpunct}
{\mcitedefaultendpunct}{\mcitedefaultseppunct}\relax
\EndOfBibitem
\bibitem[Yam \emph{et~al.}(2015)Yam, Meng, Zhang, and Chen]{yam2015multiscale}
C.~Yam, L.~Meng, Y.~Zhang and G.~Chen, \emph{Chem. Soc. Rev.}, 2015,
  \textbf{44}, 1763--1776\relax
\mciteBstWouldAddEndPuncttrue
\mciteSetBstMidEndSepPunct{\mcitedefaultmidpunct}
{\mcitedefaultendpunct}{\mcitedefaultseppunct}\relax
\EndOfBibitem
\bibitem[Meng \emph{et~al.}(2015)Meng, Yam, Zhang, Wang, and
  Chen]{meng2015multiscale}
L.~Meng, C.~Yam, Y.~Zhang, R.~Wang and G.~Chen, \emph{J. Phys. Chem. Lett.},
  2015, \textbf{6}, 4410--4416\relax
\mciteBstWouldAddEndPuncttrue
\mciteSetBstMidEndSepPunct{\mcitedefaultmidpunct}
{\mcitedefaultendpunct}{\mcitedefaultseppunct}\relax
\EndOfBibitem
\bibitem[Meng \emph{et~al.}(2017)Meng, Zhang, and Yam]{meng2017multiscale}
L.~Meng, Y.~Zhang and C.~Yam, \emph{J. Phys. Chem. Lett.}, 2017, \textbf{8},
  571--575\relax
\mciteBstWouldAddEndPuncttrue
\mciteSetBstMidEndSepPunct{\mcitedefaultmidpunct}
{\mcitedefaultendpunct}{\mcitedefaultseppunct}\relax
\EndOfBibitem
\bibitem[Wang \emph{et~al.}(2015)Wang, Zhang, Chen, and Yam]{wang2015quantum}
R.~Wang, Y.~Zhang, G.~H. Chen and C.~Y. Yam, \emph{Prog. Electromagn. Res.},
  2015, \textbf{154}, 163--170\relax
\mciteBstWouldAddEndPuncttrue
\mciteSetBstMidEndSepPunct{\mcitedefaultmidpunct}
{\mcitedefaultendpunct}{\mcitedefaultseppunct}\relax
\EndOfBibitem
\bibitem[Degen \emph{et~al.}(2017)Degen, Reinhard, and
  Cappellaro]{Degen2017rmp}
C.~L. Degen, F.~Reinhard and P.~Cappellaro, \emph{Rev. Mod. Phys.}, 2017,
  \textbf{89}, 035002\relax
\mciteBstWouldAddEndPuncttrue
\mciteSetBstMidEndSepPunct{\mcitedefaultmidpunct}
{\mcitedefaultendpunct}{\mcitedefaultseppunct}\relax
\EndOfBibitem
\bibitem[Flamini \emph{et~al.}(2018)Flamini, Spagnolo, and
  Sciarrino]{Flamini_2018}
F.~Flamini, N.~Spagnolo and F.~Sciarrino, \emph{Rep. Prog. Phys.}, 2018,
  \textbf{82}, 016001\relax
\mciteBstWouldAddEndPuncttrue
\mciteSetBstMidEndSepPunct{\mcitedefaultmidpunct}
{\mcitedefaultendpunct}{\mcitedefaultseppunct}\relax
\EndOfBibitem
\bibitem[Zhou \emph{et~al.}(2018)Zhou, Zhang, Zhuo, Neukirch, and
  Tretiak]{zhou2018interlayer}
L.~Zhou, Y.~Zhang, Z.~Zhuo, A.~J. Neukirch and S.~Tretiak, \emph{J. Phys. Chem.
  Lett.}, 2018, \textbf{9}, 6915--6920\relax
\mciteBstWouldAddEndPuncttrue
\mciteSetBstMidEndSepPunct{\mcitedefaultmidpunct}
{\mcitedefaultendpunct}{\mcitedefaultseppunct}\relax
\EndOfBibitem
\bibitem[Mirkovic \emph{et~al.}(2017)Mirkovic, Ostroumov, Anna, van Grondelle,
  Govindjee, and Scholes]{Mirkovic2018cr}
T.~Mirkovic, E.~E. Ostroumov, J.~M. Anna, R.~van Grondelle, Govindjee and G.~D.
  Scholes, \emph{Chem. Rev.}, 2017, \textbf{117}, 249--293\relax
\mciteBstWouldAddEndPuncttrue
\mciteSetBstMidEndSepPunct{\mcitedefaultmidpunct}
{\mcitedefaultendpunct}{\mcitedefaultseppunct}\relax
\EndOfBibitem
\bibitem[Forn-D\'{\i}az \emph{et~al.}(2019)Forn-D\'{\i}az, Lamata, Rico, Kono,
  and Solano]{Forndaz:2019rmp}
P.~Forn-D\'{\i}az, L.~Lamata, E.~Rico, J.~Kono and E.~Solano, \emph{Rev. Mod.
  Phys.}, 2019, \textbf{91}, 025005\relax
\mciteBstWouldAddEndPuncttrue
\mciteSetBstMidEndSepPunct{\mcitedefaultmidpunct}
{\mcitedefaultendpunct}{\mcitedefaultseppunct}\relax
\EndOfBibitem
\bibitem[Brongersma \emph{et~al.}(2015)Brongersma, Halas, and
  Nordlander]{Brongersma:2015vd}
M.~L. Brongersma, N.~J. Halas and P.~Nordlander, \emph{Nat. Nanotechnol.},
  2015, \textbf{10}, 25--34\relax
\mciteBstWouldAddEndPuncttrue
\mciteSetBstMidEndSepPunct{\mcitedefaultmidpunct}
{\mcitedefaultendpunct}{\mcitedefaultseppunct}\relax
\EndOfBibitem
\bibitem[Zhan \emph{et~al.}(2023)Zhan, Yi, Hu, Zhang, Wu, and
  Tian]{Zhan:2023tc}
C.~Zhan, J.~Yi, S.~Hu, X.-G. Zhang, D.-Y. Wu and Z.-Q. Tian, \emph{Nat. Rev.
  Methods Primers}, 2023, \textbf{3}, 12\relax
\mciteBstWouldAddEndPuncttrue
\mciteSetBstMidEndSepPunct{\mcitedefaultmidpunct}
{\mcitedefaultendpunct}{\mcitedefaultseppunct}\relax
\EndOfBibitem
\bibitem[Kazuma and Kim(2019)]{Kazuma2019angew}
E.~Kazuma and Y.~Kim, \emph{Angew. Chem. Int. Ed.}, 2019, \textbf{58},
  4800--4808\relax
\mciteBstWouldAddEndPuncttrue
\mciteSetBstMidEndSepPunct{\mcitedefaultmidpunct}
{\mcitedefaultendpunct}{\mcitedefaultseppunct}\relax
\EndOfBibitem
\bibitem[Baumberg(2022)]{Jeremy2022nl}
J.~J. Baumberg, \emph{Nano Lett.}, 2022, \textbf{22}, 5859--5865\relax
\mciteBstWouldAddEndPuncttrue
\mciteSetBstMidEndSepPunct{\mcitedefaultmidpunct}
{\mcitedefaultendpunct}{\mcitedefaultseppunct}\relax
\EndOfBibitem
\bibitem[Ribeiro \emph{et~al.}(2018)Ribeiro, Mart{\'{\i}}nez-Mart{\'{\i}}nez,
  Du, Gonzalez-Angulo, and Yuen-Zhou]{Ribeiro2018CS}
R.~F. Ribeiro, L.~A. Mart{\'{\i}}nez-Mart{\'{\i}}nez, M.~Du, J.~C.
  Gonzalez-Angulo and J.~Yuen-Zhou, \emph{Chem. Sci.}, 2018, \textbf{9},
  6325--6339\relax
\mciteBstWouldAddEndPuncttrue
\mciteSetBstMidEndSepPunct{\mcitedefaultmidpunct}
{\mcitedefaultendpunct}{\mcitedefaultseppunct}\relax
\EndOfBibitem
\bibitem[Ebbesen(2016)]{Ebbesen2016ACR}
T.~W. Ebbesen, \emph{Acc. Chem. Res.}, 2016, \textbf{49}, 2403--2412\relax
\mciteBstWouldAddEndPuncttrue
\mciteSetBstMidEndSepPunct{\mcitedefaultmidpunct}
{\mcitedefaultendpunct}{\mcitedefaultseppunct}\relax
\EndOfBibitem
\bibitem[K{\'e}na-Cohen and Forrest(2010)]{Kena-Cohen:2010ue}
S.~K{\'e}na-Cohen and S.~R. Forrest, \emph{Nat. Photon.}, 2010, \textbf{4},
  371--375\relax
\mciteBstWouldAddEndPuncttrue
\mciteSetBstMidEndSepPunct{\mcitedefaultmidpunct}
{\mcitedefaultendpunct}{\mcitedefaultseppunct}\relax
\EndOfBibitem
\bibitem[Kang \emph{et~al.}(2019)Kang, Song, Liu, Park, Agarwal, and
  Cho]{Kang:ws}
J.-W. Kang, B.~Song, W.~Liu, S.-J. Park, R.~Agarwal and C.-H. Cho, \emph{Sci.
  Adv.}, 2019, \textbf{5}, eaau9338\relax
\mciteBstWouldAddEndPuncttrue
\mciteSetBstMidEndSepPunct{\mcitedefaultmidpunct}
{\mcitedefaultendpunct}{\mcitedefaultseppunct}\relax
\EndOfBibitem
\bibitem[Orgiu \emph{et~al.}(2015)Orgiu, George, Hutchison, Devaux, Dayen,
  Doudin, Stellacci, Genet, Schachenmayer, Genes, Pupillo, Samor{\`\i}, and
  Ebbesen]{Orgiu:2015us}
E.~Orgiu, J.~George, J.~A. Hutchison, E.~Devaux, J.~F. Dayen, B.~Doudin,
  F.~Stellacci, C.~Genet, J.~Schachenmayer, C.~Genes, G.~Pupillo,
  P.~Samor{\`\i} and T.~W. Ebbesen, \emph{Nat. Mater.}, 2015, \textbf{14},
  1123--1129\relax
\mciteBstWouldAddEndPuncttrue
\mciteSetBstMidEndSepPunct{\mcitedefaultmidpunct}
{\mcitedefaultendpunct}{\mcitedefaultseppunct}\relax
\EndOfBibitem
\bibitem[Zhong \emph{et~al.}(2017)Zhong, Chervy, Zhang, Thomas, George, Genet,
  Hutchison, and Ebbesen]{Zhong:2017vq}
X.~Zhong, T.~Chervy, L.~Zhang, A.~Thomas, J.~George, C.~Genet, J.~A. Hutchison
  and T.~W. Ebbesen, \emph{Angew. Chem. Int. Ed.}, 2017, \textbf{56},
  9034--9038\relax
\mciteBstWouldAddEndPuncttrue
\mciteSetBstMidEndSepPunct{\mcitedefaultmidpunct}
{\mcitedefaultendpunct}{\mcitedefaultseppunct}\relax
\EndOfBibitem
\bibitem[Georgiou \emph{et~al.}(2021)Georgiou, Jayaprakash, Othonos, and
  Lidzey]{Georgiou2021angwew}
K.~Georgiou, R.~Jayaprakash, A.~Othonos and D.~G. Lidzey, \emph{Angew. Chem.
  Int. Ed.}, 2021, \textbf{60}, 16661--16667\relax
\mciteBstWouldAddEndPuncttrue
\mciteSetBstMidEndSepPunct{\mcitedefaultmidpunct}
{\mcitedefaultendpunct}{\mcitedefaultseppunct}\relax
\EndOfBibitem
\bibitem[Wang \emph{et~al.}(2021)Wang, Hertzog, and
  B{\"o}rjesson]{Want2021natcomm}
M.~Wang, M.~Hertzog and K.~B{\"o}rjesson, \emph{Nat. Commun.}, 2021,
  \textbf{12}, 1874\relax
\mciteBstWouldAddEndPuncttrue
\mciteSetBstMidEndSepPunct{\mcitedefaultmidpunct}
{\mcitedefaultendpunct}{\mcitedefaultseppunct}\relax
\EndOfBibitem
\bibitem[Coles \emph{et~al.}(2014)Coles, Somaschi, Michetti, Clark, Lagoudakis,
  Savvidis, and Lidzey]{Coles:2014wm}
D.~M. Coles, N.~Somaschi, P.~Michetti, C.~Clark, P.~G. Lagoudakis, P.~G.
  Savvidis and D.~G. Lidzey, \emph{Nat. Mater.}, 2014, \textbf{13},
  712--719\relax
\mciteBstWouldAddEndPuncttrue
\mciteSetBstMidEndSepPunct{\mcitedefaultmidpunct}
{\mcitedefaultendpunct}{\mcitedefaultseppunct}\relax
\EndOfBibitem
\bibitem[Dusel \emph{et~al.}(2020)Dusel, Betzold, Egorov, Klembt, Ohmer,
  Fischer, H{\"o}fling, and Schneider]{Dusel:2020wt}
M.~Dusel, S.~Betzold, O.~A. Egorov, S.~Klembt, J.~Ohmer, U.~Fischer,
  S.~H{\"o}fling and C.~Schneider, \emph{Nat. Commun.}, 2020, \textbf{11},
  2863\relax
\mciteBstWouldAddEndPuncttrue
\mciteSetBstMidEndSepPunct{\mcitedefaultmidpunct}
{\mcitedefaultendpunct}{\mcitedefaultseppunct}\relax
\EndOfBibitem
\bibitem[Zasedatelev \emph{et~al.}(2019)Zasedatelev, Baranikov, Urbonas,
  Scafirimuto, Scherf, St{\"o}ferle, Mahrt, and Lagoudakis]{Zasedatelev:2019um}
A.~V. Zasedatelev, A.~V. Baranikov, D.~Urbonas, F.~Scafirimuto, U.~Scherf,
  T.~St{\"o}ferle, R.~F. Mahrt and P.~G. Lagoudakis, \emph{Nat. Photon.}, 2019,
  \textbf{13}, 378--383\relax
\mciteBstWouldAddEndPuncttrue
\mciteSetBstMidEndSepPunct{\mcitedefaultmidpunct}
{\mcitedefaultendpunct}{\mcitedefaultseppunct}\relax
\EndOfBibitem
\bibitem[Kavokin \emph{et~al.}(2022)Kavokin, Liew, Schneider, Lagoudakis,
  Klembt, and Hoefling]{Kavokin:2022tn}
A.~Kavokin, T.~C.~H. Liew, C.~Schneider, P.~G. Lagoudakis, S.~Klembt and
  S.~Hoefling, \emph{Nat. Rev. Phys.}, 2022, \textbf{4}, 435--451\relax
\mciteBstWouldAddEndPuncttrue
\mciteSetBstMidEndSepPunct{\mcitedefaultmidpunct}
{\mcitedefaultendpunct}{\mcitedefaultseppunct}\relax
\EndOfBibitem
\bibitem[Pavo{\v s}evi{\'c} \emph{et~al.}(2022)Pavo{\v s}evi{\'c}, Smith, and
  Rubio]{pavosevic_ClickChem_Arxiv2022}
F.~Pavo{\v s}evi{\'c}, R.~L. Smith and A.~Rubio, \emph{Catalysis in {Click}
  {Chemistry} {Reactions} {Controlled} by {Cavity} {Quantum} {Vacuum}
  {Fluctuations}: {The} {Case} of endo/exo {Diels}-{Alder} {Reaction}}, 2022,
  \url{http://arxiv.org/abs/2208.06925}, arXiv:2208.06925 [physics]\relax
\mciteBstWouldAddEndPuncttrue
\mciteSetBstMidEndSepPunct{\mcitedefaultmidpunct}
{\mcitedefaultendpunct}{\mcitedefaultseppunct}\relax
\EndOfBibitem
\bibitem[Pavo{\v s}evi{\'c} \emph{et~al.}(2022)Pavo{\v s}evi{\'c},
  Hammes-Schiffer, Rubio, and Flick]{pavosevic_PTQED_JACS2022}
F.~Pavo{\v s}evi{\'c}, S.~Hammes-Schiffer, A.~Rubio and J.~Flick, \emph{J. Am.
  Chem. Soc.}, 2022\relax
\mciteBstWouldAddEndPuncttrue
\mciteSetBstMidEndSepPunct{\mcitedefaultmidpunct}
{\mcitedefaultendpunct}{\mcitedefaultseppunct}\relax
\EndOfBibitem
\bibitem[Sch{\"a}fer \emph{et~al.}(2022)Sch{\"a}fer, Flick, Ronca, Narang, and
  Rubio]{schafer_shining_NatComm2022}
C.~Sch{\"a}fer, J.~Flick, E.~Ronca, P.~Narang and A.~Rubio, \emph{Nat.
  Commun.}, 2022, \textbf{13}, 7817\relax
\mciteBstWouldAddEndPuncttrue
\mciteSetBstMidEndSepPunct{\mcitedefaultmidpunct}
{\mcitedefaultendpunct}{\mcitedefaultseppunct}\relax
\EndOfBibitem
\bibitem[Sch{\"a}fer(2022)]{schafer_EmbeddingRadReaction_JPCL2022}
C.~Sch{\"a}fer, \emph{J. Phys. Chem. Lett.}, 2022, \textbf{13},
  6905--6911\relax
\mciteBstWouldAddEndPuncttrue
\mciteSetBstMidEndSepPunct{\mcitedefaultmidpunct}
{\mcitedefaultendpunct}{\mcitedefaultseppunct}\relax
\EndOfBibitem
\bibitem[Cave and Newton(1997)]{Cave1997JCP}
R.~J. Cave and M.~D. Newton, \emph{J. Chem. Phys.}, 1997, \textbf{106},
  9213--9226\relax
\mciteBstWouldAddEndPuncttrue
\mciteSetBstMidEndSepPunct{\mcitedefaultmidpunct}
{\mcitedefaultendpunct}{\mcitedefaultseppunct}\relax
\EndOfBibitem
\bibitem[Mart{\'{i}}nez-Mart{\'{i}}nez
  \emph{et~al.}(2017)Mart{\'{i}}nez-Mart{\'{i}}nez, Ribeiro,
  Gonz{\'{a}}lez-Angulo, and Yuen-Zhou]{MartinezMartinez2017AP}
L.~A. Mart{\'{i}}nez-Mart{\'{i}}nez, R.~F. Ribeiro, J.~C. Gonz{\'{a}}lez-Angulo
  and J.~Yuen-Zhou, \emph{{ACS} Photonics}, 2017, \textbf{5}, 167--176\relax
\mciteBstWouldAddEndPuncttrue
\mciteSetBstMidEndSepPunct{\mcitedefaultmidpunct}
{\mcitedefaultendpunct}{\mcitedefaultseppunct}\relax
\EndOfBibitem
\bibitem[Yang and Cao(2021)]{Yang2021JPCL}
P.-Y. Yang and J.~Cao, \emph{J. Phys. Chem. Lett.}, 2021, \textbf{12},
  9531--9538\relax
\mciteBstWouldAddEndPuncttrue
\mciteSetBstMidEndSepPunct{\mcitedefaultmidpunct}
{\mcitedefaultendpunct}{\mcitedefaultseppunct}\relax
\EndOfBibitem
\bibitem[Climent and Feist(2020)]{Climent2020PCCP}
C.~Climent and J.~Feist, \emph{Phys. Chem. Chem. Phys.}, 2020, \textbf{22},
  23545--23552\relax
\mciteBstWouldAddEndPuncttrue
\mciteSetBstMidEndSepPunct{\mcitedefaultmidpunct}
{\mcitedefaultendpunct}{\mcitedefaultseppunct}\relax
\EndOfBibitem
\bibitem[Wang \emph{et~al.}(2022)Wang, Neuman, Yelin, and Flick]{Wang2022JPCL}
D.~S. Wang, T.~Neuman, S.~F. Yelin and J.~Flick, \emph{J. Phys. Chem. Lett.},
  2022, \textbf{13}, 3317--3324\relax
\mciteBstWouldAddEndPuncttrue
\mciteSetBstMidEndSepPunct{\mcitedefaultmidpunct}
{\mcitedefaultendpunct}{\mcitedefaultseppunct}\relax
\EndOfBibitem
\bibitem[Imperatore \emph{et~al.}(2021)Imperatore, Asbury, and
  Giebink]{Imperatore2021JCP}
M.~V. Imperatore, J.~B. Asbury and N.~C. Giebink, \emph{J. Chem. Phys.}, 2021,
  \textbf{154}, 191103\relax
\mciteBstWouldAddEndPuncttrue
\mciteSetBstMidEndSepPunct{\mcitedefaultmidpunct}
{\mcitedefaultendpunct}{\mcitedefaultseppunct}\relax
\EndOfBibitem
\bibitem[Galego \emph{et~al.}(2016)Galego, Garcia-Vidal, and
  Feist]{Galego2016NC}
J.~Galego, F.~J. Garcia-Vidal and J.~Feist, \emph{Nat. Commun.}, 2016,
  \textbf{7}, 13841\relax
\mciteBstWouldAddEndPuncttrue
\mciteSetBstMidEndSepPunct{\mcitedefaultmidpunct}
{\mcitedefaultendpunct}{\mcitedefaultseppunct}\relax
\EndOfBibitem
\bibitem[Campos-Gonzalez-Angulo \emph{et~al.}(2019)Campos-Gonzalez-Angulo,
  Ribeiro, and Yuen-Zhou]{CamposGonzalezAngulo2019NC}
J.~A. Campos-Gonzalez-Angulo, R.~F. Ribeiro and J.~Yuen-Zhou, \emph{Nat.
  Commun.}, 2019, \textbf{10}, 4685\relax
\mciteBstWouldAddEndPuncttrue
\mciteSetBstMidEndSepPunct{\mcitedefaultmidpunct}
{\mcitedefaultendpunct}{\mcitedefaultseppunct}\relax
\EndOfBibitem
\bibitem[Philbin \emph{et~al.}(2022)Philbin, Wang, Narang, and
  Dou]{Philbin2022JPCC}
J.~P. Philbin, Y.~Wang, P.~Narang and W.~Dou, \emph{J. Phys. Chem. C}, 2022,
  \textbf{126}, 14908--14913\relax
\mciteBstWouldAddEndPuncttrue
\mciteSetBstMidEndSepPunct{\mcitedefaultmidpunct}
{\mcitedefaultendpunct}{\mcitedefaultseppunct}\relax
\EndOfBibitem
\bibitem[Efrima and Bixon(1974)]{Efrima1974CPL}
S.~Efrima and M.~Bixon, \emph{Chem. Phys. Lett.}, 1974, \textbf{25},
  34--37\relax
\mciteBstWouldAddEndPuncttrue
\mciteSetBstMidEndSepPunct{\mcitedefaultmidpunct}
{\mcitedefaultendpunct}{\mcitedefaultseppunct}\relax
\EndOfBibitem
\bibitem[Phuc \emph{et~al.}(2020)Phuc, Trung, and Ishizaki]{Phuc2020SP}
N.~T. Phuc, P.~Q. Trung and A.~Ishizaki, \emph{Sci. Rep.}, 2020, \textbf{10},
  7318\relax
\mciteBstWouldAddEndPuncttrue
\mciteSetBstMidEndSepPunct{\mcitedefaultmidpunct}
{\mcitedefaultendpunct}{\mcitedefaultseppunct}\relax
\EndOfBibitem
\bibitem[Davidsson and Kowalewski(2020)]{Davidsson2020JCP}
E.~Davidsson and M.~Kowalewski, \emph{J. Chem. Phys.}, 2020, \textbf{153},
  234304\relax
\mciteBstWouldAddEndPuncttrue
\mciteSetBstMidEndSepPunct{\mcitedefaultmidpunct}
{\mcitedefaultendpunct}{\mcitedefaultseppunct}\relax
\EndOfBibitem
\bibitem[Galego \emph{et~al.}(2017)Galego, Garcia-Vidal, and
  Feist]{Galego2017PRL}
J.~Galego, F.~J. Garcia-Vidal and J.~Feist, \emph{Phys. Rev. Lett.}, 2017,
  \textbf{119}, 136001\relax
\mciteBstWouldAddEndPuncttrue
\mciteSetBstMidEndSepPunct{\mcitedefaultmidpunct}
{\mcitedefaultendpunct}{\mcitedefaultseppunct}\relax
\EndOfBibitem
\bibitem[Mauro \emph{et~al.}(2021)Mauro, Caicedo, Jonusauskas, and
  Avriller]{Mauro2021PRB}
L.~Mauro, K.~Caicedo, G.~Jonusauskas and R.~Avriller, \emph{Phys. Rev. B},
  2021, \textbf{103}, 165412\relax
\mciteBstWouldAddEndPuncttrue
\mciteSetBstMidEndSepPunct{\mcitedefaultmidpunct}
{\mcitedefaultendpunct}{\mcitedefaultseppunct}\relax
\EndOfBibitem
\bibitem[Vurgaftman \emph{et~al.}(2020)Vurgaftman, Simpkins, Dunkelberger, and
  Owrutsky]{Vurgaftman2020JPCL}
I.~Vurgaftman, B.~S. Simpkins, A.~D. Dunkelberger and J.~C. Owrutsky, \emph{J.
  Phys. Chem. Lett.}, 2020, \textbf{11}, 3557--3562\relax
\mciteBstWouldAddEndPuncttrue
\mciteSetBstMidEndSepPunct{\mcitedefaultmidpunct}
{\mcitedefaultendpunct}{\mcitedefaultseppunct}\relax
\EndOfBibitem
\bibitem[Hiura and Shalabney(2021)]{Hiura2021C}
H.~Hiura and A.~Shalabney, \emph{ChemRxiv}, 2021\relax
\mciteBstWouldAddEndPuncttrue
\mciteSetBstMidEndSepPunct{\mcitedefaultmidpunct}
{\mcitedefaultendpunct}{\mcitedefaultseppunct}\relax
\EndOfBibitem
\bibitem[Hiura \emph{et~al.}(2019)Hiura, Shalabney, and George]{Hiura2019-rate}
H.~Hiura, A.~Shalabney and J.~George, \emph{ChemRxiv}, 2019\relax
\mciteBstWouldAddEndPuncttrue
\mciteSetBstMidEndSepPunct{\mcitedefaultmidpunct}
{\mcitedefaultendpunct}{\mcitedefaultseppunct}\relax
\EndOfBibitem
\bibitem[Weight \emph{et~al.}(2023)Weight, Krauss, and
  Huo]{weight_abQED_JPCL2023}
B.~M. Weight, T.~Krauss and P.~Huo, \emph{J. Phys. Chem. Lett.}, 2023,
  \textbf{14}, 5901--5913\relax
\mciteBstWouldAddEndPuncttrue
\mciteSetBstMidEndSepPunct{\mcitedefaultmidpunct}
{\mcitedefaultendpunct}{\mcitedefaultseppunct}\relax
\EndOfBibitem
\bibitem[Curchod \emph{et~al.}(2016)Curchod, Rauer, Marquetand, Gonz{\'{a}}lez,
  and Mart{\'{i}}nez]{Curchod2016JCP}
B.~F.~E. Curchod, C.~Rauer, P.~Marquetand, L.~Gonz{\'{a}}lez and T.~J.
  Mart{\'{i}}nez, \emph{J. Chem. Phys.}, 2016, \textbf{144}, 101102\relax
\mciteBstWouldAddEndPuncttrue
\mciteSetBstMidEndSepPunct{\mcitedefaultmidpunct}
{\mcitedefaultendpunct}{\mcitedefaultseppunct}\relax
\EndOfBibitem
\bibitem[Wu \emph{et~al.}(2022)Wu, Sifain, Delpo, and Scholes]{Wu2022JCP}
W.~Wu, A.~E. Sifain, C.~A. Delpo and G.~D. Scholes, \emph{J. Chem. Phys.},
  2022, \textbf{157}, 161102\relax
\mciteBstWouldAddEndPuncttrue
\mciteSetBstMidEndSepPunct{\mcitedefaultmidpunct}
{\mcitedefaultendpunct}{\mcitedefaultseppunct}\relax
\EndOfBibitem
\bibitem[Avramenko and Rury(2020)]{Avramenko:2020uz}
A.~G. Avramenko and A.~S. Rury, \emph{The Journal of Physical Chemistry
  Letters}, 2020, \textbf{11}, 1013--1021\relax
\mciteBstWouldAddEndPuncttrue
\mciteSetBstMidEndSepPunct{\mcitedefaultmidpunct}
{\mcitedefaultendpunct}{\mcitedefaultseppunct}\relax
\EndOfBibitem
\bibitem[Climent \emph{et~al.}(2022)Climent, Casanova, Feist, and
  Garcia-Vidal]{Climent2022CRPS}
C.~Climent, D.~Casanova, J.~Feist and F.~J. Garcia-Vidal, \emph{Cell Rep. Phys.
  Sci.}, 2022, \textbf{3}, 100841\relax
\mciteBstWouldAddEndPuncttrue
\mciteSetBstMidEndSepPunct{\mcitedefaultmidpunct}
{\mcitedefaultendpunct}{\mcitedefaultseppunct}\relax
\EndOfBibitem
\bibitem[Gu and Mukamel(2021)]{Gu2021JPCL}
B.~Gu and S.~Mukamel, \emph{J. Phys. Chem. Lett.}, 2021, \textbf{12},
  2052--2056\relax
\mciteBstWouldAddEndPuncttrue
\mciteSetBstMidEndSepPunct{\mcitedefaultmidpunct}
{\mcitedefaultendpunct}{\mcitedefaultseppunct}\relax
\EndOfBibitem
\bibitem[Mart{\'\i}nez-Mart{\'\i}nez
  \emph{et~al.}(2018)Mart{\'\i}nez-Mart{\'\i}nez, Du, Ribeiro, K{\'e}na-Cohen,
  and Yuen-Zhou]{MartinezMartinez2018JPCL}
L.~A. Mart{\'\i}nez-Mart{\'\i}nez, M.~Du, R.~F. Ribeiro, S.~K{\'e}na-Cohen and
  J.~Yuen-Zhou, \emph{J. Phys. Chem. Lett.}, 2018, \textbf{9}, 1951--1957\relax
\mciteBstWouldAddEndPuncttrue
\mciteSetBstMidEndSepPunct{\mcitedefaultmidpunct}
{\mcitedefaultendpunct}{\mcitedefaultseppunct}\relax
\EndOfBibitem
\bibitem[Cho \emph{et~al.}(2022)Cho, Gu, and Mukamel]{Cho2022JACS}
D.~Cho, B.~Gu and S.~Mukamel, \emph{J. Am. Chem. Soc.}, 2022, \textbf{144},
  7758--7767\relax
\mciteBstWouldAddEndPuncttrue
\mciteSetBstMidEndSepPunct{\mcitedefaultmidpunct}
{\mcitedefaultendpunct}{\mcitedefaultseppunct}\relax
\EndOfBibitem
\bibitem[Csehi \emph{et~al.}(2022)Csehi, Vendrell, Hal{\'a}sz, and
  Vib{\'o}k]{Csehi2022NJP}
A.~Csehi, O.~Vendrell, G.~J. Hal{\'a}sz and {\'A}.~Vib{\'o}k, \emph{New J.
  Phys.}, 2022, \textbf{24}, 073022\relax
\mciteBstWouldAddEndPuncttrue
\mciteSetBstMidEndSepPunct{\mcitedefaultmidpunct}
{\mcitedefaultendpunct}{\mcitedefaultseppunct}\relax
\EndOfBibitem
\bibitem[Gu and Mukamel(2020)]{Gu2020JPCL}
B.~Gu and S.~Mukamel, \emph{J. Phys. Chem. Lett.}, 2020, \textbf{11},
  5555--5562\relax
\mciteBstWouldAddEndPuncttrue
\mciteSetBstMidEndSepPunct{\mcitedefaultmidpunct}
{\mcitedefaultendpunct}{\mcitedefaultseppunct}\relax
\EndOfBibitem
\bibitem[Gu and Mukamel(2020)]{Gu2020CS}
B.~Gu and S.~Mukamel, \emph{Chem. Sci.}, 2020, \textbf{11}, 1290--1298\relax
\mciteBstWouldAddEndPuncttrue
\mciteSetBstMidEndSepPunct{\mcitedefaultmidpunct}
{\mcitedefaultendpunct}{\mcitedefaultseppunct}\relax
\EndOfBibitem
\bibitem[Szidarovszky \emph{et~al.}(2018)Szidarovszky, Hal{\'{a}}sz,
  Cs{\'{a}}sz{\'{a}}r, Cederbaum, and Vib{\'{o}}k]{Szidarovszky2018JPCL}
T.~Szidarovszky, G.~J. Hal{\'{a}}sz, A.~G. Cs{\'{a}}sz{\'{a}}r, L.~S. Cederbaum
  and {\'{A}}.~Vib{\'{o}}k, \emph{J. Phys. Chem. Lett.}, 2018, \textbf{9},
  6215--6223\relax
\mciteBstWouldAddEndPuncttrue
\mciteSetBstMidEndSepPunct{\mcitedefaultmidpunct}
{\mcitedefaultendpunct}{\mcitedefaultseppunct}\relax
\EndOfBibitem
\bibitem[F{\'{a}}bri \emph{et~al.}(2022)F{\'{a}}bri, Hal{\'{a}}sz, and
  Vib{\'{o}}k]{Fabri2022JPCL}
C.~F{\'{a}}bri, G.~J. Hal{\'{a}}sz and {\'{A}}.~Vib{\'{o}}k, \emph{J. Phys.
  Chem. Lett.}, 2022, \textbf{13}, 1172--1179\relax
\mciteBstWouldAddEndPuncttrue
\mciteSetBstMidEndSepPunct{\mcitedefaultmidpunct}
{\mcitedefaultendpunct}{\mcitedefaultseppunct}\relax
\EndOfBibitem
\bibitem[Natan \emph{et~al.}(2016)Natan, Ware, Prabhudesai, Lev, Bruner, Heber,
  and Bucksbaum]{Natan2016PRL}
A.~Natan, M.~R. Ware, V.~S. Prabhudesai, U.~Lev, B.~D. Bruner, O.~Heber and
  P.~H. Bucksbaum, \emph{Phys. Rev. Lett.}, 2016, \textbf{116}, 143004\relax
\mciteBstWouldAddEndPuncttrue
\mciteSetBstMidEndSepPunct{\mcitedefaultmidpunct}
{\mcitedefaultendpunct}{\mcitedefaultseppunct}\relax
\EndOfBibitem
\bibitem[Hofmann and de~Vivie-Riedle(2001)]{Hofmann2001CPL}
A.~Hofmann and R.~de~Vivie-Riedle, \emph{Chem. Phys. Lett.}, 2001,
  \textbf{346}, 299--304\relax
\mciteBstWouldAddEndPuncttrue
\mciteSetBstMidEndSepPunct{\mcitedefaultmidpunct}
{\mcitedefaultendpunct}{\mcitedefaultseppunct}\relax
\EndOfBibitem
\bibitem[Farag \emph{et~al.}(2021)Farag, Mandal, and Huo]{Farag2021PCCP}
M.~H. Farag, A.~Mandal and P.~Huo, \emph{Phys. Chem. Chem. Phys.}, 2021,
  \textbf{23}, 16868--16879\relax
\mciteBstWouldAddEndPuncttrue
\mciteSetBstMidEndSepPunct{\mcitedefaultmidpunct}
{\mcitedefaultendpunct}{\mcitedefaultseppunct}\relax
\EndOfBibitem
\bibitem[Arnold \emph{et~al.}(2018)Arnold, Vendrell, Welsch, and
  Santra]{Arnold:2018vz}
C.~Arnold, O.~Vendrell, R.~Welsch and R.~Santra, \emph{Physical Review
  Letters}, 2018, \textbf{120}, 123001--\relax
\mciteBstWouldAddEndPuncttrue
\mciteSetBstMidEndSepPunct{\mcitedefaultmidpunct}
{\mcitedefaultendpunct}{\mcitedefaultseppunct}\relax
\EndOfBibitem
\bibitem[Ulusoy \emph{et~al.}(2019)Ulusoy, Gomez, and Vendrell]{Ulusoy:2019vy}
I.~S. Ulusoy, J.~A. Gomez and O.~Vendrell, \emph{The Journal of Physical
  Chemistry A}, 2019, \textbf{123}, 8832--8844\relax
\mciteBstWouldAddEndPuncttrue
\mciteSetBstMidEndSepPunct{\mcitedefaultmidpunct}
{\mcitedefaultendpunct}{\mcitedefaultseppunct}\relax
\EndOfBibitem
\bibitem[Craig and Thirunamachandran(1998)]{craig1998molecular}
D.~Craig and T.~Thirunamachandran, \emph{Molecular Quantum Electrodynamics: An
  Introduction to Radiation-molecule Interactions}, Dover Publications,
  1998\relax
\mciteBstWouldAddEndPuncttrue
\mciteSetBstMidEndSepPunct{\mcitedefaultmidpunct}
{\mcitedefaultendpunct}{\mcitedefaultseppunct}\relax
\EndOfBibitem
\bibitem[Archambault \emph{et~al.}(2010)Archambault, Marquier, Greffet, and
  Arnold]{Archambault2010prb}
A.~Archambault, F.~m.~c. Marquier, J.-J. Greffet and C.~Arnold, \emph{Phys.
  Rev. B}, 2010, \textbf{82}, 035411\relax
\mciteBstWouldAddEndPuncttrue
\mciteSetBstMidEndSepPunct{\mcitedefaultmidpunct}
{\mcitedefaultendpunct}{\mcitedefaultseppunct}\relax
\EndOfBibitem
\bibitem[Andrews \emph{et~al.}(1989)Andrews, Craig, and
  Thirunamachandran]{andrew1989irpc}
D.~L. Andrews, D.~P. Craig and T.~Thirunamachandran, \emph{Int. Rev. Phys.
  Chem.}, 1989, \textbf{8}, 339--383\relax
\mciteBstWouldAddEndPuncttrue
\mciteSetBstMidEndSepPunct{\mcitedefaultmidpunct}
{\mcitedefaultendpunct}{\mcitedefaultseppunct}\relax
\EndOfBibitem
\bibitem[Snyder and Love(1983)]{SnyderLove:1983}
A.~W. Snyder and J.~Love, \emph{Optical Waveguide Theory}, Springer, 1st edn,
  1983\relax
\mciteBstWouldAddEndPuncttrue
\mciteSetBstMidEndSepPunct{\mcitedefaultmidpunct}
{\mcitedefaultendpunct}{\mcitedefaultseppunct}\relax
\EndOfBibitem
\bibitem[Woolley(2020)]{Woolley:2020prr}
R.~G. Woolley, \emph{Phys. Rev. Res.}, 2020, \textbf{2}, 013206\relax
\mciteBstWouldAddEndPuncttrue
\mciteSetBstMidEndSepPunct{\mcitedefaultmidpunct}
{\mcitedefaultendpunct}{\mcitedefaultseppunct}\relax
\EndOfBibitem
\bibitem[Chikkaraddy \emph{et~al.}(2016)Chikkaraddy, de~Nijs, Benz, Barrow,
  Scherman, Rosta, Demetriadou, Fox, Hess, and Baumberg]{Chikkaraddy2016nature}
R.~Chikkaraddy, B.~de~Nijs, F.~Benz, S.~J. Barrow, O.~A. Scherman, E.~Rosta,
  A.~Demetriadou, P.~Fox, O.~Hess and J.~J. Baumberg, \emph{Nature}, 2016,
  \textbf{535}, 127--30\relax
\mciteBstWouldAddEndPuncttrue
\mciteSetBstMidEndSepPunct{\mcitedefaultmidpunct}
{\mcitedefaultendpunct}{\mcitedefaultseppunct}\relax
\EndOfBibitem
\bibitem[Benz \emph{et~al.}(2016)Benz, Schmidt, Dreismann, Chikkaraddy, Zhang,
  Demetriadou, Carnegie, Ohadi, de~Nijs, Esteban, Aizpurua, and
  Baumberg]{Benz2016science}
F.~Benz, M.~K. Schmidt, A.~Dreismann, R.~Chikkaraddy, Y.~Zhang, A.~Demetriadou,
  C.~Carnegie, H.~Ohadi, B.~de~Nijs, R.~Esteban, J.~Aizpurua and J.~J.
  Baumberg, \emph{Science}, 2016, \textbf{354}, 726--729\relax
\mciteBstWouldAddEndPuncttrue
\mciteSetBstMidEndSepPunct{\mcitedefaultmidpunct}
{\mcitedefaultendpunct}{\mcitedefaultseppunct}\relax
\EndOfBibitem
\bibitem[Neuman \emph{et~al.}(2018)Neuman, Esteban, Casanova,
  Garc{\'\i}a-Vidal, and Aizpurua]{Neuman2018nl}
T.~Neuman, R.~Esteban, D.~Casanova, F.~J. Garc{\'\i}a-Vidal and J.~Aizpurua,
  \emph{Nano Lett.}, 2018, \textbf{18}, 2358--2364\relax
\mciteBstWouldAddEndPuncttrue
\mciteSetBstMidEndSepPunct{\mcitedefaultmidpunct}
{\mcitedefaultendpunct}{\mcitedefaultseppunct}\relax
\EndOfBibitem
\bibitem[Wang \emph{et~al.}(2021)Wang, Neuman, and Narang]{Wang:2021jpcc}
D.~S. Wang, T.~Neuman and P.~Narang, \emph{J. Phys. Chem. C}, 2021,
  \textbf{125}, 6222--6228\relax
\mciteBstWouldAddEndPuncttrue
\mciteSetBstMidEndSepPunct{\mcitedefaultmidpunct}
{\mcitedefaultendpunct}{\mcitedefaultseppunct}\relax
\EndOfBibitem
\bibitem[Kelly \emph{et~al.}(2003)Kelly, Coronado, Zhao, and
  Schatz]{Kelly:2002jpcb}
K.~L. Kelly, E.~Coronado, L.~L. Zhao and G.~C. Schatz, \emph{J. Phys. Chem. B},
  2003, \textbf{107}, 668--677\relax
\mciteBstWouldAddEndPuncttrue
\mciteSetBstMidEndSepPunct{\mcitedefaultmidpunct}
{\mcitedefaultendpunct}{\mcitedefaultseppunct}\relax
\EndOfBibitem
\bibitem[Varas \emph{et~al.}(2016)Varas, Garc{\'{\i}}a-Gonz{\'{a}}lez, Feist,
  Garc{\'{\i}}a-Vidal, and Rubio]{Varas:2016tf}
A.~Varas, P.~Garc{\'{\i}}a-Gonz{\'{a}}lez, J.~Feist, F.~Garc{\'{\i}}a-Vidal and
  A.~Rubio, \emph{Nanophotonics}, 2016, \textbf{5}, 409--426\relax
\mciteBstWouldAddEndPuncttrue
\mciteSetBstMidEndSepPunct{\mcitedefaultmidpunct}
{\mcitedefaultendpunct}{\mcitedefaultseppunct}\relax
\EndOfBibitem
\bibitem[Tame \emph{et~al.}(2013)Tame, McEnery, Ozdemir, Lee, Maier, and
  Kim]{Tame2013np}
M.~S. Tame, K.~R. McEnery, S.~K. Ozdemir, J.~Lee, S.~A. Maier and M.~S. Kim,
  \emph{Nat. Phys.}, 2013, \textbf{9}, 329--340\relax
\mciteBstWouldAddEndPuncttrue
\mciteSetBstMidEndSepPunct{\mcitedefaultmidpunct}
{\mcitedefaultendpunct}{\mcitedefaultseppunct}\relax
\EndOfBibitem
\bibitem[Marinica \emph{et~al.}(2012)Marinica, Kazansky, Nordlander, Aizpurua,
  and Borisov]{Marinica2012nl}
D.~Marinica, A.~Kazansky, P.~Nordlander, J.~Aizpurua and A.~G. Borisov,
  \emph{Nano Lett.}, 2012, \textbf{12}, 1333--1339\relax
\mciteBstWouldAddEndPuncttrue
\mciteSetBstMidEndSepPunct{\mcitedefaultmidpunct}
{\mcitedefaultendpunct}{\mcitedefaultseppunct}\relax
\EndOfBibitem
\bibitem[Schuller \emph{et~al.}(2010)Schuller, Barnard, Cai, Jun, White, and
  Brongersma]{Schuller2010NatMat}
J.~A. Schuller, E.~S. Barnard, W.~Cai, Y.~C. Jun, J.~S. White and M.~L.
  Brongersma, \emph{Nat. Mater.}, 2010, \textbf{9}, 193--204\relax
\mciteBstWouldAddEndPuncttrue
\mciteSetBstMidEndSepPunct{\mcitedefaultmidpunct}
{\mcitedefaultendpunct}{\mcitedefaultseppunct}\relax
\EndOfBibitem
\bibitem[Jiang \emph{et~al.}(2003)Jiang, Bosnick, Maillard, and
  Brus]{Jiang2003jpcb}
Jiang, K.~Bosnick, M.~Maillard and L.~Brus, \emph{J. Phys. Chem. B}, 2003,
  \textbf{107}, 9964--9972\relax
\mciteBstWouldAddEndPuncttrue
\mciteSetBstMidEndSepPunct{\mcitedefaultmidpunct}
{\mcitedefaultendpunct}{\mcitedefaultseppunct}\relax
\EndOfBibitem
\bibitem[Brus(2008)]{Brus2008acr}
L.~Brus, \emph{Acc. Chem. Res.}, 2008, \textbf{41}, 1742--1749\relax
\mciteBstWouldAddEndPuncttrue
\mciteSetBstMidEndSepPunct{\mcitedefaultmidpunct}
{\mcitedefaultendpunct}{\mcitedefaultseppunct}\relax
\EndOfBibitem
\bibitem[Nie and Emory(1997)]{Nie1997sci}
S.~Nie and S.~R. Emory, \emph{Science}, 1997, \textbf{275}, 1102--1106\relax
\mciteBstWouldAddEndPuncttrue
\mciteSetBstMidEndSepPunct{\mcitedefaultmidpunct}
{\mcitedefaultendpunct}{\mcitedefaultseppunct}\relax
\EndOfBibitem
\bibitem[Zhan \emph{et~al.}(2018)Zhan, Chen, Yi, Li, Wu, and Tian]{Zhan:2018tv}
C.~Zhan, X.-J. Chen, J.~Yi, J.-F. Li, D.-Y. Wu and Z.-Q. Tian, \emph{Nat. Rev.
  Chem.}, 2018, \textbf{2}, 216--230\relax
\mciteBstWouldAddEndPuncttrue
\mciteSetBstMidEndSepPunct{\mcitedefaultmidpunct}
{\mcitedefaultendpunct}{\mcitedefaultseppunct}\relax
\EndOfBibitem
\bibitem[Atwater and Polman(2010)]{NatMater2010}
H.~A. Atwater and A.~Polman, \emph{Nat. Mater.}, 2010, \textbf{9},
  205--213\relax
\mciteBstWouldAddEndPuncttrue
\mciteSetBstMidEndSepPunct{\mcitedefaultmidpunct}
{\mcitedefaultendpunct}{\mcitedefaultseppunct}\relax
\EndOfBibitem
\bibitem[Zhang \emph{et~al.}(2016)Zhang, Yam, and Schatz]{zhang2016fundamental}
Y.~Zhang, C.~Yam and G.~C. Schatz, \emph{J. Phys. Chem. Lett.}, 2016,
  \textbf{7}, 1852--1858\relax
\mciteBstWouldAddEndPuncttrue
\mciteSetBstMidEndSepPunct{\mcitedefaultmidpunct}
{\mcitedefaultendpunct}{\mcitedefaultseppunct}\relax
\EndOfBibitem
\bibitem[Ma \emph{et~al.}(2013)Ma, Oulton, Sorger, and Zhang]{Ma2013laser}
R.-M. Ma, R.~F. Oulton, V.~J. Sorger and X.~Zhang, \emph{Laser Photonics Rev.},
  2013, \textbf{7}, 1--21\relax
\mciteBstWouldAddEndPuncttrue
\mciteSetBstMidEndSepPunct{\mcitedefaultmidpunct}
{\mcitedefaultendpunct}{\mcitedefaultseppunct}\relax
\EndOfBibitem
\bibitem[Berini and De~Leon(2012)]{Berini2012natphotonics}
P.~Berini and I.~De~Leon, \emph{Nat. Photon.}, 2012, \textbf{6}, 16--24\relax
\mciteBstWouldAddEndPuncttrue
\mciteSetBstMidEndSepPunct{\mcitedefaultmidpunct}
{\mcitedefaultendpunct}{\mcitedefaultseppunct}\relax
\EndOfBibitem
\bibitem[Guan \emph{et~al.}(2021)Guan, Li, Juarez, Sample, Wang, Schatz, and
  Odom]{Guan2021am}
J.~Guan, R.~Li, X.~G. Juarez, A.~D. Sample, Y.~Wang, G.~C. Schatz and T.~W.
  Odom, \emph{Adv. Mater.}, 2021, \textbf{n/a}, 2103262\relax
\mciteBstWouldAddEndPuncttrue
\mciteSetBstMidEndSepPunct{\mcitedefaultmidpunct}
{\mcitedefaultendpunct}{\mcitedefaultseppunct}\relax
\EndOfBibitem
\bibitem[Guan \emph{et~al.}(2022)Guan, Park, Deng, Tan, Hu, and
  Odom]{Guan2022cr}
J.~Guan, J.-E. Park, S.~Deng, M.~J.~H. Tan, J.~Hu and T.~W. Odom, \emph{Chem.
  Rev.}, 2022, \textbf{122}, 15177--15203\relax
\mciteBstWouldAddEndPuncttrue
\mciteSetBstMidEndSepPunct{\mcitedefaultmidpunct}
{\mcitedefaultendpunct}{\mcitedefaultseppunct}\relax
\EndOfBibitem
\bibitem[Cushing and Wu(2016)]{Cushing016jpcl}
S.~K. Cushing and N.~Wu, \emph{J. Phys. Chem. Lett.}, 2016, \textbf{7},
  666--675\relax
\mciteBstWouldAddEndPuncttrue
\mciteSetBstMidEndSepPunct{\mcitedefaultmidpunct}
{\mcitedefaultendpunct}{\mcitedefaultseppunct}\relax
\EndOfBibitem
\bibitem[Li \emph{et~al.}(2015)Li, Cushing, Meng, Senty, Bristow, and
  Wu]{Li2015natphotonics}
J.~Li, S.~K. Cushing, F.~Meng, T.~R. Senty, A.~D. Bristow and N.~Wu, \emph{Nat.
  Photon.}, 2015, \textbf{9}, 601--607\relax
\mciteBstWouldAddEndPuncttrue
\mciteSetBstMidEndSepPunct{\mcitedefaultmidpunct}
{\mcitedefaultendpunct}{\mcitedefaultseppunct}\relax
\EndOfBibitem
\bibitem[Zhang(2021)]{Zhang2021JPCA}
Y.~Zhang, \emph{J. Phys. Chem. A}, 2021, \textbf{125}, 9201--9208\relax
\mciteBstWouldAddEndPuncttrue
\mciteSetBstMidEndSepPunct{\mcitedefaultmidpunct}
{\mcitedefaultendpunct}{\mcitedefaultseppunct}\relax
\EndOfBibitem
\bibitem[Clavero(2014)]{Clavero2014}
C.~Clavero, \emph{Nat. Photon.}, 2014, \textbf{8}, 95--103\relax
\mciteBstWouldAddEndPuncttrue
\mciteSetBstMidEndSepPunct{\mcitedefaultmidpunct}
{\mcitedefaultendpunct}{\mcitedefaultseppunct}\relax
\EndOfBibitem
\bibitem[Bernardi \emph{et~al.}(2015)Bernardi, Mustafa, Neaton, and
  Louie]{bernardi2015}
M.~Bernardi, J.~Mustafa, J.~B. Neaton and S.~G. Louie, \emph{Nat. Commun.},
  2015, \textbf{6}, 7044\relax
\mciteBstWouldAddEndPuncttrue
\mciteSetBstMidEndSepPunct{\mcitedefaultmidpunct}
{\mcitedefaultendpunct}{\mcitedefaultseppunct}\relax
\EndOfBibitem
\bibitem[Brown \emph{et~al.}(2016)Brown, Sundararaman, Narang, William
  A.~Goddard, and Atwater]{acsnano5b06199}
A.~M. Brown, R.~Sundararaman, P.~Narang, I.~William A.~Goddard and H.~A.
  Atwater, \emph{ACS Nano}, 2016, \textbf{10}, 957--966\relax
\mciteBstWouldAddEndPuncttrue
\mciteSetBstMidEndSepPunct{\mcitedefaultmidpunct}
{\mcitedefaultendpunct}{\mcitedefaultseppunct}\relax
\EndOfBibitem
\bibitem[Manjavacas \emph{et~al.}(2014)Manjavacas, Liu, Kulkarni, and
  Nordlander]{nn502445f}
A.~Manjavacas, J.~G. Liu, V.~Kulkarni and P.~Nordlander, \emph{ACS Nano}, 2014,
  \textbf{8}, 7630--7638\relax
\mciteBstWouldAddEndPuncttrue
\mciteSetBstMidEndSepPunct{\mcitedefaultmidpunct}
{\mcitedefaultendpunct}{\mcitedefaultseppunct}\relax
\EndOfBibitem
\bibitem[Yuan \emph{et~al.}(2022)Yuan, Zhou, Zhang, Wen, Martirez, Robatjazi,
  Zhou, Carter, Nordlander, and Halas]{Yuan:2022tb}
L.~Yuan, J.~Zhou, M.~Zhang, X.~Wen, J.~M.~P. Martirez, H.~Robatjazi, L.~Zhou,
  E.~A. Carter, P.~Nordlander and N.~J. Halas, \emph{ACS Nano}, 2022,
  \textbf{16}, 17365--17375\relax
\mciteBstWouldAddEndPuncttrue
\mciteSetBstMidEndSepPunct{\mcitedefaultmidpunct}
{\mcitedefaultendpunct}{\mcitedefaultseppunct}\relax
\EndOfBibitem
\bibitem[Zhou \emph{et~al.}(2018)Zhou, Swearer, Zhang, Robatjazi, Zhao,
  Henderson, Dong, Christopher, Carter, Nordlander, and Halas]{Zhou:2018tu}
L.~Zhou, D.~F. Swearer, C.~Zhang, H.~Robatjazi, H.~Zhao, L.~Henderson, L.~Dong,
  P.~Christopher, E.~A. Carter, P.~Nordlander and N.~J. Halas, \emph{Science},
  2018, \textbf{362}, 69--72\relax
\mciteBstWouldAddEndPuncttrue
\mciteSetBstMidEndSepPunct{\mcitedefaultmidpunct}
{\mcitedefaultendpunct}{\mcitedefaultseppunct}\relax
\EndOfBibitem
\bibitem[Wu \emph{et~al.}(2023)Wu, van~der Heide, Wen, Frauenheim, Tretiak,
  Yam, and Zhang]{wu2023chemsci}
X.~Wu, T.~van~der Heide, S.~Wen, T.~Frauenheim, S.~Tretiak, C.~Yam and
  Y.~Zhang, \emph{Chem. Sci.}, 2023, \textbf{14}, 4714--4723\relax
\mciteBstWouldAddEndPuncttrue
\mciteSetBstMidEndSepPunct{\mcitedefaultmidpunct}
{\mcitedefaultendpunct}{\mcitedefaultseppunct}\relax
\EndOfBibitem
\bibitem[Zhang \emph{et~al.}(2018)Zhang, Nelson, Tretiak, Guo, and
  Schatz]{zhang2018plasmonic}
Y.~Zhang, T.~Nelson, S.~Tretiak, H.~Guo and G.~C. Schatz, \emph{Acs Nano},
  2018, \textbf{12}, 8415--8422\relax
\mciteBstWouldAddEndPuncttrue
\mciteSetBstMidEndSepPunct{\mcitedefaultmidpunct}
{\mcitedefaultendpunct}{\mcitedefaultseppunct}\relax
\EndOfBibitem
\bibitem[Wu \emph{et~al.}(2020)Wu, Zhou, Schatz, Zhang, and
  Guo]{wu2020mechanistic}
Q.~Wu, L.~Zhou, G.~C. Schatz, Y.~Zhang and H.~Guo, \emph{J. Am. Chem. Soc.},
  2020, \textbf{142}, 13090--13101\relax
\mciteBstWouldAddEndPuncttrue
\mciteSetBstMidEndSepPunct{\mcitedefaultmidpunct}
{\mcitedefaultendpunct}{\mcitedefaultseppunct}\relax
\EndOfBibitem
\bibitem[Kazuma \emph{et~al.}(2018)Kazuma, Jung, Ueba, Trenary, and
  Kim]{Kazuma2018science}
E.~Kazuma, J.~Jung, H.~Ueba, M.~Trenary and Y.~Kim, \emph{Science}, 2018,
  \textbf{360}, 521--526\relax
\mciteBstWouldAddEndPuncttrue
\mciteSetBstMidEndSepPunct{\mcitedefaultmidpunct}
{\mcitedefaultendpunct}{\mcitedefaultseppunct}\relax
\EndOfBibitem
\bibitem[Zhang \emph{et~al.}(2018)Zhang, He, Guo, Hu, Huang, Mulcahy, and
  Wei]{ZhangYuchao2018chemrev}
Y.~Zhang, S.~He, W.~Guo, Y.~Hu, J.~Huang, J.~R. Mulcahy and W.~D. Wei,
  \emph{Chem. Rev.}, 2018, \textbf{118}, 2927--2954\relax
\mciteBstWouldAddEndPuncttrue
\mciteSetBstMidEndSepPunct{\mcitedefaultmidpunct}
{\mcitedefaultendpunct}{\mcitedefaultseppunct}\relax
\EndOfBibitem
\bibitem[Tesema \emph{et~al.}(2019)Tesema, Kafle, and Habteyes]{Tesema2019jpcc}
T.~E. Tesema, B.~Kafle and T.~G. Habteyes, \emph{J. Phys. Chem. C}, 2019,
  \textbf{123}, 8469--8483\relax
\mciteBstWouldAddEndPuncttrue
\mciteSetBstMidEndSepPunct{\mcitedefaultmidpunct}
{\mcitedefaultendpunct}{\mcitedefaultseppunct}\relax
\EndOfBibitem
\bibitem[Adleman \emph{et~al.}(2009)Adleman, Boyd, Goodwin, and
  Psaltis]{Adleman2009nl}
J.~R. Adleman, D.~A. Boyd, D.~G. Goodwin and D.~Psaltis, \emph{Nano Lett.},
  2009, \textbf{9}, 4417--4423\relax
\mciteBstWouldAddEndPuncttrue
\mciteSetBstMidEndSepPunct{\mcitedefaultmidpunct}
{\mcitedefaultendpunct}{\mcitedefaultseppunct}\relax
\EndOfBibitem
\bibitem[Boerigter \emph{et~al.}(2016)Boerigter, Campana, Morabito, and
  Linic]{Boergter2016natcomm}
C.~Boerigter, R.~Campana, M.~Morabito and S.~Linic, \emph{Nat. Commun.}, 2016,
  \textbf{7}, 10545--10545\relax
\mciteBstWouldAddEndPuncttrue
\mciteSetBstMidEndSepPunct{\mcitedefaultmidpunct}
{\mcitedefaultendpunct}{\mcitedefaultseppunct}\relax
\EndOfBibitem
\bibitem[Zhang \emph{et~al.}(2019)Zhang, Nelson, and
  Tretiak]{zhang2019atomistic}
Y.~Zhang, T.~Nelson and S.~Tretiak, \emph{Computational Photocatalysis:
  Modeling of Photophysics and Photochemistry at Interfaces}, ACS Publications,
  2019, pp. 239--256\relax
\mciteBstWouldAddEndPuncttrue
\mciteSetBstMidEndSepPunct{\mcitedefaultmidpunct}
{\mcitedefaultendpunct}{\mcitedefaultseppunct}\relax
\EndOfBibitem
\bibitem[Gell{\'e} \emph{et~al.}(2020)Gell{\'e}, Jin, de~la Garza, Price,
  Besteiro, and Moores]{Gelle:2020uh}
A.~Gell{\'e}, T.~Jin, L.~de~la Garza, G.~D. Price, L.~V. Besteiro and
  A.~Moores, \emph{Chemical Reviews}, 2020, \textbf{120}, 986--1041\relax
\mciteBstWouldAddEndPuncttrue
\mciteSetBstMidEndSepPunct{\mcitedefaultmidpunct}
{\mcitedefaultendpunct}{\mcitedefaultseppunct}\relax
\EndOfBibitem
\bibitem[Zhan \emph{et~al.}(2019)Zhan, Wang, Zhang, Chen, Huang, Hu, Li, Wu,
  Moskovits, and Tian]{Zhan:2019wp}
C.~Zhan, Z.-Y. Wang, X.-G. Zhang, X.-J. Chen, Y.-F. Huang, S.~Hu, J.-F. Li,
  D.-Y. Wu, M.~Moskovits and Z.-Q. Tian, \emph{J. Am. Chem. Soc.}, 2019,
  \textbf{141}, 8053--8057\relax
\mciteBstWouldAddEndPuncttrue
\mciteSetBstMidEndSepPunct{\mcitedefaultmidpunct}
{\mcitedefaultendpunct}{\mcitedefaultseppunct}\relax
\EndOfBibitem
\bibitem[Li \emph{et~al.}(2023)Li, Lei, Li, Liu, Zheng, Liu, Guo, Liu, Hao, and
  He]{Li:2023ux}
Y.~Li, Y.~Lei, D.~Li, A.~Liu, Z.~Zheng, H.~Liu, J.~Guo, S.~Liu, C.~Hao and
  D.~He, \emph{ACS Catal.}, 2023, \textbf{13}, 10177--10204\relax
\mciteBstWouldAddEndPuncttrue
\mciteSetBstMidEndSepPunct{\mcitedefaultmidpunct}
{\mcitedefaultendpunct}{\mcitedefaultseppunct}\relax
\EndOfBibitem
\bibitem[Mubeen \emph{et~al.}(2013)Mubeen, Lee, Singh, Kr{\"a}mer, Stucky, and
  Moskovits]{Mubeen:2013tk}
S.~Mubeen, J.~Lee, N.~Singh, S.~Kr{\"a}mer, G.~D. Stucky and M.~Moskovits,
  \emph{Nat. Nano.}, 2013, \textbf{8}, 247--251\relax
\mciteBstWouldAddEndPuncttrue
\mciteSetBstMidEndSepPunct{\mcitedefaultmidpunct}
{\mcitedefaultendpunct}{\mcitedefaultseppunct}\relax
\EndOfBibitem
\bibitem[Zhou \emph{et~al.}(2020)Zhou, Martirez, Finzel, Zhang, Swearer, Tian,
  Robatjazi, Lou, Dong, Henderson, Christopher, Carter, Nordlander, and
  Halas]{Zhou:2020tn}
L.~Zhou, J.~M.~P. Martirez, J.~Finzel, C.~Zhang, D.~F. Swearer, S.~Tian,
  H.~Robatjazi, M.~Lou, L.~Dong, L.~Henderson, P.~Christopher, E.~A. Carter,
  P.~Nordlander and N.~J. Halas, \emph{Nat. Energy}, 2020, \textbf{5},
  61--70\relax
\mciteBstWouldAddEndPuncttrue
\mciteSetBstMidEndSepPunct{\mcitedefaultmidpunct}
{\mcitedefaultendpunct}{\mcitedefaultseppunct}\relax
\EndOfBibitem
\bibitem[Wang \emph{et~al.}(2015)Wang, Ando, and Camargo]{Wang:2015uk}
J.~Wang, R.~A. Ando and P.~H.~C. Camargo, \emph{Angew. Chem. Int. Ed.}, 2015,
  \textbf{54}, 6909--6912\relax
\mciteBstWouldAddEndPuncttrue
\mciteSetBstMidEndSepPunct{\mcitedefaultmidpunct}
{\mcitedefaultendpunct}{\mcitedefaultseppunct}\relax
\EndOfBibitem
\bibitem[Zhan \emph{et~al.}(2019)Zhan, Liu, Huang, Hu, Ren, Moskovits, and
  Tian]{Zhan:2019wc}
C.~Zhan, B.-W. Liu, Y.-F. Huang, S.~Hu, B.~Ren, M.~Moskovits and Z.-Q. Tian,
  \emph{Nat. Commun.}, 2019, \textbf{10}, 2671\relax
\mciteBstWouldAddEndPuncttrue
\mciteSetBstMidEndSepPunct{\mcitedefaultmidpunct}
{\mcitedefaultendpunct}{\mcitedefaultseppunct}\relax
\EndOfBibitem
\bibitem[Xiao \emph{et~al.}(2017)Xiao, Hu, Zhang, and MacFarlane]{Xiao:2017uw}
C.~Xiao, H.~Hu, X.~Zhang and D.~R. MacFarlane, \emph{ACS Sustainable Chem.
  Eng.}, 2017, \textbf{5}, 10858--10863\relax
\mciteBstWouldAddEndPuncttrue
\mciteSetBstMidEndSepPunct{\mcitedefaultmidpunct}
{\mcitedefaultendpunct}{\mcitedefaultseppunct}\relax
\EndOfBibitem
\bibitem[Yu and Jain(2019)]{Yu:2019ut}
S.~Yu and P.~K. Jain, \emph{ACS Energy Lett.}, 2019, \textbf{4},
  2295--2300\relax
\mciteBstWouldAddEndPuncttrue
\mciteSetBstMidEndSepPunct{\mcitedefaultmidpunct}
{\mcitedefaultendpunct}{\mcitedefaultseppunct}\relax
\EndOfBibitem
\bibitem[Singh \emph{et~al.}(2023)Singh, Verma, Kaul, Sa, Punjal, Prabhu, and
  Polshettiwar]{Singh:2023wt}
S.~Singh, R.~Verma, N.~Kaul, J.~Sa, A.~Punjal, S.~Prabhu and V.~Polshettiwar,
  \emph{Nat. Commun.}, 2023, \textbf{14}, 2551\relax
\mciteBstWouldAddEndPuncttrue
\mciteSetBstMidEndSepPunct{\mcitedefaultmidpunct}
{\mcitedefaultendpunct}{\mcitedefaultseppunct}\relax
\EndOfBibitem
\bibitem[Dhiman \emph{et~al.}(2019)Dhiman, Maity, Das, Belgamwar, Chalke, Lee,
  Sim, Nam, and Polshettiwar]{Dhiman2023chemsci}
M.~Dhiman, A.~Maity, A.~Das, R.~Belgamwar, B.~Chalke, Y.~Lee, K.~Sim, J.-M. Nam
  and V.~Polshettiwar, \emph{Chem. Sci.}, 2019, \textbf{10}, 6594--6603\relax
\mciteBstWouldAddEndPuncttrue
\mciteSetBstMidEndSepPunct{\mcitedefaultmidpunct}
{\mcitedefaultendpunct}{\mcitedefaultseppunct}\relax
\EndOfBibitem
\bibitem[Aslam \emph{et~al.}(2017)Aslam, Chavez, and Linic]{Aslam:2017tw}
U.~Aslam, S.~Chavez and S.~Linic, \emph{Nat. Nanotechnol.}, 2017, \textbf{12},
  1000--1005\relax
\mciteBstWouldAddEndPuncttrue
\mciteSetBstMidEndSepPunct{\mcitedefaultmidpunct}
{\mcitedefaultendpunct}{\mcitedefaultseppunct}\relax
\EndOfBibitem
\bibitem[Zhang \emph{et~al.}(2016)Zhang, Zhao, Zhou, Schlather, Dong, McClain,
  Swearer, Nordlander, and Halas]{Zhang:2016vx}
C.~Zhang, H.~Zhao, L.~Zhou, A.~E. Schlather, L.~Dong, M.~J. McClain, D.~F.
  Swearer, P.~Nordlander and N.~J. Halas, \emph{Nano Lett.}, 2016, \textbf{16},
  6677--6682\relax
\mciteBstWouldAddEndPuncttrue
\mciteSetBstMidEndSepPunct{\mcitedefaultmidpunct}
{\mcitedefaultendpunct}{\mcitedefaultseppunct}\relax
\EndOfBibitem
\bibitem[Swearer \emph{et~al.}(2016)Swearer, Zhao, Zhou, Zhang, Robatjazi,
  Martirez, Krauter, Yazdi, McClain, Ringe, Carter, Nordlander, and
  Halas]{Swearer:2016pnas}
D.~F. Swearer, H.~Zhao, L.~Zhou, C.~Zhang, H.~Robatjazi, J.~M.~P. Martirez,
  C.~M. Krauter, S.~Yazdi, M.~J. McClain, E.~Ringe, E.~A. Carter, P.~Nordlander
  and N.~J. Halas, \emph{Proc. Natl. Acad. Sci.}, 2016, \textbf{113},
  8916--8920\relax
\mciteBstWouldAddEndPuncttrue
\mciteSetBstMidEndSepPunct{\mcitedefaultmidpunct}
{\mcitedefaultendpunct}{\mcitedefaultseppunct}\relax
\EndOfBibitem
\bibitem[da~Silva \emph{et~al.}(2016)da~Silva, Rodrigues, Correia, Alves,
  Alves, Ando, Ornellas, Wang, Andrade, and Camargo]{Silva:2016tp}
A.~G.~M. da~Silva, T.~S. Rodrigues, V.~G. Correia, T.~V. Alves, R.~S. Alves,
  R.~A. Ando, F.~R. Ornellas, J.~Wang, L.~H. Andrade and P.~H.~C. Camargo,
  \emph{Angew. Chem. Int. Ed.}, 2016, \textbf{55}, 7111--7115\relax
\mciteBstWouldAddEndPuncttrue
\mciteSetBstMidEndSepPunct{\mcitedefaultmidpunct}
{\mcitedefaultendpunct}{\mcitedefaultseppunct}\relax
\EndOfBibitem
\bibitem[Huang \emph{et~al.}(2010)Huang, Zhu, Liu, Wu, Ren, and
  Tian]{Huang:2010uf}
Y.-F. Huang, H.-P. Zhu, G.-K. Liu, D.-Y. Wu, B.~Ren and Z.-Q. Tian, \emph{J.
  Am. Chem. Soc.}, 2010, \textbf{132}, 9244--9246\relax
\mciteBstWouldAddEndPuncttrue
\mciteSetBstMidEndSepPunct{\mcitedefaultmidpunct}
{\mcitedefaultendpunct}{\mcitedefaultseppunct}\relax
\EndOfBibitem
\bibitem[Huang \emph{et~al.}(2014)Huang, Zhang, Zhao, Feng, Wu, Ren, and
  Tian]{Huang:2014ux}
Y.-F. Huang, M.~Zhang, L.-B. Zhao, J.-M. Feng, D.-Y. Wu, B.~Ren and Z.-Q. Tian,
  \emph{Angew. Chem. Int. Ed.}, 2014, \textbf{53}, 2353--2357\relax
\mciteBstWouldAddEndPuncttrue
\mciteSetBstMidEndSepPunct{\mcitedefaultmidpunct}
{\mcitedefaultendpunct}{\mcitedefaultseppunct}\relax
\EndOfBibitem
\bibitem[Zhao \emph{et~al.}(2014)Zhao, Zhang, Huang, Williams, Wu, Ren, and
  Tian]{Zhao:2014vj}
L.-B. Zhao, M.~Zhang, Y.-F. Huang, C.~T. Williams, D.-Y. Wu, B.~Ren and Z.-Q.
  Tian, \emph{The Journal of Physical Chemistry Letters}, 2014, \textbf{5},
  1259--1266\relax
\mciteBstWouldAddEndPuncttrue
\mciteSetBstMidEndSepPunct{\mcitedefaultmidpunct}
{\mcitedefaultendpunct}{\mcitedefaultseppunct}\relax
\EndOfBibitem
\bibitem[Yu \emph{et~al.}(2018)Yu, Sundaresan, and Willets]{Yu:2018vh}
Y.~Yu, V.~Sundaresan and K.~A. Willets, \emph{J. Phys. Chem. C}, 2018,
  \textbf{122}, 5040--5048\relax
\mciteBstWouldAddEndPuncttrue
\mciteSetBstMidEndSepPunct{\mcitedefaultmidpunct}
{\mcitedefaultendpunct}{\mcitedefaultseppunct}\relax
\EndOfBibitem
\bibitem[Zhou \emph{et~al.}(2018)Zhou, Swearer, Zhang, Robatjazi, Zhao,
  Henderson, Dong, Christopher, Carter, Nordlander, and Halas]{Zhou:2018aa}
L.~Zhou, D.~F. Swearer, C.~Zhang, H.~Robatjazi, H.~Zhao, L.~Henderson, L.~Dong,
  P.~Christopher, E.~A. Carter, P.~Nordlander and N.~J. Halas, \emph{Science},
  2018, \textbf{362}, 69--72\relax
\mciteBstWouldAddEndPuncttrue
\mciteSetBstMidEndSepPunct{\mcitedefaultmidpunct}
{\mcitedefaultendpunct}{\mcitedefaultseppunct}\relax
\EndOfBibitem
\bibitem[Zhang \emph{et~al.}(2018)Zhang, Li, Reish, Zhang, Su, Guti{\'e}rrez,
  Moreno, Yang, Everitt, and Liu]{Zhang:2018ug}
X.~Zhang, X.~Li, M.~E. Reish, D.~Zhang, N.~Q. Su, Y.~Guti{\'e}rrez, F.~Moreno,
  W.~Yang, H.~O. Everitt and J.~Liu, \emph{Nano Lett.}, 2018, \textbf{18},
  1714--1723\relax
\mciteBstWouldAddEndPuncttrue
\mciteSetBstMidEndSepPunct{\mcitedefaultmidpunct}
{\mcitedefaultendpunct}{\mcitedefaultseppunct}\relax
\EndOfBibitem
\bibitem[Ou \emph{et~al.}(2020)Ou, Zhou, Shen, Lo, Lei, Li, Zhong, Li, and
  Lu]{Ou:2020vk}
W.~Ou, B.~Zhou, J.~Shen, T.~W. Lo, D.~Lei, S.~Li, J.~Zhong, Y.~Y. Li and J.~Lu,
  \emph{Angew. Chem. Int. Ed.}, 2020, \textbf{59}, 6790--6793\relax
\mciteBstWouldAddEndPuncttrue
\mciteSetBstMidEndSepPunct{\mcitedefaultmidpunct}
{\mcitedefaultendpunct}{\mcitedefaultseppunct}\relax
\EndOfBibitem
\bibitem[Baffou \emph{et~al.}(2020)Baffou, Bordacchini, Baldi, and
  Quidant]{Baffou:2020ts}
G.~Baffou, I.~Bordacchini, A.~Baldi and R.~Quidant, \emph{Light: Sci. Appl.},
  2020, \textbf{9}, 108\relax
\mciteBstWouldAddEndPuncttrue
\mciteSetBstMidEndSepPunct{\mcitedefaultmidpunct}
{\mcitedefaultendpunct}{\mcitedefaultseppunct}\relax
\EndOfBibitem
\bibitem[Perera \emph{et~al.}(2020)Perera, Gunapala, Stockman, and
  Premaratne]{Perera:2020jpcc}
T.~Perera, S.~D. Gunapala, M.~I. Stockman and M.~Premaratne, \emph{J. Phys.
  Chem. C}, 2020, \textbf{124}, 27694--27708\relax
\mciteBstWouldAddEndPuncttrue
\mciteSetBstMidEndSepPunct{\mcitedefaultmidpunct}
{\mcitedefaultendpunct}{\mcitedefaultseppunct}\relax
\EndOfBibitem
\bibitem[Rossi \emph{et~al.}(2017)Rossi, Kuisma, Puska, Nieminen, and
  Erhart]{Rossi:2017jctc}
T.~P. Rossi, M.~Kuisma, M.~J. Puska, R.~M. Nieminen and P.~Erhart, \emph{J.
  Chem. Theory Comput.}, 2017, \textbf{13}, 4779--4790\relax
\mciteBstWouldAddEndPuncttrue
\mciteSetBstMidEndSepPunct{\mcitedefaultmidpunct}
{\mcitedefaultendpunct}{\mcitedefaultseppunct}\relax
\EndOfBibitem
\bibitem[Asadi-Aghbolaghi \emph{et~al.}(2020)Asadi-Aghbolaghi, R{\"u}ger,
  Jamshidi, and Visscher]{Asadi:2020jpcc}
N.~Asadi-Aghbolaghi, R.~R{\"u}ger, Z.~Jamshidi and L.~Visscher, \emph{J. Phys.
  Chem. C}, 2020, \textbf{124}, 7946--7955\relax
\mciteBstWouldAddEndPuncttrue
\mciteSetBstMidEndSepPunct{\mcitedefaultmidpunct}
{\mcitedefaultendpunct}{\mcitedefaultseppunct}\relax
\EndOfBibitem
\bibitem[Pandeya and Aikens(2021)]{Pandeya:2021jpca}
P.~Pandeya and C.~M. Aikens, \emph{J. Phys. Chem. A}, 2021, \textbf{125},
  4847--4860\relax
\mciteBstWouldAddEndPuncttrue
\mciteSetBstMidEndSepPunct{\mcitedefaultmidpunct}
{\mcitedefaultendpunct}{\mcitedefaultseppunct}\relax
\EndOfBibitem
\bibitem[Alkan and Aikens(2018)]{Alkan:2018ug}
F.~Alkan and C.~M. Aikens, \emph{J. Phys. Chem. C}, 2018, \textbf{122},
  23639--23650\relax
\mciteBstWouldAddEndPuncttrue
\mciteSetBstMidEndSepPunct{\mcitedefaultmidpunct}
{\mcitedefaultendpunct}{\mcitedefaultseppunct}\relax
\EndOfBibitem
\bibitem[Alkan and Aikens(2021)]{Alkan:2021vq}
F.~Alkan and C.~M. Aikens, \emph{J. Phys. Chem. C}, 2021, \textbf{125},
  12198--12206\relax
\mciteBstWouldAddEndPuncttrue
\mciteSetBstMidEndSepPunct{\mcitedefaultmidpunct}
{\mcitedefaultendpunct}{\mcitedefaultseppunct}\relax
\EndOfBibitem
\bibitem[Della~Sala(2022)]{DellaSala:2022vo}
F.~Della~Sala, \emph{J. Chem. Phys.}, 2022, \textbf{157}, 104101\relax
\mciteBstWouldAddEndPuncttrue
\mciteSetBstMidEndSepPunct{\mcitedefaultmidpunct}
{\mcitedefaultendpunct}{\mcitedefaultseppunct}\relax
\EndOfBibitem
\bibitem[D'Agostino \emph{et~al.}(2018)D'Agostino, Rinaldi, Cuniberti, and
  Della~Sala]{DAgostino:2018tg}
S.~D'Agostino, R.~Rinaldi, G.~Cuniberti and F.~Della~Sala, \emph{J. Phys. Chem.
  C}, 2018, \textbf{122}, 19756--19766\relax
\mciteBstWouldAddEndPuncttrue
\mciteSetBstMidEndSepPunct{\mcitedefaultmidpunct}
{\mcitedefaultendpunct}{\mcitedefaultseppunct}\relax
\EndOfBibitem
\bibitem[You and Panoiu(2019)]{You:2019vn}
J.~W. You and N.~C. Panoiu, \emph{IEEE J. Multiscale and Multiphys. Comput.
  Techn.}, 2019, \textbf{4}, 111--118\relax
\mciteBstWouldAddEndPuncttrue
\mciteSetBstMidEndSepPunct{\mcitedefaultmidpunct}
{\mcitedefaultendpunct}{\mcitedefaultseppunct}\relax
\EndOfBibitem
\bibitem[Hohenberg and Kohn(1964)]{HK1964}
P.~Hohenberg and W.~Kohn, \emph{Phys. Rev.}, 1964, \textbf{136},
  B864--B871\relax
\mciteBstWouldAddEndPuncttrue
\mciteSetBstMidEndSepPunct{\mcitedefaultmidpunct}
{\mcitedefaultendpunct}{\mcitedefaultseppunct}\relax
\EndOfBibitem
\bibitem[Parr and Yang(1995)]{Parr1995annuphys}
R.~G. Parr and W.~Yang, \emph{Annu. Rev. Phys. Chem.}, 1995, \textbf{46},
  701--728\relax
\mciteBstWouldAddEndPuncttrue
\mciteSetBstMidEndSepPunct{\mcitedefaultmidpunct}
{\mcitedefaultendpunct}{\mcitedefaultseppunct}\relax
\EndOfBibitem
\bibitem[Runge and Gross(1984)]{RG1984}
E.~Runge and E.~K.~U. Gross, \emph{Phys. Rev. Lett.}, 1984, \textbf{52},
  997--1000\relax
\mciteBstWouldAddEndPuncttrue
\mciteSetBstMidEndSepPunct{\mcitedefaultmidpunct}
{\mcitedefaultendpunct}{\mcitedefaultseppunct}\relax
\EndOfBibitem
\bibitem[Casida and Huix-Rotllant(2012)]{Casida:2012ti}
M.~E. Casida and M.~Huix-Rotllant, \emph{Annu. Rev. Phys. Chem.}, 2012,
  \textbf{63}, 287--323\relax
\mciteBstWouldAddEndPuncttrue
\mciteSetBstMidEndSepPunct{\mcitedefaultmidpunct}
{\mcitedefaultendpunct}{\mcitedefaultseppunct}\relax
\EndOfBibitem
\bibitem[Tiago and Chelikowsky(2006)]{PRB73205334}
M.~L. Tiago and J.~R. Chelikowsky, \emph{Phys. Rev. B}, 2006, \textbf{73},
  205334\relax
\mciteBstWouldAddEndPuncttrue
\mciteSetBstMidEndSepPunct{\mcitedefaultmidpunct}
{\mcitedefaultendpunct}{\mcitedefaultseppunct}\relax
\EndOfBibitem
\bibitem[Rom{\'a}n~Castellanos \emph{et~al.}(2019)Rom{\'a}n~Castellanos, Hess,
  and Lischner]{Castellanos}
L.~Rom{\'a}n~Castellanos, O.~Hess and J.~Lischner, \emph{Commun. Phys.}, 2019,
  \textbf{2}, 47\relax
\mciteBstWouldAddEndPuncttrue
\mciteSetBstMidEndSepPunct{\mcitedefaultmidpunct}
{\mcitedefaultendpunct}{\mcitedefaultseppunct}\relax
\EndOfBibitem
\bibitem[Wu \emph{et~al.}(2022)Wu, Liu, Frauenheim, Tretiak, Yam, and
  Zhang]{wu2022jcp_relaxation}
X.~Wu, B.~Liu, T.~Frauenheim, S.~Tretiak, C.~Yam and Y.~Zhang, \emph{J. Chem.
  Phys.}, 2022, \textbf{157}, 214201\relax
\mciteBstWouldAddEndPuncttrue
\mciteSetBstMidEndSepPunct{\mcitedefaultmidpunct}
{\mcitedefaultendpunct}{\mcitedefaultseppunct}\relax
\EndOfBibitem
\bibitem[Liu \emph{et~al.}(2018)Liu, Zhang, Link, and Nordlander]{acsph7b00881}
J.~G. Liu, H.~Zhang, S.~Link and P.~Nordlander, \emph{ACS Photonics}, 2018,
  \textbf{5}, 2584--2595\relax
\mciteBstWouldAddEndPuncttrue
\mciteSetBstMidEndSepPunct{\mcitedefaultmidpunct}
{\mcitedefaultendpunct}{\mcitedefaultseppunct}\relax
\EndOfBibitem
\bibitem[Breuer \emph{et~al.}(2002)Breuer, Petruccione, and
  Petruccione]{breuer2002theory}
H.~Breuer, F.~Petruccione and S.~Petruccione, \emph{The Theory of Open Quantum
  Systems}, Oxford University Press, 2002\relax
\mciteBstWouldAddEndPuncttrue
\mciteSetBstMidEndSepPunct{\mcitedefaultmidpunct}
{\mcitedefaultendpunct}{\mcitedefaultseppunct}\relax
\EndOfBibitem
\bibitem[Haug and Jauho(1998)]{haug1998}
H.~Haug and A.~Jauho, \emph{Quantum Kinetics in Transport and Optics of
  Semiconductors}, Springer Berlin Heidelberg, 1998\relax
\mciteBstWouldAddEndPuncttrue
\mciteSetBstMidEndSepPunct{\mcitedefaultmidpunct}
{\mcitedefaultendpunct}{\mcitedefaultseppunct}\relax
\EndOfBibitem
\bibitem[Tesema \emph{et~al.}(2017)Tesema, Kafle, Tadesse, and
  Habteyes]{Tesema2017jpcc}
T.~E. Tesema, B.~Kafle, M.~G. Tadesse and T.~G. Habteyes, \emph{J. Phys. Chem.
  C}, 2017, \textbf{121}, 7421--7428\relax
\mciteBstWouldAddEndPuncttrue
\mciteSetBstMidEndSepPunct{\mcitedefaultmidpunct}
{\mcitedefaultendpunct}{\mcitedefaultseppunct}\relax
\EndOfBibitem
\bibitem[Boerigter \emph{et~al.}(2016)Boerigter, Aslam, and
  Linic]{Boerigter2016acsnano}
C.~Boerigter, U.~Aslam and S.~Linic, \emph{ACS Nano}, 2016, \textbf{10},
  6108--6115\relax
\mciteBstWouldAddEndPuncttrue
\mciteSetBstMidEndSepPunct{\mcitedefaultmidpunct}
{\mcitedefaultendpunct}{\mcitedefaultseppunct}\relax
\EndOfBibitem
\bibitem[Long and Prezhdo(2014)]{Longrun2014jacs}
R.~Long and O.~V. Prezhdo, \emph{J. Am. Chem. Soc.}, 2014, \textbf{136},
  4343--4354\relax
\mciteBstWouldAddEndPuncttrue
\mciteSetBstMidEndSepPunct{\mcitedefaultmidpunct}
{\mcitedefaultendpunct}{\mcitedefaultseppunct}\relax
\EndOfBibitem
\bibitem[Long \emph{et~al.}(2012)Long, English, and Prezhdo]{Long2012jacs}
R.~Long, N.~J. English and O.~V. Prezhdo, \emph{J. Am. Chem. Soc.}, 2012,
  \textbf{134}, 14238--14248\relax
\mciteBstWouldAddEndPuncttrue
\mciteSetBstMidEndSepPunct{\mcitedefaultmidpunct}
{\mcitedefaultendpunct}{\mcitedefaultseppunct}\relax
\EndOfBibitem
\bibitem[Bora \emph{et~al.}(2016)Bora, Zoepfl, and Dutta]{Bora2016scirep}
T.~Bora, D.~Zoepfl and J.~Dutta, \emph{Sci. Rep.}, 2016, \textbf{6},
  26913\relax
\mciteBstWouldAddEndPuncttrue
\mciteSetBstMidEndSepPunct{\mcitedefaultmidpunct}
{\mcitedefaultendpunct}{\mcitedefaultseppunct}\relax
\EndOfBibitem
\bibitem[Sivan \emph{et~al.}(2019)Sivan, Un, and Dubi]{Sivan2019farady}
Y.~Sivan, I.~W. Un and Y.~Dubi, \emph{Faraday Discuss.}, 2019, \textbf{214},
  215--233\relax
\mciteBstWouldAddEndPuncttrue
\mciteSetBstMidEndSepPunct{\mcitedefaultmidpunct}
{\mcitedefaultendpunct}{\mcitedefaultseppunct}\relax
\EndOfBibitem
\bibitem[Fusco \emph{et~al.}(2022)Fusco, Catchpole, and Beck]{Fusco:2022jmcc}
Z.~Fusco, K.~Catchpole and F.~J. Beck, \emph{J. Mater. Chem. C}, 2022,
  \textbf{10}, 7511--7524\relax
\mciteBstWouldAddEndPuncttrue
\mciteSetBstMidEndSepPunct{\mcitedefaultmidpunct}
{\mcitedefaultendpunct}{\mcitedefaultseppunct}\relax
\EndOfBibitem
\bibitem[Chen \emph{et~al.}(2012)Chen, Pu, Chang, Liang, Liu, Yeh, Shih, and
  Hsu]{Chen:2012ur}
K.-H. Chen, Y.-C. Pu, K.-D. Chang, Y.-F. Liang, C.-M. Liu, J.-W. Yeh, H.-C.
  Shih and Y.-J. Hsu, \emph{The Journal of Physical Chemistry C}, 2012,
  \textbf{116}, 19039--19045\relax
\mciteBstWouldAddEndPuncttrue
\mciteSetBstMidEndSepPunct{\mcitedefaultmidpunct}
{\mcitedefaultendpunct}{\mcitedefaultseppunct}\relax
\EndOfBibitem
\bibitem[Golubev \emph{et~al.}(2018)Golubev, Khlebtsov, Rodriguez, Chen, and
  Zahn]{Golubev:2018vr}
A.~A. Golubev, B.~N. Khlebtsov, R.~D. Rodriguez, Y.~Chen and D.~R.~T. Zahn,
  \emph{The Journal of Physical Chemistry C}, 2018, \textbf{122},
  5657--5663\relax
\mciteBstWouldAddEndPuncttrue
\mciteSetBstMidEndSepPunct{\mcitedefaultmidpunct}
{\mcitedefaultendpunct}{\mcitedefaultseppunct}\relax
\EndOfBibitem
\bibitem[Keller and Frontiera(2018)]{Keller:2018tl}
E.~L. Keller and R.~R. Frontiera, \emph{ACS Nano}, 2018, \textbf{12},
  5848--5855\relax
\mciteBstWouldAddEndPuncttrue
\mciteSetBstMidEndSepPunct{\mcitedefaultmidpunct}
{\mcitedefaultendpunct}{\mcitedefaultseppunct}\relax
\EndOfBibitem
\bibitem[Brown \emph{et~al.}(2017)Brown, Sundararaman, Narang, Schwartzberg,
  Goddard, and Atwater]{Brown2017prl}
A.~M. Brown, R.~Sundararaman, P.~Narang, A.~M. Schwartzberg, W.~A. Goddard and
  H.~A. Atwater, \emph{Phys. Rev. Lett.}, 2017, \textbf{118}, 087401\relax
\mciteBstWouldAddEndPuncttrue
\mciteSetBstMidEndSepPunct{\mcitedefaultmidpunct}
{\mcitedefaultendpunct}{\mcitedefaultseppunct}\relax
\EndOfBibitem
\bibitem[Nelson \emph{et~al.}(2020)Nelson, White, Bjorgaard, Sifain, Zhang,
  Nebgen, Fernandez-Alberti, Mozyrsky, Roitberg, and
  Tretiak]{Nelson2020ChemRev}
T.~R. Nelson, A.~J. White, J.~A. Bjorgaard, A.~E. Sifain, Y.~Zhang, B.~Nebgen,
  S.~Fernandez-Alberti, D.~Mozyrsky, A.~E. Roitberg and S.~Tretiak, \emph{Chem.
  Rev.}, 2020, \textbf{120}, 2215--2287\relax
\mciteBstWouldAddEndPuncttrue
\mciteSetBstMidEndSepPunct{\mcitedefaultmidpunct}
{\mcitedefaultendpunct}{\mcitedefaultseppunct}\relax
\EndOfBibitem
\bibitem[Tully(1990)]{tully1990molecular}
J.~C. Tully, \emph{J. Chem. Phys.}, 1990, \textbf{93}, 1061--1071\relax
\mciteBstWouldAddEndPuncttrue
\mciteSetBstMidEndSepPunct{\mcitedefaultmidpunct}
{\mcitedefaultendpunct}{\mcitedefaultseppunct}\relax
\EndOfBibitem
\bibitem[Tully(2012)]{Tully2012jcp}
J.~C. Tully, \emph{J. Chem. Phys.}, 2012, \textbf{137}, 22A301\relax
\mciteBstWouldAddEndPuncttrue
\mciteSetBstMidEndSepPunct{\mcitedefaultmidpunct}
{\mcitedefaultendpunct}{\mcitedefaultseppunct}\relax
\EndOfBibitem
\bibitem[Curchod \emph{et~al.}(2013)Curchod, Rothlisberger, and
  Tavernelli]{curchod2013trajectory}
B.~F. Curchod, U.~Rothlisberger and I.~Tavernelli, \emph{Chem. Phys. Chem.},
  2013, \textbf{14}, 1314--1340\relax
\mciteBstWouldAddEndPuncttrue
\mciteSetBstMidEndSepPunct{\mcitedefaultmidpunct}
{\mcitedefaultendpunct}{\mcitedefaultseppunct}\relax
\EndOfBibitem
\bibitem[Malone \emph{et~al.}(2020)Malone, Nebgen, White, Zhang, Song,
  Bjorgaard, Sifain, Rodriguez-Hernandez, Freixas,
  Fernandez-Alberti,\emph{et~al.}]{malone2020nexmd}
W.~Malone, B.~Nebgen, A.~White, Y.~Zhang, H.~Song, J.~A. Bjorgaard, A.~E.
  Sifain, B.~Rodriguez-Hernandez, V.~M. Freixas, S.~Fernandez-Alberti
  \emph{et~al.}, \emph{J. Chem. Theory Comput.}, 2020, \textbf{16},
  5771--5783\relax
\mciteBstWouldAddEndPuncttrue
\mciteSetBstMidEndSepPunct{\mcitedefaultmidpunct}
{\mcitedefaultendpunct}{\mcitedefaultseppunct}\relax
\EndOfBibitem
\bibitem[Fabiano \emph{et~al.}(2008)Fabiano, Keal, and
  Thiel]{Fabiano2008chemphys}
E.~Fabiano, T.~W. Keal and W.~Thiel, \emph{Chem. Phys.}, 2008, \textbf{349},
  334--347\relax
\mciteBstWouldAddEndPuncttrue
\mciteSetBstMidEndSepPunct{\mcitedefaultmidpunct}
{\mcitedefaultendpunct}{\mcitedefaultseppunct}\relax
\EndOfBibitem
\bibitem[Makhov \emph{et~al.}(2017)Makhov, Symonds, Fernandez-Alberti, and
  Shalashilin]{makhov_MCEh_ChemPhys2017}
D.~V. Makhov, C.~Symonds, S.~Fernandez-Alberti and D.~V. Shalashilin,
  \emph{Chem. Phys.}, 2017, \textbf{493}, 200--218\relax
\mciteBstWouldAddEndPuncttrue
\mciteSetBstMidEndSepPunct{\mcitedefaultmidpunct}
{\mcitedefaultendpunct}{\mcitedefaultseppunct}\relax
\EndOfBibitem
\bibitem[Makhov \emph{et~al.}(2014)Makhov, Glover, Martinez, and
  Shalashilin]{makhov_AIMS_JCP2014}
D.~V. Makhov, W.~J. Glover, T.~J. Martinez and D.~V. Shalashilin, \emph{J.
  Chem. Phys.}, 2014, \textbf{141}, 054110\relax
\mciteBstWouldAddEndPuncttrue
\mciteSetBstMidEndSepPunct{\mcitedefaultmidpunct}
{\mcitedefaultendpunct}{\mcitedefaultseppunct}\relax
\EndOfBibitem
\bibitem[Makhov \emph{et~al.}(2015)Makhov, Saita, Martinez, and
  Shalashilin]{makhov_AIMS_PCCP2015}
D.~V. Makhov, K.~Saita, T.~J. Martinez and D.~V. Shalashilin, \emph{Phys. Chem.
  Chem. Phys.}, 2015, \textbf{17}, 3316--3325\relax
\mciteBstWouldAddEndPuncttrue
\mciteSetBstMidEndSepPunct{\mcitedefaultmidpunct}
{\mcitedefaultendpunct}{\mcitedefaultseppunct}\relax
\EndOfBibitem
\bibitem[V.~Makhov \emph{et~al.}(2016)V.~Makhov, J.~Martinez, and
  V.~Shalashilin]{makhov_AIMS_FaraDisc2016}
D.~V.~Makhov, T.~J.~Martinez and D.~V.~Shalashilin, \emph{Faraday Discuss.},
  2016, \textbf{194}, 81--94\relax
\mciteBstWouldAddEndPuncttrue
\mciteSetBstMidEndSepPunct{\mcitedefaultmidpunct}
{\mcitedefaultendpunct}{\mcitedefaultseppunct}\relax
\EndOfBibitem
\bibitem[Song \emph{et~al.}(2021)Song, Freixas, Fernandez-Alberti, White,
  Zhang, Mukamel, Govind, and Tretiak]{song2021ab}
H.~Song, V.~M. Freixas, S.~Fernandez-Alberti, A.~J. White, Y.~Zhang,
  S.~Mukamel, N.~Govind and S.~Tretiak, \emph{J. Chem. Theory Comput.}, 2021,
  \textbf{17}, 3629--3643\relax
\mciteBstWouldAddEndPuncttrue
\mciteSetBstMidEndSepPunct{\mcitedefaultmidpunct}
{\mcitedefaultendpunct}{\mcitedefaultseppunct}\relax
\EndOfBibitem
\bibitem[Ben-Nun \emph{et~al.}(2000)Ben-Nun, Quenneville, and
  Mart{\'\i}nez]{bennun_AIMS_JPCA2000}
M.~Ben-Nun, J.~Quenneville and T.~J. Mart{\'\i}nez, \emph{J. Phys. Chem. A},
  2000, \textbf{104}, 5161--5175\relax
\mciteBstWouldAddEndPuncttrue
\mciteSetBstMidEndSepPunct{\mcitedefaultmidpunct}
{\mcitedefaultendpunct}{\mcitedefaultseppunct}\relax
\EndOfBibitem
\bibitem[Besteiro \emph{et~al.}(2017)Besteiro, Kong, Wang, Hartland, and
  Govorov]{Besterio2017acsphotonics}
L.~V. Besteiro, X.-T. Kong, Z.~Wang, G.~Hartland and A.~O. Govorov, \emph{ACS
  Photonics}, 2017, \textbf{4}, 2759--2781\relax
\mciteBstWouldAddEndPuncttrue
\mciteSetBstMidEndSepPunct{\mcitedefaultmidpunct}
{\mcitedefaultendpunct}{\mcitedefaultseppunct}\relax
\EndOfBibitem
\bibitem[Chang \emph{et~al.}(2019)Chang, Besteiro, Sun, Santiago, Gray, Wang,
  and Govorov]{Chang2019acs}
L.~Chang, L.~V. Besteiro, J.~Sun, E.~Y. Santiago, S.~K. Gray, Z.~Wang and A.~O.
  Govorov, \emph{ACS Energy Lett.}, 2019, \textbf{4}, 2552--2568\relax
\mciteBstWouldAddEndPuncttrue
\mciteSetBstMidEndSepPunct{\mcitedefaultmidpunct}
{\mcitedefaultendpunct}{\mcitedefaultseppunct}\relax
\EndOfBibitem
\bibitem[Yan \emph{et~al.}(2016)Yan, Wang, and Meng]{Yan:2016uz}
L.~Yan, F.~Wang and S.~Meng, \emph{ACS Nano}, 2016, \textbf{10},
  5452--5458\relax
\mciteBstWouldAddEndPuncttrue
\mciteSetBstMidEndSepPunct{\mcitedefaultmidpunct}
{\mcitedefaultendpunct}{\mcitedefaultseppunct}\relax
\EndOfBibitem
\bibitem[Akimov \emph{et~al.}(2013)Akimov, Neukirch, and
  Prezhdo]{akimov_theoretical_ChemRev2013}
A.~V. Akimov, A.~J. Neukirch and O.~V. Prezhdo, \emph{Chem. Rev.}, 2013,
  \textbf{113}, 4496--4565\relax
\mciteBstWouldAddEndPuncttrue
\mciteSetBstMidEndSepPunct{\mcitedefaultmidpunct}
{\mcitedefaultendpunct}{\mcitedefaultseppunct}\relax
\EndOfBibitem
\bibitem[Trani \emph{et~al.}(2011)Trani, Scalmani, Zheng, Carnimeo, Frisch, and
  Barone]{trani_TDDFTB_JCTC2011}
F.~Trani, G.~Scalmani, G.~Zheng, I.~Carnimeo, M.~J. Frisch and V.~Barone,
  \emph{J. Chem. Theory Comput.}, 2011, \textbf{7}, 3304--3313\relax
\mciteBstWouldAddEndPuncttrue
\mciteSetBstMidEndSepPunct{\mcitedefaultmidpunct}
{\mcitedefaultendpunct}{\mcitedefaultseppunct}\relax
\EndOfBibitem
\bibitem[Niehaus \emph{et~al.}(2001)Niehaus, Suhai, Della~Sala, Lugli, Elstner,
  Seifert, and Frauenheim]{niehaus2001tight}
T.~A. Niehaus, S.~Suhai, F.~Della~Sala, P.~Lugli, M.~Elstner, G.~Seifert and
  T.~Frauenheim, \emph{Phys. Rev. B}, 2001, \textbf{63}, 085108\relax
\mciteBstWouldAddEndPuncttrue
\mciteSetBstMidEndSepPunct{\mcitedefaultmidpunct}
{\mcitedefaultendpunct}{\mcitedefaultseppunct}\relax
\EndOfBibitem
\bibitem[Song \emph{et~al.}(2020)Song, Fischer, Zhang, Cramer, Mukamel, Govind,
  and Tretiak]{song2020first}
H.~Song, S.~A. Fischer, Y.~Zhang, C.~J. Cramer, S.~Mukamel, N.~Govind and
  S.~Tretiak, \emph{J. Chem. Theory Comput.}, 2020, \textbf{16},
  6418--6427\relax
\mciteBstWouldAddEndPuncttrue
\mciteSetBstMidEndSepPunct{\mcitedefaultmidpunct}
{\mcitedefaultendpunct}{\mcitedefaultseppunct}\relax
\EndOfBibitem
\bibitem[Wu \emph{et~al.}(2022)Wu, Wen, Song, Frauenheim, Tretiak, Yam, and
  Zhang]{wu2022nonadiabatic}
X.~Wu, S.~Wen, H.~Song, T.~Frauenheim, S.~Tretiak, C.~Yam and Y.~Zhang,
  \emph{J. Chem. Phys.}, 2022, \textbf{157}, 084114\relax
\mciteBstWouldAddEndPuncttrue
\mciteSetBstMidEndSepPunct{\mcitedefaultmidpunct}
{\mcitedefaultendpunct}{\mcitedefaultseppunct}\relax
\EndOfBibitem
\bibitem[Hourahine \emph{et~al.}(2020)Hourahine, Aradi, Blum, Bonaf{\'e},
  Buccheri, Camacho, Cevallos, Deshaye, Dumitric{\u{a}},
  Dominguez,\emph{et~al.}]{hourahine2020dftb+}
B.~Hourahine, B.~Aradi, V.~Blum, F.~Bonaf{\'e}, A.~Buccheri, C.~Camacho,
  C.~Cevallos, M.~Deshaye, T.~Dumitric{\u{a}}, A.~Dominguez \emph{et~al.},
  \emph{J. Chem. Phys.}, 2020, \textbf{152}, 124101\relax
\mciteBstWouldAddEndPuncttrue
\mciteSetBstMidEndSepPunct{\mcitedefaultmidpunct}
{\mcitedefaultendpunct}{\mcitedefaultseppunct}\relax
\EndOfBibitem
\bibitem[Douglas-Gallardo \emph{et~al.}(2019)Douglas-Gallardo, Berdakin,
  Frauenheim, and S{\'a}nchez]{douglas2019plasmon}
O.~A. Douglas-Gallardo, M.~Berdakin, T.~Frauenheim and C.~G. S{\'a}nchez,
  \emph{Nanoscale}, 2019, \textbf{11}, 8604--8615\relax
\mciteBstWouldAddEndPuncttrue
\mciteSetBstMidEndSepPunct{\mcitedefaultmidpunct}
{\mcitedefaultendpunct}{\mcitedefaultseppunct}\relax
\EndOfBibitem
\bibitem[Berdakin \emph{et~al.}(2019)Berdakin, Douglas-Gallardo, and
  Sanchez]{berdakin2019interplay}
M.~Berdakin, O.~A. Douglas-Gallardo and C.~G. Sanchez, \emph{J. Phys. Chem. C},
  2019, \textbf{124}, 1631--1639\relax
\mciteBstWouldAddEndPuncttrue
\mciteSetBstMidEndSepPunct{\mcitedefaultmidpunct}
{\mcitedefaultendpunct}{\mcitedefaultseppunct}\relax
\EndOfBibitem
\bibitem[Mandal \emph{et~al.}(2023)Mandal, Taylor, Weight, Koessler, Li, and
  Huo]{mandal_ChemRev2022}
A.~Mandal, M.~A. Taylor, B.~M. Weight, E.~R. Koessler, X.~Li and P.~Huo,
  \emph{Chem. Rev.}, 2023, \textbf{123}, 9786--9879\relax
\mciteBstWouldAddEndPuncttrue
\mciteSetBstMidEndSepPunct{\mcitedefaultmidpunct}
{\mcitedefaultendpunct}{\mcitedefaultseppunct}\relax
\EndOfBibitem
\bibitem[Foley~IV \emph{et~al.}(2023)Foley~IV, McTague, and
  DePrince~III]{foley:2023arxiv}
J.~J. Foley~IV, J.~F. McTague and A.~E. DePrince~III, \emph{Ab initio methods
  for polariton chemistry}, 2023, \url{http://arxiv.org/abs/2307.04881},
  arXiv:2307.04881 [physics]\relax
\mciteBstWouldAddEndPuncttrue
\mciteSetBstMidEndSepPunct{\mcitedefaultmidpunct}
{\mcitedefaultendpunct}{\mcitedefaultseppunct}\relax
\EndOfBibitem
\bibitem[Herrera and Owrutsky(2020)]{Herrera:2020jcp}
F.~Herrera and J.~Owrutsky, \emph{The Journal of Chemical Physics}, 2020,
  \textbf{152}, 100902\relax
\mciteBstWouldAddEndPuncttrue
\mciteSetBstMidEndSepPunct{\mcitedefaultmidpunct}
{\mcitedefaultendpunct}{\mcitedefaultseppunct}\relax
\EndOfBibitem
\bibitem[Simpkins \emph{et~al.}(2023)Simpkins, Dunkelberger, and
  Vurgaftman]{Simpkins:2023tm}
B.~S. Simpkins, A.~D. Dunkelberger and I.~Vurgaftman, \emph{Chemical Reviews},
  2023, \textbf{123}, 5020--5048\relax
\mciteBstWouldAddEndPuncttrue
\mciteSetBstMidEndSepPunct{\mcitedefaultmidpunct}
{\mcitedefaultendpunct}{\mcitedefaultseppunct}\relax
\EndOfBibitem
\bibitem[Dunkelberger \emph{et~al.}(2022)Dunkelberger, Simpkins, Vurgaftman,
  and Owrutsky]{Dunkelberger:2022uo}
A.~D. Dunkelberger, B.~S. Simpkins, I.~Vurgaftman and J.~C. Owrutsky,
  \emph{Annual Review of Physical Chemistry}, 2022, \textbf{73}, 429--451\relax
\mciteBstWouldAddEndPuncttrue
\mciteSetBstMidEndSepPunct{\mcitedefaultmidpunct}
{\mcitedefaultendpunct}{\mcitedefaultseppunct}\relax
\EndOfBibitem
\bibitem[Hutchison \emph{et~al.}(2012)Hutchison, Schwartz, Genet, Devaux, and
  Ebbesen]{Hutchison2012ACIE}
J.~A. Hutchison, T.~Schwartz, C.~Genet, E.~Devaux and T.~W. Ebbesen,
  \emph{Angew. Chem. Int. Ed.}, 2012, \textbf{51}, 1592--1596\relax
\mciteBstWouldAddEndPuncttrue
\mciteSetBstMidEndSepPunct{\mcitedefaultmidpunct}
{\mcitedefaultendpunct}{\mcitedefaultseppunct}\relax
\EndOfBibitem
\bibitem[George \emph{et~al.}(2016)George, Chervy, Shalabney, Devaux, Hiura,
  Genet, and Ebbesen]{George2016PRL}
J.~George, T.~Chervy, A.~Shalabney, E.~Devaux, H.~Hiura, C.~Genet and T.~W.
  Ebbesen, \emph{Phys. Rev. Lett.}, 2016, \textbf{117}, 153601\relax
\mciteBstWouldAddEndPuncttrue
\mciteSetBstMidEndSepPunct{\mcitedefaultmidpunct}
{\mcitedefaultendpunct}{\mcitedefaultseppunct}\relax
\EndOfBibitem
\bibitem[Vergauwe \emph{et~al.}(2019)Vergauwe, Thomas, Nagarajan, Shalabney,
  George, Chervy, Seidel, Devaux, Torbeev, and Ebbesen]{Vergauwe2019ACIE}
R.~M.~A. Vergauwe, A.~Thomas, K.~Nagarajan, A.~Shalabney, J.~George, T.~Chervy,
  M.~Seidel, E.~Devaux, V.~Torbeev and T.~W. Ebbesen, \emph{Angew. Chem. Int.
  Ed.}, 2019, \textbf{58}, 15324--15328\relax
\mciteBstWouldAddEndPuncttrue
\mciteSetBstMidEndSepPunct{\mcitedefaultmidpunct}
{\mcitedefaultendpunct}{\mcitedefaultseppunct}\relax
\EndOfBibitem
\bibitem[Lather \emph{et~al.}(2019)Lather, Bhatt, Thomas, Ebbesen, and
  George]{Lather2019ACIE}
J.~Lather, P.~Bhatt, A.~Thomas, T.~W. Ebbesen and J.~George, \emph{Angew. Chem.
  Int. Ed.}, 2019, \textbf{58}, 10635--10638\relax
\mciteBstWouldAddEndPuncttrue
\mciteSetBstMidEndSepPunct{\mcitedefaultmidpunct}
{\mcitedefaultendpunct}{\mcitedefaultseppunct}\relax
\EndOfBibitem
\bibitem[Sau \emph{et~al.}(2021)Sau, Nagarajan, Patrahau, Lethuillier-Karl,
  Vergauwe, Thomas, Moran, Genet, and Ebbesen]{Sau2021ACIE}
A.~Sau, K.~Nagarajan, B.~Patrahau, L.~Lethuillier-Karl, R.~M.~A. Vergauwe,
  A.~Thomas, J.~Moran, C.~Genet and T.~W. Ebbesen, \emph{Angew. Chem. Int.
  Ed.}, 2021, \textbf{60}, 5712--5717\relax
\mciteBstWouldAddEndPuncttrue
\mciteSetBstMidEndSepPunct{\mcitedefaultmidpunct}
{\mcitedefaultendpunct}{\mcitedefaultseppunct}\relax
\EndOfBibitem
\bibitem[Hirai \emph{et~al.}(2021)Hirai, Ishikawa, Chervy, Hutchison, and
  Uji-i]{Hirai2021CS}
K.~Hirai, H.~Ishikawa, T.~Chervy, J.~A. Hutchison and H.~Uji-i, \emph{Chem.
  Sci.}, 2021, \textbf{12}, 11986--11994\relax
\mciteBstWouldAddEndPuncttrue
\mciteSetBstMidEndSepPunct{\mcitedefaultmidpunct}
{\mcitedefaultendpunct}{\mcitedefaultseppunct}\relax
\EndOfBibitem
\bibitem[Takele \emph{et~al.}(2021)Takele, Wackenhut, Liu, Piatkowski, Waluk,
  and Meixner]{Takele2021JPCC}
W.~M. Takele, F.~Wackenhut, Q.~Liu, L.~Piatkowski, J.~Waluk and A.~J. Meixner,
  \emph{J. Phys. Chem. C}, 2021, \textbf{125}, 14932--14939\relax
\mciteBstWouldAddEndPuncttrue
\mciteSetBstMidEndSepPunct{\mcitedefaultmidpunct}
{\mcitedefaultendpunct}{\mcitedefaultseppunct}\relax
\EndOfBibitem
\bibitem[Thomas \emph{et~al.}(2019)Thomas, Lethuillier-Karl, Nagarajan,
  Vergauwe, George, Chervy, Shalabney, Devaux, Genet, Moran, and
  Ebbesen]{Thomas2019S}
A.~Thomas, L.~Lethuillier-Karl, K.~Nagarajan, R.~M.~A. Vergauwe, J.~George,
  T.~Chervy, A.~Shalabney, E.~Devaux, C.~Genet, J.~Moran and T.~W. Ebbesen,
  \emph{Science}, 2019, \textbf{363}, 615--619\relax
\mciteBstWouldAddEndPuncttrue
\mciteSetBstMidEndSepPunct{\mcitedefaultmidpunct}
{\mcitedefaultendpunct}{\mcitedefaultseppunct}\relax
\EndOfBibitem
\bibitem[Lather \emph{et~al.}(2022)Lather, Thabassum, Singh, and
  George]{Lather2022CS}
J.~Lather, A.~N.~K. Thabassum, J.~Singh and J.~George, \emph{Chem. Sci.}, 2022,
  \textbf{13}, 195--202\relax
\mciteBstWouldAddEndPuncttrue
\mciteSetBstMidEndSepPunct{\mcitedefaultmidpunct}
{\mcitedefaultendpunct}{\mcitedefaultseppunct}\relax
\EndOfBibitem
\bibitem[Tretiak and Mukamel(2002)]{tretiak2002CR}
S.~Tretiak and S.~Mukamel, \emph{Chem. Rev.}, 2002, \textbf{102},
  3171--3212\relax
\mciteBstWouldAddEndPuncttrue
\mciteSetBstMidEndSepPunct{\mcitedefaultmidpunct}
{\mcitedefaultendpunct}{\mcitedefaultseppunct}\relax
\EndOfBibitem
\bibitem[Timmer \emph{et~al.}(2023)Timmer, Gittinger, Quenzel, Stephan, Zhang,
  Schumacher, Lützen, Silies, Tretiak, Zhong, De~Sio, and
  Lienau]{timmer_plasmon_2023}
D.~Timmer, M.~Gittinger, T.~Quenzel, S.~Stephan, Y.~Zhang, M.~F. Schumacher,
  A.~Lützen, M.~Silies, S.~Tretiak, J.-H. Zhong, A.~De~Sio and C.~Lienau,
  2023\relax
\mciteBstWouldAddEndPuncttrue
\mciteSetBstMidEndSepPunct{\mcitedefaultmidpunct}
{\mcitedefaultendpunct}{\mcitedefaultseppunct}\relax
\EndOfBibitem
\bibitem[Curchod and Mart{\'{i}}nez(2018)]{Curchod2018CR}
B.~F.~E. Curchod and T.~J. Mart{\'{i}}nez, \emph{Chem. Rev.}, 2018,
  \textbf{118}, 3305--3336\relax
\mciteBstWouldAddEndPuncttrue
\mciteSetBstMidEndSepPunct{\mcitedefaultmidpunct}
{\mcitedefaultendpunct}{\mcitedefaultseppunct}\relax
\EndOfBibitem
\bibitem[Crespo-Otero and Barbatti(2018)]{Crespo-Otero:2018wa}
R.~Crespo-Otero and M.~Barbatti, \emph{Chemical Reviews}, 2018, \textbf{118},
  7026--7068\relax
\mciteBstWouldAddEndPuncttrue
\mciteSetBstMidEndSepPunct{\mcitedefaultmidpunct}
{\mcitedefaultendpunct}{\mcitedefaultseppunct}\relax
\EndOfBibitem
\bibitem[Yang \emph{et~al.}(2021)Yang, Ou, Pei, Wang, Weng, Shuai, Mullen, and
  Shao]{yang2021JCP}
J.~Yang, Q.~Ou, Z.~Pei, H.~Wang, B.~Weng, Z.~Shuai, K.~Mullen and Y.~Shao,
  \emph{J. Chem. Phys.}, 2021, \textbf{155}, 064107\relax
\mciteBstWouldAddEndPuncttrue
\mciteSetBstMidEndSepPunct{\mcitedefaultmidpunct}
{\mcitedefaultendpunct}{\mcitedefaultseppunct}\relax
\EndOfBibitem
\bibitem[Zhang \emph{et~al.}(2019)Zhang, Nelson, and Tretiak]{zhang2019jcp}
Y.~Zhang, T.~Nelson and S.~Tretiak, \emph{J. Chem. Phys.}, 2019, \textbf{151},
  154109\relax
\mciteBstWouldAddEndPuncttrue
\mciteSetBstMidEndSepPunct{\mcitedefaultmidpunct}
{\mcitedefaultendpunct}{\mcitedefaultseppunct}\relax
\EndOfBibitem
\bibitem[Tichauer \emph{et~al.}(2021)Tichauer, Feist, and
  Groenhof]{Tichauer2021JCP}
R.~H. Tichauer, J.~Feist and G.~Groenhof, \emph{J. Chem. Phys.}, 2021,
  \textbf{154}, 104112\relax
\mciteBstWouldAddEndPuncttrue
\mciteSetBstMidEndSepPunct{\mcitedefaultmidpunct}
{\mcitedefaultendpunct}{\mcitedefaultseppunct}\relax
\EndOfBibitem
\bibitem[Groenhof and Toppari(2018)]{Groenhof2018JPCL}
G.~Groenhof and J.~J. Toppari, \emph{J. Phys. Chem. Lett.}, 2018, \textbf{9},
  4848--4851\relax
\mciteBstWouldAddEndPuncttrue
\mciteSetBstMidEndSepPunct{\mcitedefaultmidpunct}
{\mcitedefaultendpunct}{\mcitedefaultseppunct}\relax
\EndOfBibitem
\bibitem[Luk \emph{et~al.}(2017)Luk, Feist, Toppari, and Groenhof]{Luk2017JCTC}
H.~L. Luk, J.~Feist, J.~J. Toppari and G.~Groenhof, \emph{J. Chem. Theory
  Comput.}, 2017, \textbf{13}, 4324--4335\relax
\mciteBstWouldAddEndPuncttrue
\mciteSetBstMidEndSepPunct{\mcitedefaultmidpunct}
{\mcitedefaultendpunct}{\mcitedefaultseppunct}\relax
\EndOfBibitem
\bibitem[Groenhof \emph{et~al.}(2019)Groenhof, Climent, Feist, Morozov, and
  Toppari]{Groenhof2019JPCL}
G.~Groenhof, C.~Climent, J.~Feist, D.~Morozov and J.~J. Toppari, \emph{J. Phys.
  Chem. Lett.}, 2019, \textbf{10}, 5476--5483\relax
\mciteBstWouldAddEndPuncttrue
\mciteSetBstMidEndSepPunct{\mcitedefaultmidpunct}
{\mcitedefaultendpunct}{\mcitedefaultseppunct}\relax
\EndOfBibitem
\bibitem[Tichauer \emph{et~al.}(2022)Tichauer, Morozov, Sokolovskii, Toppari,
  and Groenhof]{Tichauer2022JPCL}
R.~H. Tichauer, D.~Morozov, I.~Sokolovskii, J.~J. Toppari and G.~Groenhof,
  \emph{J. Phys. Chem. Lett.}, 2022, \textbf{13}, 6259--6267\relax
\mciteBstWouldAddEndPuncttrue
\mciteSetBstMidEndSepPunct{\mcitedefaultmidpunct}
{\mcitedefaultendpunct}{\mcitedefaultseppunct}\relax
\EndOfBibitem
\bibitem[Tavis and Cummings(1968)]{Tavis1968PR}
M.~Tavis and F.~W. Cummings, \emph{Phys. Rev.}, 1968, \textbf{170},
  379--384\relax
\mciteBstWouldAddEndPuncttrue
\mciteSetBstMidEndSepPunct{\mcitedefaultmidpunct}
{\mcitedefaultendpunct}{\mcitedefaultseppunct}\relax
\EndOfBibitem
\bibitem[Rokaj \emph{et~al.}(2018)Rokaj, Welakuh, Ruggenthaler, and
  Rubio]{Rokaj2018JPB}
V.~Rokaj, D.~M. Welakuh, M.~Ruggenthaler and A.~Rubio, \emph{J. Phys. B: At.
  Mol. Opt. Phys.}, 2018, \textbf{51}, 034005\relax
\mciteBstWouldAddEndPuncttrue
\mciteSetBstMidEndSepPunct{\mcitedefaultmidpunct}
{\mcitedefaultendpunct}{\mcitedefaultseppunct}\relax
\EndOfBibitem
\bibitem[Mandal \emph{et~al.}(2020)Mandal, Krauss, and
  Huo]{Mandal_QED_eT_JPCB2020}
A.~Mandal, T.~D. Krauss and P.~Huo, \emph{J. Phys. Chem. B}, 2020,
  \textbf{124}, 6321--6340\relax
\mciteBstWouldAddEndPuncttrue
\mciteSetBstMidEndSepPunct{\mcitedefaultmidpunct}
{\mcitedefaultendpunct}{\mcitedefaultseppunct}\relax
\EndOfBibitem
\bibitem[Flick and Narang(2020)]{Flick2020JCP}
J.~Flick and P.~Narang, \emph{J. Chem. Phys.}, 2020, \textbf{153}, 094116\relax
\mciteBstWouldAddEndPuncttrue
\mciteSetBstMidEndSepPunct{\mcitedefaultmidpunct}
{\mcitedefaultendpunct}{\mcitedefaultseppunct}\relax
\EndOfBibitem
\bibitem[Wang \emph{et~al.}(2021)Wang, Neuman, Flick, and
  Narang]{wang_dissipative_2021JCP}
D.~S. Wang, T.~Neuman, J.~Flick and P.~Narang, \emph{J. Chem. Phys.}, 2021,
  \textbf{154}, 104109\relax
\mciteBstWouldAddEndPuncttrue
\mciteSetBstMidEndSepPunct{\mcitedefaultmidpunct}
{\mcitedefaultendpunct}{\mcitedefaultseppunct}\relax
\EndOfBibitem
\bibitem[McTague and Foley(2022)]{mctague_nhCIS_JCP2022}
J.~McTague and J.~J. Foley, \emph{J. Chem. Phys.}, 2022, \textbf{156},
  154103\relax
\mciteBstWouldAddEndPuncttrue
\mciteSetBstMidEndSepPunct{\mcitedefaultmidpunct}
{\mcitedefaultendpunct}{\mcitedefaultseppunct}\relax
\EndOfBibitem
\bibitem[Vu \emph{et~al.}(2022)Vu, McLeod, Hanson, and
  DePrince]{Vu_enhanced_JPCA2022}
N.~Vu, G.~M. McLeod, K.~Hanson and A.~E. DePrince, \emph{J. Phys. Chem. A},
  2022, \textbf{126}, 9303--9312\relax
\mciteBstWouldAddEndPuncttrue
\mciteSetBstMidEndSepPunct{\mcitedefaultmidpunct}
{\mcitedefaultendpunct}{\mcitedefaultseppunct}\relax
\EndOfBibitem
\bibitem[Haugland \emph{et~al.}(2020)Haugland, Ronca, Kj{\o}nstad, Rubio, and
  Koch]{haugland2020PRX}
T.~S. Haugland, E.~Ronca, E.~F. Kj{\o}nstad, A.~Rubio and H.~Koch, \emph{Phys.
  Rev. X}, 2020, \textbf{10}, 041043\relax
\mciteBstWouldAddEndPuncttrue
\mciteSetBstMidEndSepPunct{\mcitedefaultmidpunct}
{\mcitedefaultendpunct}{\mcitedefaultseppunct}\relax
\EndOfBibitem
\bibitem[Haugland \emph{et~al.}(2021)Haugland, Sch{\"{a}}fer, Ronca, Rubio, and
  Koch]{Haugland_intermolecular_JCP2021}
T.~S. Haugland, C.~Sch{\"{a}}fer, E.~Ronca, A.~Rubio and H.~Koch, \emph{J.
  Chem. Phys.}, 2021, \textbf{154}, 094113\relax
\mciteBstWouldAddEndPuncttrue
\mciteSetBstMidEndSepPunct{\mcitedefaultmidpunct}
{\mcitedefaultendpunct}{\mcitedefaultseppunct}\relax
\EndOfBibitem
\bibitem[DePrince(2022)]{deprince_IP_EA_2021JCP}
A.~E. DePrince, \emph{J. Chem. Phys.}, 2022, \textbf{154}, 094112\relax
\mciteBstWouldAddEndPuncttrue
\mciteSetBstMidEndSepPunct{\mcitedefaultmidpunct}
{\mcitedefaultendpunct}{\mcitedefaultseppunct}\relax
\EndOfBibitem
\bibitem[Liebenthal \emph{et~al.}(2022)Liebenthal, Vu, and
  DePrince]{liebenthal_EOMCC_2022JCP}
M.~D. Liebenthal, N.~Vu and A.~E. DePrince, \emph{J. Chem. Phys.}, 2022,
  \textbf{156}, 054105\relax
\mciteBstWouldAddEndPuncttrue
\mciteSetBstMidEndSepPunct{\mcitedefaultmidpunct}
{\mcitedefaultendpunct}{\mcitedefaultseppunct}\relax
\EndOfBibitem
\bibitem[Flick \emph{et~al.}(2017)Flick, Appel, Ruggenthaler, and
  Rubio]{Flick2017JCTC}
J.~Flick, H.~Appel, M.~Ruggenthaler and A.~Rubio, \emph{J. Chem. Theory
  Comput.}, 2017, \textbf{13}, 1616--1625\relax
\mciteBstWouldAddEndPuncttrue
\mciteSetBstMidEndSepPunct{\mcitedefaultmidpunct}
{\mcitedefaultendpunct}{\mcitedefaultseppunct}\relax
\EndOfBibitem
\bibitem[Philbin(2014)]{philbin2014AmerJourPhys}
T.~G. Philbin, \emph{Amer. J. Phys.}, 2014, \textbf{82}, 742--748\relax
\mciteBstWouldAddEndPuncttrue
\mciteSetBstMidEndSepPunct{\mcitedefaultmidpunct}
{\mcitedefaultendpunct}{\mcitedefaultseppunct}\relax
\EndOfBibitem
\bibitem[Riso \emph{et~al.}(2022)Riso, Haugland, Ronca, and
  Koch]{riso_molecularQED_NatComm2022}
R.~R. Riso, T.~S. Haugland, E.~Ronca and H.~Koch, \emph{Nat. Commun.}, 2022,
  \textbf{13}, 1368\relax
\mciteBstWouldAddEndPuncttrue
\mciteSetBstMidEndSepPunct{\mcitedefaultmidpunct}
{\mcitedefaultendpunct}{\mcitedefaultseppunct}\relax
\EndOfBibitem
\bibitem[Mandal and Huo(2019)]{Mandal2019JPCL}
A.~Mandal and P.~Huo, \emph{J. Phys. Chem. Lett.}, 2019, \textbf{10},
  5519--5529\relax
\mciteBstWouldAddEndPuncttrue
\mciteSetBstMidEndSepPunct{\mcitedefaultmidpunct}
{\mcitedefaultendpunct}{\mcitedefaultseppunct}\relax
\EndOfBibitem
\bibitem[Riso \emph{et~al.}(2022)Riso, Haugland, Ronca, and
  Koch]{riso_QEDionization_JCP2022}
R.~R. Riso, T.~S. Haugland, E.~Ronca and H.~Koch, \emph{J. Chem. Phys.}, 2022,
  \textbf{156}, 234103\relax
\mciteBstWouldAddEndPuncttrue
\mciteSetBstMidEndSepPunct{\mcitedefaultmidpunct}
{\mcitedefaultendpunct}{\mcitedefaultseppunct}\relax
\EndOfBibitem
\bibitem[Pellegrini \emph{et~al.}(2015)Pellegrini, Flick, Tokatly, Appel, and
  Rubio]{pellegrini_OEP_PRL2015}
C.~Pellegrini, J.~Flick, I.~V. Tokatly, H.~Appel and A.~Rubio, \emph{Phys. Rev.
  Lett.}, 2015, \textbf{115}, 093001\relax
\mciteBstWouldAddEndPuncttrue
\mciteSetBstMidEndSepPunct{\mcitedefaultmidpunct}
{\mcitedefaultendpunct}{\mcitedefaultseppunct}\relax
\EndOfBibitem
\bibitem[Ruggenthaler \emph{et~al.}(2014)Ruggenthaler, Flick, Pellegrini,
  Appel, Tokatly, and Rubio]{ruggenthaler_QEDFT_2014}
M.~Ruggenthaler, J.~Flick, C.~Pellegrini, H.~Appel, I.~V. Tokatly and A.~Rubio,
  \emph{Phys. Rev. A}, 2014, \textbf{90}, 012508\relax
\mciteBstWouldAddEndPuncttrue
\mciteSetBstMidEndSepPunct{\mcitedefaultmidpunct}
{\mcitedefaultendpunct}{\mcitedefaultseppunct}\relax
\EndOfBibitem
\bibitem[Ruggenthaler \emph{et~al.}(2018)Ruggenthaler, Tancogne-Dejean, Flick,
  Appel, and Rubio]{ruggenthaler_QEDSpectra_NatRevChem2018}
M.~Ruggenthaler, N.~Tancogne-Dejean, J.~Flick, H.~Appel and A.~Rubio,
  \emph{Nat. Rev. Chem.}, 2018, \textbf{2}, 1--16\relax
\mciteBstWouldAddEndPuncttrue
\mciteSetBstMidEndSepPunct{\mcitedefaultmidpunct}
{\mcitedefaultendpunct}{\mcitedefaultseppunct}\relax
\EndOfBibitem
\bibitem[Flick \emph{et~al.}(2015)Flick, Ruggenthaler, Appel, and
  Rubio]{flick_KSQED_PNAS2015}
J.~Flick, M.~Ruggenthaler, H.~Appel and A.~Rubio, \emph{Proc. Nat. Acad. Sci.},
  2015, \textbf{112}, 15285--15290\relax
\mciteBstWouldAddEndPuncttrue
\mciteSetBstMidEndSepPunct{\mcitedefaultmidpunct}
{\mcitedefaultendpunct}{\mcitedefaultseppunct}\relax
\EndOfBibitem
\bibitem[Li \emph{et~al.}(2022)Li, Tao, and
  Hammes-Schiffer]{Li_QEDNEO_JCTC2022}
T.~E. Li, Z.~Tao and S.~Hammes-Schiffer, \emph{J. Chem. Theory Comput.}, 2022,
  \textbf{18}, 2774--2784\relax
\mciteBstWouldAddEndPuncttrue
\mciteSetBstMidEndSepPunct{\mcitedefaultmidpunct}
{\mcitedefaultendpunct}{\mcitedefaultseppunct}\relax
\EndOfBibitem
\bibitem[Li and Hammes-Schiffer(2023)]{li_QEDNEO_2023}
T.~E. Li and S.~Hammes-Schiffer, \emph{J. Chem. Phys.}, 2023, \textbf{158},
  114118\relax
\mciteBstWouldAddEndPuncttrue
\mciteSetBstMidEndSepPunct{\mcitedefaultmidpunct}
{\mcitedefaultendpunct}{\mcitedefaultseppunct}\relax
\EndOfBibitem
\bibitem[Mordovina \emph{et~al.}(2020)Mordovina, Bungey, Appel, Knowles, Rubio,
  and Manby]{mordovina_QEDCC_PRR2020}
U.~Mordovina, C.~Bungey, H.~Appel, P.~J. Knowles, A.~Rubio and F.~R. Manby,
  \emph{Phys. Rev. Research}, 2020, \textbf{2}, 023262\relax
\mciteBstWouldAddEndPuncttrue
\mciteSetBstMidEndSepPunct{\mcitedefaultmidpunct}
{\mcitedefaultendpunct}{\mcitedefaultseppunct}\relax
\EndOfBibitem
\bibitem[Fregoni \emph{et~al.}(2021)Fregoni, Haugland, Pipolo, Giovannini,
  Koch, and Corni]{fregoni_QEDCCPlasmon_NanoLett2021}
J.~Fregoni, T.~S. Haugland, S.~Pipolo, T.~Giovannini, H.~Koch and S.~Corni,
  \emph{Nano Lett.}, 2021, \textbf{21}, 6664--6670\relax
\mciteBstWouldAddEndPuncttrue
\mciteSetBstMidEndSepPunct{\mcitedefaultmidpunct}
{\mcitedefaultendpunct}{\mcitedefaultseppunct}\relax
\EndOfBibitem
\bibitem[Philbin(2014)]{Philbin:2014jcp}
T.~G. Philbin, \emph{Amer. J. Phys.}, 2014, \textbf{82}, 742--748\relax
\mciteBstWouldAddEndPuncttrue
\mciteSetBstMidEndSepPunct{\mcitedefaultmidpunct}
{\mcitedefaultendpunct}{\mcitedefaultseppunct}\relax
\EndOfBibitem
\bibitem[White \emph{et~al.}(2020)White, Gao, Minnich, and Chan]{jcp0033132}
A.~F. White, Y.~Gao, A.~J. Minnich and G.~K.-L. Chan, \emph{J. Chem. Phys.},
  2020, \textbf{153}, 224112\relax
\mciteBstWouldAddEndPuncttrue
\mciteSetBstMidEndSepPunct{\mcitedefaultmidpunct}
{\mcitedefaultendpunct}{\mcitedefaultseppunct}\relax
\EndOfBibitem
\bibitem[Li \emph{et~al.}(2021)Li, Cui, Subotnik, and
  Nitzan]{Li_MolPolRev_ARPC2021}
T.~E. Li, B.~Cui, J.~E. Subotnik and A.~Nitzan, \emph{Annu. Rev. Phys. Chem.},
  2021, \textbf{73}, 43--71\relax
\mciteBstWouldAddEndPuncttrue
\mciteSetBstMidEndSepPunct{\mcitedefaultmidpunct}
{\mcitedefaultendpunct}{\mcitedefaultseppunct}\relax
\EndOfBibitem
\bibitem[Taflove and Hagness(2005)]{taflove2005computational}
A.~Taflove and S.~C. Hagness, \emph{Computational Electrodynamics: The
  Finite-Difference Time-Domain Method}, Artech House, 3rd edn, 2005\relax
\mciteBstWouldAddEndPuncttrue
\mciteSetBstMidEndSepPunct{\mcitedefaultmidpunct}
{\mcitedefaultendpunct}{\mcitedefaultseppunct}\relax
\EndOfBibitem
\bibitem[Sch{\"a}fer and Johansson(2022)]{Christian2022PRL}
C.~Sch{\"a}fer and G.~Johansson, \emph{Phys. Rev. Lett.}, 2022, \textbf{128},
  156402\relax
\mciteBstWouldAddEndPuncttrue
\mciteSetBstMidEndSepPunct{\mcitedefaultmidpunct}
{\mcitedefaultendpunct}{\mcitedefaultseppunct}\relax
\EndOfBibitem
\bibitem[Sch{\"a}fer \emph{et~al.}(2021)Sch{\"a}fer, Buchholz, Penz,
  Ruggenthaler, and Rubio]{schafer_makingQEDFTFunc_PNAS2021}
C.~Sch{\"a}fer, F.~Buchholz, M.~Penz, M.~Ruggenthaler and A.~Rubio, \emph{Proc.
  Nat. Acad. Sci.}, 2021, \textbf{118}, e2110464118\relax
\mciteBstWouldAddEndPuncttrue
\mciteSetBstMidEndSepPunct{\mcitedefaultmidpunct}
{\mcitedefaultendpunct}{\mcitedefaultseppunct}\relax
\EndOfBibitem
\bibitem[Peterson \emph{et~al.}(1998)Peterson, Ray, and
  Mittra]{peterson1998computational}
A.~F. Peterson, S.~L. Ray and R.~Mittra, \emph{IEEE Transactions on Antennas
  and Propagation}, 1998, \textbf{46}, 276--294\relax
\mciteBstWouldAddEndPuncttrue
\mciteSetBstMidEndSepPunct{\mcitedefaultmidpunct}
{\mcitedefaultendpunct}{\mcitedefaultseppunct}\relax
\EndOfBibitem
\bibitem[Gaita-Ari{\~n}o \emph{et~al.}(2019)Gaita-Ari{\~n}o, Luis, Hill, and
  Coronado]{Gaita-Arino:2019ti}
A.~Gaita-Ari{\~n}o, F.~Luis, S.~Hill and E.~Coronado, \emph{Nat. Chem.}, 2019,
  \textbf{11}, 301--309\relax
\mciteBstWouldAddEndPuncttrue
\mciteSetBstMidEndSepPunct{\mcitedefaultmidpunct}
{\mcitedefaultendpunct}{\mcitedefaultseppunct}\relax
\EndOfBibitem
\bibitem[Sanvitto and K{\'e}na-Cohen(2016)]{Sanvitto:2016td}
D.~Sanvitto and S.~K{\'e}na-Cohen, \emph{Nat. Mater.}, 2016, \textbf{15},
  1061--1073\relax
\mciteBstWouldAddEndPuncttrue
\mciteSetBstMidEndSepPunct{\mcitedefaultmidpunct}
{\mcitedefaultendpunct}{\mcitedefaultseppunct}\relax
\EndOfBibitem
\bibitem[Zheng \emph{et~al.}(2007)Zheng, Wang, Yam, Mo, and Chen]{Zheng2007prb}
X.~Zheng, F.~Wang, C.~Y. Yam, Y.~Mo and G.~Chen, \emph{Phys. Rev. B}, 2007,
  \textbf{75}, 195127\relax
\mciteBstWouldAddEndPuncttrue
\mciteSetBstMidEndSepPunct{\mcitedefaultmidpunct}
{\mcitedefaultendpunct}{\mcitedefaultseppunct}\relax
\EndOfBibitem
\bibitem[Zhang \emph{et~al.}(2013)Zhang, Chen, and Chen]{zhang2013first}
Y.~Zhang, S.~Chen and G.~Chen, \emph{Phys. Rev. B}, 2013, \textbf{87},
  085110\relax
\mciteBstWouldAddEndPuncttrue
\mciteSetBstMidEndSepPunct{\mcitedefaultmidpunct}
{\mcitedefaultendpunct}{\mcitedefaultseppunct}\relax
\EndOfBibitem
\bibitem[Zhang \emph{et~al.}(2013)Zhang, Yam, and Chen]{zhang2013dissipative}
Y.~Zhang, C.~Y. Yam and G.~Chen, \emph{J. Chem. Phys.}, 2013, \textbf{138},
  164121\relax
\mciteBstWouldAddEndPuncttrue
\mciteSetBstMidEndSepPunct{\mcitedefaultmidpunct}
{\mcitedefaultendpunct}{\mcitedefaultseppunct}\relax
\EndOfBibitem
\bibitem[Chen \emph{et~al.}(2018)Chen, Zhang, Chen, and Franco]{chen2018stark}
L.~Chen, Y.~Zhang, G.~Chen and I.~Franco, \emph{Nat. Commun.}, 2018,
  \textbf{9}, 1--12\relax
\mciteBstWouldAddEndPuncttrue
\mciteSetBstMidEndSepPunct{\mcitedefaultmidpunct}
{\mcitedefaultendpunct}{\mcitedefaultseppunct}\relax
\EndOfBibitem
\bibitem[Boolakee \emph{et~al.}(2022)Boolakee, Heide, Garz{\'o}n-Ram{\'\i}rez,
  Weber, Franco, and Hommelhoff]{Boolakee:2022tm}
T.~Boolakee, C.~Heide, A.~Garz{\'o}n-Ram{\'\i}rez, H.~B. Weber, I.~Franco and
  P.~Hommelhoff, \emph{Nature}, 2022, \textbf{605}, 251--255\relax
\mciteBstWouldAddEndPuncttrue
\mciteSetBstMidEndSepPunct{\mcitedefaultmidpunct}
{\mcitedefaultendpunct}{\mcitedefaultseppunct}\relax
\EndOfBibitem
\end{mcitethebibliography}
\bibliographystyle{rsc} 

\end{document}